**JAN KOCHANOWSKI UNIVERSITY**

**DOCTORAL SCHOOL**

**FACULTY OF NATURAL SCIENCES**

**PHYSICAL SCIENCES**

**Shahriyar Jafarzade**

# Phenomenology of light mesons with $J = 2, 3$

**Doctoral dissertation written under the supervision of**

**Prof. dr. hab. Francesco Giacosa**

Kielce 2023

This doctoral thesis was prepared within the research project 'Development Accelerator of Jan Kochanowski University in Kielce,' No. POWR.03.05.00-00-Z212/18, co-financed by the European Union funds under the European Social Fund.

# Abstract

In this thesis, I present the results for light mesons with $J = 2, 3$

$$J^{PC} = 2^{++} = \{a_2(1320), K_2^*(1430), f_2(1270), f_2'(1525), G_2(???)\},$$
$$J^{PC} = 2^{--} = \{\rho_2(???), K_2(1820)/K_2(1770), \phi_2(???), \omega_2(???)\},$$
$$J^{PC} = 3^{--} = \{\rho_3(1690), K_3^*(1780), \phi_3(1850), \omega_3(1670), G_3(???)\},$$

within an effective hadronic model, the so-called extended Linear Sigma Model (eLSM). This model is based on the approximate chiral symmetry of QCD. Qualitative agreement between model results and PDG as well as LQCD data is obtained. Various predictions for the radiative decays can be tested in e.g., GlueX and CLAS12 experiments at Jefferson Lab.

After applying the chiral model for well-established spin-2 mesons (with $J^{PC} = 2^{++}$), I move on to their chiral partners (the axial-tensor mesons with $J^{PC} = 2^{++}$), where the resonances are still missing. Large decay widths are predicted in the chiral model as well as in LQCD simulations.

Furthermore, I describe the decays of the spin-2 tensor glueball $G_2(???)$ within the eLSM, which can be helpful in future experimental searches e.g., BESIII and LHCb experiments. Some isoscalar tensor resonances of PDG with a mass of around 2 GeV are studied, and at present, the best tensor glueball candidate turns out to be the broad resonance $f_2(1950)$.

In the case of the ground-state spin-3 mesons ($J^{PC} = 3^{--}$), our chiral model is limited to the $SU(3)$ flavor symmetry because of the yet unknown chiral partners ($J^{PC} = 3^{++}$). The effective model results are in good agreement with the PDG data and LQCD results, and the predictions for the radiative decays can be interesting for photoproduction experiments such as GlueX and CLAS12. Additional estimates for the decay ratios of the $G_3(???)$ tensor glueball with $J^{PC} = 3^{--}$ are also presented.

The role of the glueball spectrum at non-zero temperature is explored within the Glueball Resonance Gas model below the critical temperature of the pure Yang-Mills sector of the QCD phase diagram. It turns out that the latest glueball spectrum from LQCD can well describe the pressure up to almost the critical temperature.

# Phenomenologia lekkich mezonòw z J=2,3

## Streszczenie

W niniejszej rozprawie przedstawiam wyniki dla lekkich mezonów z J = 2, 3

$$
\begin{aligned}
J^{PC} &= 2^{++} = \{a_2(1320), K_2^*(1430), f_2(1270), f_2'(1525), G_2(???)\}\,, \\
J^{PC} &= 2^{--} = \{\rho_2(???), K_2(1820)/K_2(1770), \phi_2(???), \omega_2(???)\}\,, \\
J^{PC} &= 3^{--} = \{\rho_3(1690), K_3^*(1780), \phi_3(1850), \omega_3(1670), G_3(???)\}\,,
\end{aligned}
$$

w ramach efektywnego modelu hadronowego, tzw. rozszerzonego liniowego modelu Sigma. Model ten oparty jest na przybliżonej symetrii chiralnej QCD. Uzyskano jakościową zgodność pomiędzy wynikami modelu a danymi PDG oraz LQCD. Liczne przewidywania dla rozpadów radiacyjnych mogą być przetestowane w przyszłych eksperymentach, m.in. GlueX i CLAS12 w Jefferson Lab.

Po weryfikacji działania modelu chiralnego na dobrze poznanych mezonach o spinie 2, przechodzę do sektora ich chiralnych partnerów (mezony aksjalne tensorowe o $J^{PC} = 2^{++}$), gdzie rezonansów wciąż brakuje. Duże szerokości rozpadów zostały przewidziane zarówno w modelu chiralnym jak i w symulacjach LQCD.

Ponadto opisuję brakujący glueball tensorowy $G_2(???)$ o spinie 2, w ramach modelu chiralnego, który może być użyteczny m. in. dla eksperymentów BESIII i LHCb. Badane są niektóre izoskalarne tensorowe rezonanse, zestawione w PDG, o spinie 2 i masie około 2 GeV. Dotychczasowe wyniki pokazują, że najlepszym kandydatem na glueball tensorowy okazuje się być szeroki rezonans$f_2(1950)$.

W przypadku mezonów w stanie podstawowym o spinie 3 ($J^{PC} = 3^{--}$), nasz model chiralny ma ograniczenia symetrii zapachu SU(3) ze względu na ciagle niepoznanych partnerów chiralnych ($J^{PC} = 3^{++}$). Wyniki modelu efektywnego pozostają w zgodzie z danymi PDG i wynikami LQCD. Przewidywania dla rozpadów radiacyjnych mogą być interesujace dla eksperymentów opartych na fotoprodukcji, tj.

GlueX i CLAS12. W pracy przedstawiono również dodatkowe oszacowania dla stosunków rozpadów gluaballa tensorowego $G_3(???)$ o $J^{PC} = 3^{--}$.

Rola spektrum glueballi w niezerowej temperaturze jest badana w ramach modelu Glueballowego rezonansowego gazu w obszarze poniżej krytycznej temperatury czystego sektora Yang-Millsa na diagramie fazowym QCD. Okazuje się, że najnowsze widmo glueballi z LQCD może dobrze opisywać ciśnienie prawie do samej temperatury krytycznej.

# Acknowledgements

There are a lot of people who deserve my thanks, but some are particularly important.

First and foremost, I am thankful to Francesco Giacosa, my supervisor, working with me on Refs [1–4], sharing his insights with me, and providing me a lot of support throughout my Ph.D. studies.

Robert Pisarski's acceptance of my doctoral internship, his hospitality, our work on Instantons, and his continuing support of me are all things for which I am thankful.

I am thankful to Adrian Königstein for collaboration in [1], Milena Piotrowska, Arthur Vereijken in [2,3] and Enrico Trotti in [4]. I appreciate Shahin Mamedov and Zeynab Hashimli's cooperation on the holographic QCD.

Special thanks to Wojciech Broniowski and Stanislaw Mrówczyski, two of my teachers, for their support and advice. I am grateful for the interesting discussions that I had with Maciej Rybczynski, Subhasis Samanta, Vanamali Shastry, and Leonardo Tinti. Haradhan Adhikary, Ali Bazgir, Valeria Renya, Uzair Shah, Regina Stachura, Enrico Trotti and many others from the Institute of Physics deserve special acknowledgment. Enrico Trotti taught me how to cook, and I'm grateful that he was a kind flatmate.

I want to express my gratitude to my teachers Azar Ahmadov, Gurban Ahmedov, Emil Akhmedov, and Ilmar Gahramanov for their guidance until my doctoral studies. Special thanks to Ilmar Gahramanov, who inspired me to become a physicist and provided me with unending support during my undergraduate years.

My mother Gulshan, sister Arasta, and wife Zeynab have a lot to be thankful for.

# Contents

# Chapter 1

# Introduction

## 1.1 QCD: a brief recall

The discovery of various short-living resonances in the 1960s motivated the postulation of quarks and gluons in order to understand that, at first sight, complicated "zoo" of particles [5, 6]. However, another concept, the so-called color of quarks and gluons, is introduced to explain the resonance $\Delta^{++}(1232)$, which is made of three quarks of the type up (with aligned spin). Because of the Pauli principle, the wave function has to be anti-symmetric, which makes the color part of the wave function necessary. According to the hypothesis about color confinement, only color singlet objects (i.e. white) can be seen in particle detectors.

Quarks are fermions with half spin and carry $\pm\frac{1}{3}$ or $\pm\frac{2}{3}$ units of the electric charge. Mathematically, the quark fields take the form

$$q_i = \begin{pmatrix} q_{i,r} \\ q_{i,g} \\ q_{i,b} \end{pmatrix} , \tag{1.1}$$

where $r, g, b$ stands for "red", "green" and "blue" and each $q_{i,c}$ is a Dirac spinor. Moreover, $i = u, d, s, c, b, t$ represents the so-called quark flavor.

As we shall see later on, the bound states of quark and antiquark states exist in nature because they appear as a color singlet:

$$3 \otimes \overline{3} = 1 \oplus 8 \,. \tag{1.2}$$

Gluons are spin-1 massless gauge bosons, which mediate the color in the strong interaction where the color has to be conserved.

Quantum Chromodynamics (QCD) is a renormalizable quantum field theory of the strong interaction that describes the interaction of quarks and gluons. The QCD Lagrangian for $N_f$ number of flavors involves the quark $q_i$ and anti-quark $\bar{q}_i$ fields

$$\mathcal{L}_{QCD} = \sum_{i=1}^{N_f} \text{tr}\Big(\bar{q}_i(x)(i\gamma_\mu D^\mu - m_i)q_i(x)\Big) + \mathcal{L}_{YM}, \tag{1.3}$$

where the covariant derivative with the QCD coupling $g$ has the following form:

$$D_\mu := \partial_\mu - igA_\mu(x)\,. \tag{1.4}$$

The gluonic fields

$$A_\mu(x) := \sum_{a=1}^{8} A_\mu^a(x)\frac{\lambda^a}{2}\,, \tag{1.5}$$

are represented in terms of the eight Gell-Mann matrices $\lambda_a$. The pure Yang-Mills (YM) part of the QCD Lagrangian is defined as:

$$\mathcal{L}_{YM} = -\frac{1}{2}\text{Tr}\Big[G_{\mu\nu}(x)G^{\mu\nu}(x)\Big] \tag{1.6}$$

where

$$G_{\mu\nu}(x) := D_\mu A_\nu(x) - D_\nu A_\mu(x) - ig[A_\mu(x), A_\nu(x)]\,. \tag{1.7}$$

By considering $N_f = 3$ (thus, with quarks $u, d, s$), the QCD Lagrangian in Eq.(1.3) has the following classical symmetries:

- Chiral symmetry $U(3)_L \times U(3)_R$ and dilation symmetry in the massless quark limit[1] ($m_u = m_d = m_s = 0$);
- Flavor symmetry $SU(3)_F$ in the equal mass limit of the three light quarks ($m_u = m_d = m_s \neq 0$);

[1]This is the so-called the chiral limit of QCD. Including quantum fluctuations leads to the anomalous breaking of the axial ($U_A(1)$) and dilation symmetry, the so-called chiral and trace anomalies.

- Color symmetry $SU(3)_c$, which is an exact local symmetry and is expected to lead to confinement.

As a result of confinement, we only measure hadrons in the detectors rather than separate quarks and gluons. Hadrons can be classified as mesons with integer spins and baryons with half-integer spins. Mesons themselves are classified as conventional (quark-antiquark objects) and non-conventional mesons, such as glueballs (only gluons), and hybrids (quark-antiquark and gluons). While conventional baryons are made up of three quarks, pentaquarks are examples of non-conventional baryons.

At low energies one works with hadronic degrees of freedom, rather than quarks and gluons, in effective model approaches. Effective models that use the chiral symmetry are, for example, chiral perturbation theory (ChPT) [7–17], in which chiral symmetry is nonlinearly realized and sigma models, in which it is linearly realized [18–23]. I make use of an extension of the latter, called the extended Linear Sigma Model(eLSM), throughout my thesis.

At large distances ($\gtrsim 1$ fm) quarks are confined, and only non-perturbative methods are applicable. Lattice QCD (LQCD) simulation is a well-known example of a non-perturbative approach, which starts from QCD Lagrangian and evaluates certain quantities (i.e. masses of hadrons) on a computer grid (lattice). The comparison of the models above with LQCD is a useful way to understand QCD at low energy.

Upon quantizing QCD and considering the renormalization procedure, the coupling constant $g$ in Eq. (1.4) turns out to be a function of the energy scale $\mu$

$$g^2(\mu) = \frac{48\pi^2}{\left(11N_c - 2N_f\right)\log\left(\frac{\mu}{\Lambda_{\text{YM}}}\right)}, \tag{1.8}$$

where the YM scale is around $\Lambda_{YM} \approx 200$ MeV. At short distances, quarks behave as free particles. This is also a non-trivial feature of QCD and is so-called "asymptotic freedom". Perturbation theory is applicable in this regime.

## 1.2 Conventional mesons

Conventional mesons consist of quark and antiquark pairs. To explain them, let us first present the properties of the three light quarks of QCD. Light (anti-)quarks have the following quantum numbers: the electric charge related to the isospin $I_3$ and hypercharge $Y$:

$$Q = I_3 + \frac{Y}{2}, \tag{1.9}$$

where the hypercharge $Y$ is the sum of the baryon number $B$ and the strangeness quantum number $S$

$$Y := B + S\,. \tag{1.10}$$

We list the quantum numbers for the quarks and anti-quarks in Table 1.1. In addition,

| | $Q$ | $I$ | $I_3$ | $Y$ | $S$ | $B$ |
|---|---|---|---|---|---|---|
| u | $\frac{2}{3}$ | $\frac{1}{2}$ | $\frac{1}{2}$ | $\frac{1}{3}$ | 0 | $\frac{1}{3}$ |
| d | $-\frac{1}{3}$ | $\frac{1}{2}$ | $-\frac{1}{2}$ | $\frac{1}{3}$ | 0 | $\frac{1}{3}$ |
| s | $-\frac{1}{3}$ | 0 | 0 | $-\frac{2}{3}$ | 1 | $\frac{1}{3}$ |
| $\overline{u}$ | $-\frac{2}{3}$ | $\frac{1}{2}$ | $-\frac{1}{2}$ | $-\frac{1}{3}$ | 0 | $-\frac{1}{3}$ |
| $\overline{d}$ | $\frac{1}{3}$ | $\frac{1}{2}$ | $\frac{1}{2}$ | $-\frac{1}{3}$ | 0 | $-\frac{1}{3}$ |
| $\overline{s}$ | $\frac{1}{3}$ | 0 | 0 | $\frac{2}{3}$ | $-1$ | $-\frac{1}{3}$ |

**Table 1.1:** Quantum numbers for the light (anti-)quarks.

conventional mesons carry the total angular momentum $\vec{J}$

$$\vec{J} = \vec{L} + \vec{S}, \tag{1.11}$$

as a sum of the spin $\vec{S}$ and the angular momentum $\vec{L}$, where $S = 0$ refers to anti-parallel quarks and $S = 1$ for parallel quarks, while the angular momentum can take the values $L = 0, 1, 2, \cdots$.

The parity operator (that corresponds to the transformation $\vec{x} \to -\vec{x}$) acting on conventional mesons has eigenvalues as

$$P = (-1)^{L+1}, \tag{1.12}$$

while the charge conjugation operator (particles $\leftrightarrow$ anti-particles) has the following eigenvalues:

$$C = (-1)^{L+S} . \tag{1.13}$$

Strong interaction processes are also G-symmetry invariant. G-symmetry is the combination of $SU(2)$ isospin symmetry and the charge conjugation symmetry [24].

Based on the eigenvalues of the spin-S, we can make the following considerations:

1. In the case of anti-parallel quark-antiquark states

$$|S=0\rangle = \frac{1}{\sqrt{2}}|\uparrow_q\downarrow_{\bar{q}} - \downarrow_q\uparrow_{\bar{q}}\rangle , \tag{1.14}$$

   conventional mesons have the quantum numbers

$$J^{PC} \in \left\{ \mathbf{even}^{-+}, \mathbf{odd}^{+-} \right\} . \tag{1.15}$$

2. The parallel structure of the quark-antiquark pairs

$$|S=1\rangle = |\uparrow_q\uparrow_{\bar{q}}\rangle , \frac{1}{\sqrt{2}}|\uparrow_q\downarrow_{\bar{q}} + \downarrow_q\uparrow_{\bar{q}}\rangle , |\downarrow_q\downarrow_{\bar{q}}\rangle , \tag{1.16}$$

   leads to the following quantum numbers for the conventional mesons

$$J^{PC} \in \left\{ \{0,1,2,\cdots\}^{++}, \{1,2,\cdots\}^{--} \right\} . \tag{1.17}$$

3. The following quantum numbers correspond to so-called exotic states

$$J^{PC} \in \left\{ \mathbf{even}^{+-}, \mathbf{odd}^{-+}, 0^{--} \right\} . \tag{1.18}$$

Namely, they are impossible for $\bar{q}q$ pairs. A notable example is the resonance $\pi_1(1600)$ with $J^{PC} = 1^{-+}$, see e.g., Refs. [25,26].

We list the conventional mesons up to $J=3$ in Table 1.2, where the first row shows the spectroscopic notation of the quantum numbers with radial quantum number $n$, angular momentum $L = S, P, D, \cdots$, as well as $2S+1$ and $J$.

| $n^{2S+1}L_J$ | $J^{PC}$ | $u\bar{d}, d\bar{u}, \frac{d\bar{d}-u\bar{u}}{\sqrt{2}}$ | $u\bar{s}, d\bar{s}, s\bar{d}, s\bar{u}$ | $\approx \frac{u\bar{u}+d\bar{d}}{\sqrt{2}}$ | $\approx s\bar{s}$ | Meson names |
|---|---|---|---|---|---|---|
| $1^1S_0$ | $0^{-+}$ | $\pi$ | $K$ | $\eta(547)$ | $\eta'(958)$ | Pseudoscalar |
| $1^3P_0$ | $0^{++}$ | $a_0(1450)$ | $K_0^\star(1430)$ | $f_0(1370)$ | $f_0(1500)/f_0(1710)$ | Scalar |
| $1^3S_1$ | $1^{--}$ | $\rho(770)$ | $K^\star(892)$ | $\omega(782)$ | $\phi(1020)$ | Vector |
| $1^3P_1$ | $1^{++}$ | $a_1(1260)$ | $K_{1A}$ | $f_1(1285)$ | $f_1'(1420)$ | Axial-vector |
| $1^1P_1$ | $1^{+-}$ | $b_1(1235)$ | $K_{1B}$ | $h_1(1170)$ | $h_1(1415)$ | Pseudovector |
| $1^3D_1$ | $1^{--}$ | $\rho(1700)$ | $K^\star(1680)$ | $\omega(1650)$ | $\boldsymbol{\phi}$**(???)** | Excited-vector |
| $1^3P_2$ | $2^{++}$ | $a_2(1320)$ | $K_2^\star(1430)$ | $f_2(1270)$ | $f_2'(1525)$ | Tensor |
| $1^3D_2$ | $2^{--}$ | $\boldsymbol{\rho_2}$**(???)** | $K_{2A}$ | $\boldsymbol{\omega_2}$**(???)** | $\boldsymbol{\phi_2}$**(???)** | Axial-tensor |
| $1^1D_2$ | $2^{-+}$ | $\pi_2(1670)$ | $K_{2P}$ | $\eta_2(1645)$ | $\eta_2(1870)$ | Pseudotensor |
| $1^3D_3$ | $3^{--}$ | $\rho_3(1690)$ | $K_3^\star(1780)$ | $\omega_3(1670)$ | $\phi_3(1850)$ | Spin-3 Tensor |

**Table 1.2:** List of conventional $q\bar{q}$ mesons following the quark model review of Ref. [27] up to $J = 3$ and for the lowest radial excitation (n=1). Bold entries with question marks have not been assigned to known resonances yet.

## 1.3 Open problems with $J = 2, 3$ mesons

The last four rows of Table 1.2 are phenomenologically interesting because of various reasons:

1. Many decay channels are known for tensor ($J^{PC} = 2^{++}$) and spin-3 tensor ($J^{PC} = 3^{--}$) mesons [27], which allow for further theoretical and experimental tests of the $\bar{q}q$ assignment.
2. Predictions for not-yet measured strong and radiative decay rates are possible.
3. Decays of spin-3 tensor ($J^{PC} = 3^{--}$) and axial-tensor ($J^{PC} = 2^{--}$) mesons have been recently estimated in LQCD calculations [28].
4. Three axial-tensor mesons are still missing in this list.
5. The mixing angle between the isoscalar members of the pseudotensor ($J^{PC} = 2^{-+}$)mesons is unknown.
6. There is an unknown mixing in the kaonic sector between $K_2^\star(1820)$ and $K_2(1770)$.

The decay widths versus masses of the resonances under study are summarized in Figure 1.1.

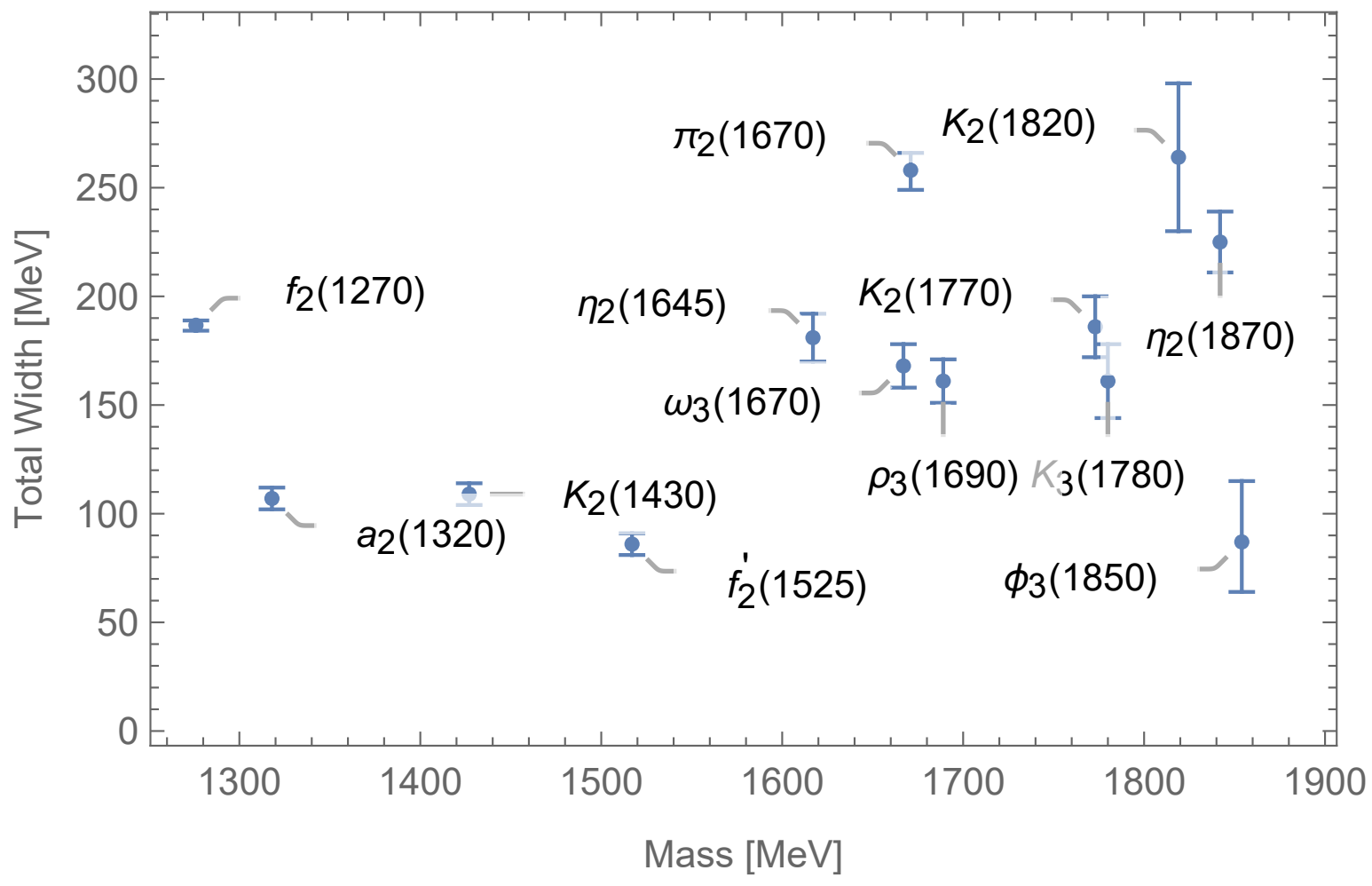

**Figure 1.1:** List of conventional mesons with $2 \leq J \leq 3$.

One type of non-conventional mesons is the so-called "glueball" which is defined as a pure gluonic state. Despite the lack of experimental data, various LQCD simulations observe glueball states in which the $2^{++}$ glueball is the second lightest state after the scalar glueball (see e.g., a recent analysis of the glueball spectrum in Ref. [29]). From a phenomenological perspective, there are several resonances in PDG [27] whose physical masses are close to the LQCD value of 2.2 GeV, such as $f_2(1910)$, $f_2(1950)$, $f_2(2010)$, $f_2(2150)$, $f_J(2220)$ , $f_2(2300)$ and $f_2(2340)$, see Figure 1.2. Which of them,

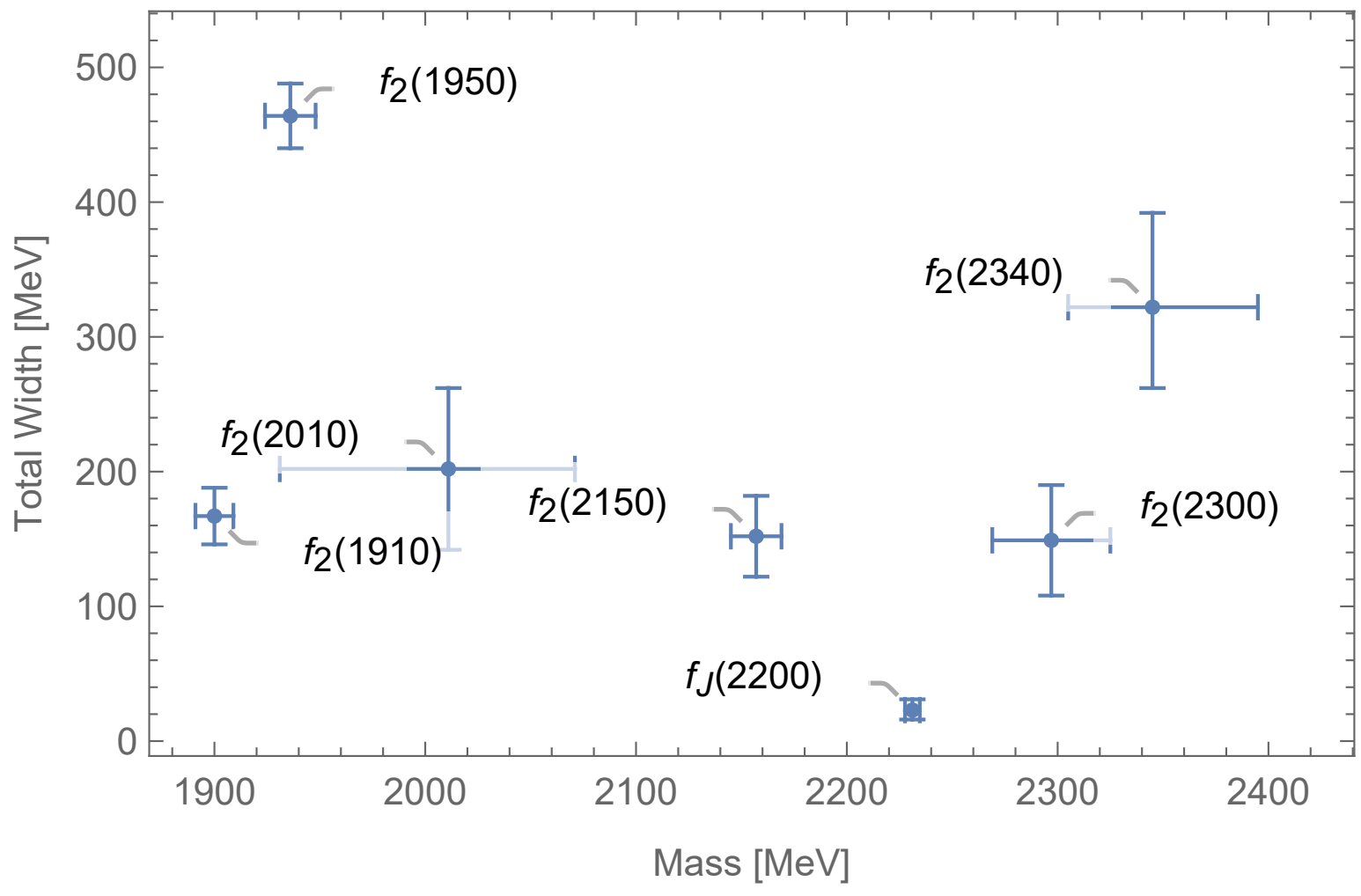

**Figure 1.2:** List of spin-2 resonances with a mass above 1.9 GeV.

if any, could be the best candidate for the tensor glueball? Exploring it is one of our goals.

Figure 1.3 shows the parts of the thesis where some of the aforementioned open questions will be investigated. I will construct effective models to investigate those resonances as follows:

1. Use chiral symmetry to construct an effective model which links tensor and axial-tensor mesons with spin-2.
2. Extend the chiral model to the spin-2 glueball sector.
3. Flavor symmetry-based model for spin-3 mesons and hypothetical spin-3 glueball.

Moreover, I will show that LQCD simulations for the thermodynamical properties of the pure YM sector [30] are explained well within the Glueball Resonance Gas (GRG) based on the spectrum obtained in [29] up to the critical temperature for deconfinement. In addition to the above-mentioned goals, I will also show the role of the spin-2 tensor glueball within the GRG.

There is, of course, a full plethora of theoretical methods devoted to the investigations of the aforementioned problems, such as the already mentioned LQCD [28,31], Holographic QCD [32,33], QCD sum rules [34], Functional Methods [35,36], the Bethe-Salpeter approach of Ref. [37], Chiral Perturbation theory [38], etc. Whenever it is possible, we will compare the results obtained within the employed chiral model (the eLSM) with those from the approaches listed above, especially with LQCD.

## 1.4 Publications

My thesis is based on the following publications [2] [1–4,39]:

- Chapter 3

1. Shahriyar Jafarzade, Arthur Vereijken, Milena Piotrowska, Francesco Giacosa
   "From well-known tensor mesons to yet unknown axial-tensor mesons"
   Phys. Rev. D 106(2022) 3, 036008
   https://doi.org/10.1103/PhysRevD.106.036008
   arXiv:2203.16585 [hep-ph]

[2] Here, we report the correct title of Refs. [1–4,39].

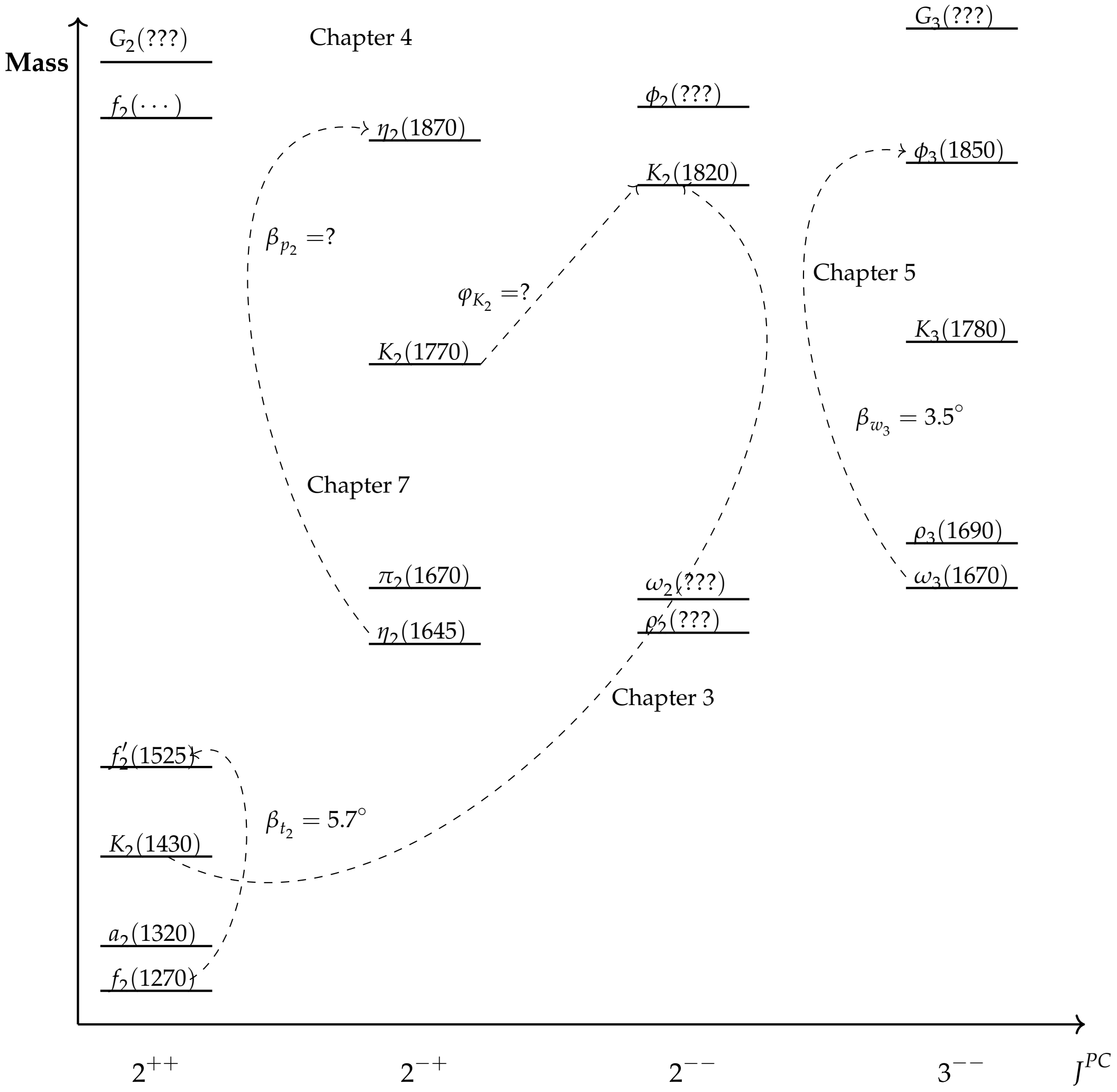

**Figure 1.3:** Ground-state light mesons with $2 \leq J \leq 3$.

2. Shahriyar Jafarzade
   "A phenomenological note on the missing $\rho_2$ meson"
   PoS ICHEP2022 (2022) 903
   https://doi.org/10.22323/1.414.0903
   arXiv:2210.12421 [hep-ph]

   - Chapter 4

3. Arthur Vereijken, Shahriyar Jafarzade, Milena Piotrowska, Francesco Giacosa
   "Is $f_2(1950)$ the tensor glueball?"
   Phys. Rev. D 108 (2023) 1, 014023
   https://doi.org/10.1103/PhysRevD.108.014023
   arXiv:2304.05225 [hep-ph]

- Chapter 5

4. Shahriyar Jafarzade, Adrian Koenigstein, Francesco Giacosa
   "Phenomenology of $J^{PC} = 3^{--}$ tensor mesons"
   Phys. Rev. D 103 (2021) 9, 096027
   https://doi.org/10.1103/PhysRevD.103.096027
   arXiv:2101.03195 [hep-ph]

- Chapter 6

5. Enrico Trotti, Shahriyar Jafarzade, Francesco Giacosa
   "Thermodynamics of the glueball resonance gas"
   Eur. Phys. J. C 83 (2023) 5, 390
   https://doi.org/10.1140/epjc/s10052-023-11557-0
   arXiv:2212.03272 [hep-ph].

## 1.5 Main Results

1. We have studied the ground state spin-2 mesons ($J^{PC} = 2^{++}$)

$$\{a_2(1320), K_2^*(1430), f_2(1270), f_2'(1525)\}$$

   within the extended Linear Sigma Model. Various predictions for hadronic and radiative decay rates are presented. Their quark-antiquark picture is confirmed. Some predicted radiative decay rates can be investigated at e.g., GlueX and CLAS12 experiments at the Jefferson Lab.

2. The chiral partners of the above-mentioned resonances (axial-tensor mesons with $J^{PC} = 2^{--}$ ) have not been yet discovered experimentally. Nevertheless, we have some information about the kaonic member of it, namely, that it is a mixture of $K_2(1820)$ and $K_2(1770)$. Under the assumption that $K_2(1820)$ belongs most dominantly to this nonet, we have obtained the masses of the other missing resonances with the same quantum numbers. While the lightest member of this meson family has a mass of around 1.7 GeV, its decay width is larger than that of its chiral partner. Decay rate results are compared to LQCD outcomes. Our results could be useful for an eventual future identification of these resonances, even though their large expected widths make them difficult to be observed.

3. Another missing resonance for the spin-2 sector is the tensor glueball ($J^{PC} = 2^{++}$). Its mass is expected to be around 2 GeV (LQCD). Indeed, various tensor resonances are listed in PDG with a mass close to that value. Therefore, we have studied the tensor glueball within the eLSM, showing that $f_2(1950)$ is at present, the best glueball candidate, despite being a quite broad resonance. We have also presented some decay ratios for the mass value of 2.21 GeV, which corresponds to an enhancement in the $\pi\pi$ invariant mass spectrum of the recent BESIII data analyses implying a pole at 2.21 GeV. The dominant decay channel for the tensor glueball turns out to be the decay into two vector mesons.

4. The ground-state spin-3 mesons ($J^{PC} = 3^{--}$) are

$$\{\rho_3(1690), K_3^*(1780), \phi_3(1850), \omega_3(1670)\}\,. \tag{1.19}$$

   Their description within the chiral model is limited to the $SU(3)$ flavor symmetry because of their unknown chiral partners. We have constructed an effective model in order to study their decay properties. Interestingly, model results are in good agreement with the PDG data and LQCD results for hadronic decays. Analogously to the spin-2 case, we made some predictions for radiative decays, which can be interesting for photoproduction experiments such as GlueX and CLAS12. We have shown additional predictions for the decay ratios of the spin-3 tensor glueball by using the mass $\sim 3$ GeV estimates from LQCD simulations.

5. The last topic is the non-zero temperature description of the pure Yang-Mills $SU(3)$ theory. The role of the recent glueball spectrum and the interaction between the lightest glueballs are investigated within the Glueball Resonance Gas Model (GRG). Results are compared to the pure Yang-Mills thermodynamic simulations for the pressure in the region below the critical temperature. Further radially excited glueball states, assuming Regge trajectory, and scalar and tensor glueball interactions lead to negligible contributions.

## 1.6 Organization of the thesis

In chapter 1 I have described the main properties of QCD and a list of the conventional mesons, as well as the open problems in mesons with $2 \leq J \leq 3$. Chapter 2 presents the LSM and mesons within certain nonets. The phenomenology of spin-2 mesons and the glueball within the eLSM is discussed in Chapter 3 and Chapter 4, respectively. Chapter 5 is devoted to the ground-state spin-3 mesons and a spin-3 glueball. The role of the glueball states at non-zero $T$ within the GRG is described in Chapter 6. Conclusions and discussions are given in Chapter 7.

# Chapter 2

# Linear Sigma Model and its extension

## 2.1 Introduction

As we mentioned in the previous chapter, effective models are based on the symmetries of QCD and involve hadronic degrees of freedom. One example of an effective model is the so-called Linear Sigma Model (LSM). Historically, it was introduced in order to study the pion-nucleon interaction [18]. The Lagrangian contains the free terms of mesons as well as the nucleon in addition to the Yukawa type interaction between them:

$$\mathcal{L}_{\text{LSM}} = \mathcal{L}_{\text{mesons}} + \mathcal{L}_{\text{nucleons}} + \mathcal{L}_{\text{Yukawa}} \,. \tag{2.1}$$

The explicit forms of the above Lagrangian in terms of the mesons $(\sigma, \vec{\pi})$ and nucleon $\Psi$ read

$$\begin{aligned}
\mathcal{L}_{\text{mesons}} &= \frac{1}{2}\left(\partial_\mu \sigma\right)^2 + \frac{1}{2}\left(\partial_\mu \vec{\pi}\right)^2 + \frac{\mu^2}{2}\left(\sigma^2 + \vec{\pi}^2\right) - \frac{\lambda^2}{4}\left(\sigma^2 + \vec{\pi}^2\right)^2 , &(2.2)\\
\mathcal{L}_{\text{nucleons}} &= \overline{\Psi}\gamma^\mu \partial_\mu \Psi \,, &(2.3)\\
\mathcal{L}_{\text{Yukawa}} &= i g_y \overline{\Psi}\gamma_5 \vec{\tau}\cdot\vec{\pi}\Psi + g_y \overline{\Psi}\sigma\Psi \,. &(2.4)
\end{aligned}$$

Chiral perturbation theory (ChPT), in which chiral symmetry is realized nonlinearly [7–17], and the LSM, in which it is realized linearly [18–23], are two examples of theories that use chiral symmetry as a guiding principle. We see that LSM keeps both the scalar $\sigma$ and the pseudoscalar $\vec{\pi}$ fields, while in the case of the chiral perturbation theory, the scalar d.o.f. are integrated out. A mass difference between chiral partner mesons results from the spontaneous breaking of chiral symmetry. In the Functional

Renormalization Group studies of the chiral symmetry dynamics, consideration of the finite temperature and chemical potential was performed in Refs. [40–45]. The framework is used to characterize chiral symmetry breaking (and phase transitions), see Refs. [46–57].

Our main focus on the LSM regards mesons. The LSM was extended to mesons with $J = 1$ in [20,21,58,59] and it will be generalized to $J = 2, 3$ in the present thesis.

## 2.2 Chiral symmetry and its breakings

According to Noether's theorem, there is a conserved current for the following infinitesimal field transformation of the fields $\phi_i(x)$ which leaves the Lagrangian density invariant

$$\begin{gathered}\phi_i(x) \longrightarrow \phi_i(x) + \delta\phi_i(x)\,, \quad \text{with} \quad \delta\phi_i(x) = i\epsilon^a t^a_{ij}\phi_j(x) \\ \mathcal{L}(\phi_i(x)) = \mathcal{L}(\phi_i(x) + \delta\phi_i(x))\,,\end{gathered} \tag{2.5}$$

where $\epsilon^a$ is the small space-time independent parameter and $t^a$'s are the generators of the Lie algebra satisfying

$$[t^a, t^b] = iC^{abc}t^c\,, \tag{2.6}$$

for some structure constants $C^{abc}$. The form of the conserved current reads

$$J^a_\mu = -i\frac{\delta\mathcal{L}}{\delta(\partial^\mu\phi_i)}t^a_{ij}\phi_j\,, \tag{2.7}$$

while the conserved charges have the form

$$Q^a = \int d^3x J^a_0(x)\,, \quad \text{with} \quad \frac{dQ^a}{dt} = 0\,. \tag{2.8}$$

In the case of the three flavors, we have the following mass term for the fermionic sector of QCD Lagrangian:

$$\mathcal{L} = i\bar{q}^i\gamma^\mu\partial_\mu q_i + m_u\bar{u}u + m_d\bar{d}d + m_s\bar{s}s\,. \tag{2.9}$$

This is basically a free fermionic Lagrangian for 3 quarks. Indeed, including gluons does not change the following considerations. The above Lagrangian is flavor symmetry in the case of $m_u = m_d = m_s \equiv m_i$. Namely, it is invariant under the infinitesimal transformation of the quark fields

$$\delta q_i(x) = i\alpha^a \frac{\lambda^a_{ij}}{2} q_j(x) \longrightarrow \mathcal{L}(m_i, q_i(x)) = \mathcal{L}(m_i, q_i(x) + \delta q_i(x)) \,, \tag{2.10}$$

for small parameters $\alpha^a$. Here $\lambda^a$'s are the eight Gell-Mann matrices satisfying the following algebra

$$\left[\frac{\lambda_a}{2}, \frac{\lambda_b}{2}\right] = i f^{abc} \frac{\lambda_c}{2} \,, \tag{2.11}$$

with the structure constants $f^{abc}$. The conserved charges ($a = 1, \cdots, 8$) for $SU(3)$ symmetry

$$Q^a(t) = \int d^3x V^a_0(x) \,, \quad \text{for} \quad V^a_\mu(x) := \bar{q}(x)\gamma_\mu \frac{\lambda_a}{2} q(x) \,, \tag{2.12}$$

satisfy also the algebra

$$\left[Q^a(t), Q^b(t)\right] = i f^{abc} Q^c(t) \,. \tag{2.13}$$

In the case of the zero quark mass limit, there is another conserved current, called $Q^a_5(t)$[1]

$$Q^a_5(t) = \int d^3x A^a_0(x) \,, \quad \text{for} \quad A^a_\mu(x) := \bar{q}(x)\gamma_5\gamma_\mu \frac{\lambda_a}{2} q(x) \,, \tag{2.14}$$

with the following algebraic structure:

$$\left[Q^a_5(t), Q^b_5(t)\right] = i f^{abc} Q^c(t) \,. \tag{2.15}$$

One can construct linear combinations of the conserved charges and end up with the following algebra:

$$\left[Q^a_L(t), Q^b_L(t)\right] = i f^{abc} Q^c_L(t) \,, \tag{2.16}$$

$$\left[Q^a_R(t), Q^b_R(t)\right] = i f^{abc} Q^c_R(t) \,, \tag{2.17}$$

[1]Subscript "5" indicates a $\gamma_5$ matrix within the conserved charge.

$$\left[Q^a_L(t), Q^b_R(t)\right] = 0\,, \quad \text{where} \quad Q^a_{L,R} := \frac{1}{2}\left(Q^a \mp Q^a_5\right). \tag{2.18}$$

They are $SU(3)_L \times SU(3)_R$ chiral algebras.

Let us now consider a certain Hamiltonian $H_0$, which is invariant under $SU(3)_L \times SU(3)_R$. Thus, $U$ being the element of this transformation group leaves the $H_0$ invariant via

$$UH_0U^\dagger = H_0\,. \tag{2.19}$$

Upon considering a state $|A\rangle$, let us define $|B\rangle$ as

$$U|A\rangle = |B\rangle\,. \tag{2.20}$$

Based on the ones above, one can write

$$\langle B|H_0|B\rangle = \langle A|H_0|A\rangle\,. \tag{2.21}$$

Considering the action of the creation operators on the vacuum $|0\rangle$

$$\phi_A|0\rangle = |A\rangle\,, \quad \text{and} \quad \phi_B|0\rangle = |B\rangle \tag{2.22}$$

we get

$$U\phi_A U^\dagger = \phi_B\,. \tag{2.23}$$

If then

$$U|0\rangle = |0\rangle\,, \tag{2.24}$$

one gets

$$U|A\rangle = U\phi_A|0\rangle = U\phi_A U^\dagger U|0\rangle = \phi_B|0\rangle = |B\rangle\,, \tag{2.25}$$

and thus $|A\rangle$ and $|B\rangle$ are degenerate; namely,

$$H|B\rangle = m_B|B\rangle = HU|A\rangle = m_A U|A\rangle = m_A|B\rangle \longrightarrow m_A = m_B\,. \tag{2.26}$$

Otherwise, the Spontaneous Symmetry Breaking (SSB) $U|0\rangle = |0'\rangle \neq |0\rangle$ appears. When we write $U$ in terms of the conserved current

$$U = e^{i\epsilon^a Q^a}\,, \tag{2.27}$$

for the parameter $\epsilon^a$, we can represent the SSB in terms of the conserved charges such that, using

$$HQ^a|0\rangle = Q^a H|0\rangle = 0|0\rangle\,, \tag{2.28}$$

then

$$Q^a|0\rangle \neq 0\,, \tag{2.29}$$

implies a massless state.

According to Goldstone's theorem, SSB of a continuous symmetry leads to particles with no mass and no spin, which are defined as Nambu-Goldstone bosons.

As a concrete example, let us consider the potential with two real fields $\sigma$ and $\pi$

$$V(\sigma,\pi) := -\frac{\mu^2}{2}\left(\sigma^2+\pi^2\right) + \frac{\lambda}{4}\left(\sigma^2+\pi^2\right). \tag{2.30}$$

It has an $U(1) \simeq SO(2)$ continuous symmetry

$$\begin{pmatrix}\pi\\ \sigma\end{pmatrix} \to \begin{pmatrix}\pi'\\ \sigma'\end{pmatrix} = \begin{pmatrix}\cos\theta & \sin\theta\\ -\sin\theta & \cos\theta\end{pmatrix}\begin{pmatrix}\pi\\ \sigma\end{pmatrix}. \tag{2.31}$$

The potential term in Eq. (2.30) has the extremum as

$$\frac{\partial V}{\partial\sigma} = \sigma\big(-\mu^2+\lambda v^2\big) = 0\,, \quad \frac{\partial V}{\partial\pi} = \pi\big(-\mu^2+\lambda v^2\big) = 0\,, \quad v^2 := \sigma^2+\pi^2\,. \tag{2.32}$$

In the case of $\mu^2 > 0$, the minima are the points of the circle with radius $v = \sqrt{\frac{\mu^2}{\lambda}}$ (left panel of Fig. 2.1). These points are equivalent, and the vacuum is infinitely degenerated. One can choose any point in this circle that is the true vacuum (e.g., $\langle 0|\sigma|0\rangle = v$ and $\langle 0|\pi|0\rangle = 0$) and spontaneously break the original symmetry of the Lagrangian.

The conserved current for the Lagrangian with U(1) symmetry has the following form:

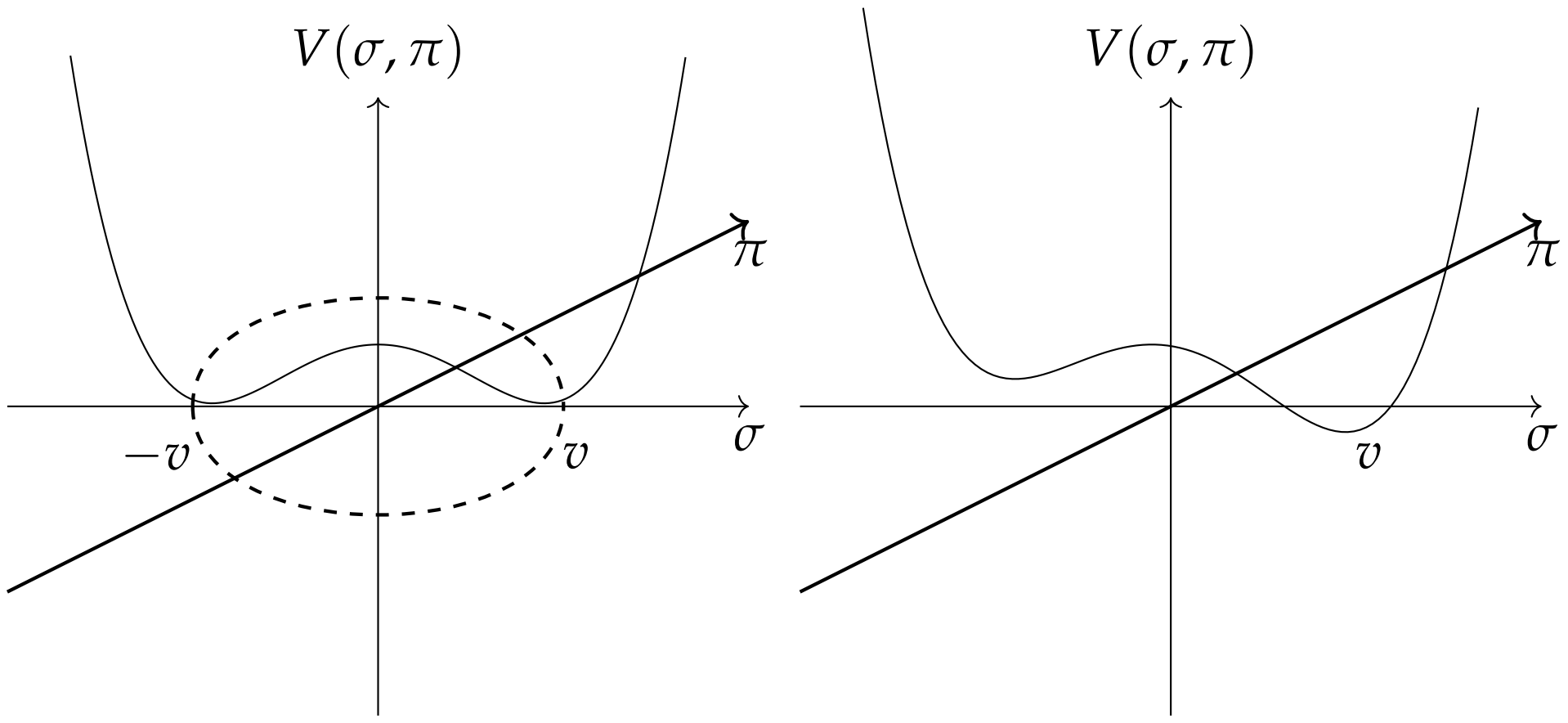

**Figure 2.1:** Mexican hat potential with minima along the chiral circle (left) and the explicit breaking of the chiral symmetry (right).

$$J_\mu = (\partial_\mu \pi)\sigma - (\partial_\mu \sigma)\pi \,, \tag{2.33}$$

which implies the following conserved charge:

$$Q = \int d^3x \Big( (\partial_0 \pi)\sigma - (\partial_0 \sigma)\pi \Big) \,. \tag{2.34}$$

Using the canonical commutation relation between the fields, we get the commutation between the charge and the fields

$$\Big[Q, \pi(0)\Big] = -i\sigma(0) \,, \quad \text{and} \quad \Big[Q, \sigma(0)\Big] = i\pi(0) \,. \tag{2.35}$$

One can show that, having a non-zero expectation value for the sigma field $\langle 0|\sigma(0)|0\rangle = v$ implies the existence of a massless pion ( see e.g., Ref. [60]).

Let us consider the shift on the $\sigma$ field, $\sigma = \overline{\sigma} + v$ and rewrite the Lagrangian in terms of the $\overline{\sigma}$

$$\mathcal{L}(\overline{\sigma}, \pi) = \frac{1}{2}\Big( (\partial_\mu \overline{\sigma})^2 + (\partial_\mu \pi)^2 \Big) - \frac{\lambda}{4}\Big( \overline{\sigma}^2 + \pi^2 \Big)^2 - \mu^2 \overline{\sigma}^2 - \lambda v \overline{\sigma} (\overline{\sigma}^2 + \pi^2) \,, \tag{2.36}$$

which leads to the massless $\pi$ and the mass for the $\sigma'$ as $m_{\overline{\sigma}} = \sqrt{2\mu^2}$. As the pion is not massless in reality, one needs to add an explicit symmetry breaking term $\mathcal{L}_{ESB}$

$$\mathcal{L}_{ESB} = -h\sigma \,. \tag{2.37}$$

The non-abelian version of LSM with $SU(2)_L \otimes SU(2)_R$ symmetry can be written as

$$\mathcal{L}_{\text{mesons}} = \frac{1}{2}\text{tr}\left(\partial_\mu \Sigma^\dagger \partial^\mu \Sigma\right) + \frac{\mu^2}{4}\text{tr}\left(\Sigma^\dagger \Sigma\right) - \frac{\lambda}{16}\left(\text{tr}(\Sigma^\dagger \Sigma)\right)^2, \tag{2.38}$$

where $\Sigma := \sigma + i\vec{\tau} \cdot \vec{\pi}$ for the three Pauli matrices $\vec{\tau}$. In this case, we will have the following commutation relation:

$$\left[Q^{5a}, \pi^b\right] = -i\sigma\delta^{ab}, \quad \text{where} \quad Q^{5a} = \int d^3x \Big((\partial_0 \pi^a)\sigma - (\partial_0 \sigma)\pi^a\Big). \tag{2.39}$$

However, the other conserved charge corresponding to the vector current annihilates the vacuum, $Q^a|0\rangle = 0$, where

$$Q^a(t) = \int d^3x \epsilon^{abc} \pi^b(x,t)(\partial_0 \pi^c(x,t)), \tag{2.40}$$

which implies the spontaneous breaking of the $SU(2)_L \otimes SU(2)_R$ to the $SU(2)$ symmetry. In the next chapter, spontaneous breaking of the chiral symmetry will be the main method to obtain the masses of spin-2 mesons.

## 2.3 Decay process of $\rho \to \pi\pi$ within effective model

In this section, I describe the decay process based on an illustrative effective Lagrangian. The decay processes in the next chapters will be extensions of this relatively simple example.

A free Proca style Lagrangian for massive vector meson reads

$$\mathcal{L} = -\frac{1}{4}\rho_{\mu\nu}\rho^{\mu\nu} + \frac{m_\rho^2}{2}\rho_\mu \rho^\mu, \quad \text{where} \quad \rho_{\mu\nu} := \partial_\mu \rho_\nu - \partial_\nu \rho_\mu. \tag{2.41}$$

Using the Euler-Lagrange equation

$$\partial_\mu \frac{\partial \mathcal{L}}{\partial(\partial_\mu \rho_\nu)} - \frac{\partial \mathcal{L}}{\partial \rho_\nu} = 0, \tag{2.42}$$

we obtain the following equation of motion:

$$\Box \rho^\mu - \partial^\mu(\partial_\nu \rho^\nu) + m^2 \rho^\mu = 0. \tag{2.43}$$

Acting by $\partial_\mu$ on all terms leads to the constraint equation

$$\partial_\mu \rho^\mu = 0\,. \tag{2.44}$$

Thus, a massive vector field has three degrees of freedom instead of four. The replacement of Eq. (2.44) within the equation of motion leads to the Klein- Gordon equation for the vector field

$$(\Box + m_\rho^2)\rho^\mu = 0\,. \tag{2.45}$$

By quantizing the solutions of the above equation we obtain the following expression for the massive vector field

$$\rho_\mu = \frac{1}{\sqrt{V}} \sum_{\vec{k}} \frac{1}{\sqrt{2\omega}} \sum_{r=1,2,3} \left[ a_{\vec{k}}^{(r)} \varepsilon_\mu^{(r)}(\vec{k}) e^{-ikx} + a_{\vec{k}}^{\dagger(r)} \varepsilon_\mu^{(r)}(\vec{k}) e^{ikx} \right], \ \omega = \sqrt{m_\rho^2 + \vec{k}^2}\,. \tag{2.46}$$

Taking into account the constraint of the massive vector field, we observe

$$k_\mu \varepsilon^{(r)\mu}(\vec{k}) = 0\,. \tag{2.47}$$

There is also a diagonality condition between the polarization vectors such that

$$\varepsilon_\mu^{(r)}(\vec{k}) \varepsilon^{(s)\mu}(\vec{k}) = -\delta^{rs}\,. \tag{2.48}$$

By using these two formulas, we get the form of the polarization vectors. The following formula describes the polarization sums for the massive vector field, which is useful for the computation of the decay rates

$$\sum_{r=1}^{3} \varepsilon_\mu^{(r)}(\vec{k}) \varepsilon_\nu^{(r)}(\vec{k}) = -g_{\mu\nu} + \frac{k_\mu k_\nu}{m_\rho^2}\,. \tag{2.49}$$

Now, let us compute the decay process $\rho(770) \to \pi\pi$ within a Lagrangian based on the $U(2)$-flavor symmetry. Considering pseudoscalar mesons within a multiplet in the case of two flavors [2]

$$\mathbb{P} = \begin{pmatrix} \pi^0 & \sqrt{2}\pi^+ \\ \sqrt{2}\pi^- & -\pi^0 \end{pmatrix} , \tag{2.50}$$

and analogously for the vector mesons

$$\mathbb{V}_\mu = \begin{pmatrix} \rho^0_\mu & \sqrt{2}\rho^+_\mu \\ \sqrt{2}\rho^-_\mu & -\rho^0_\mu \end{pmatrix} , \tag{2.51}$$

the following Lagrangian describes an isospin-invariant interaction between the vector and pseudoscalar mesons:

$$\begin{aligned} \mathcal{L}_{VPP} &= \frac{ig}{2}\mathrm{Tr}\Big(\mathbb{V}_\mu \partial^\mu \mathbb{P}\mathbb{P} - \mathbb{V}_\mu \mathbb{P}\partial^\mu \mathbb{P}\Big) \qquad (2.52) \\ = ig\rho^0_{\ \mu}\{\pi^+\partial^\mu\pi^- + \pi^-\partial^\mu\pi^+\} &+ ig\rho^+_{\ \mu}\{\pi^-\partial^\mu\pi^0 + \pi^0\partial^\mu\pi^-\} \\ &+ig\rho^-_{\ \mu}\{\pi^+\partial^\mu\pi^0 + \pi^+\partial^\mu\pi^0\} . \end{aligned}$$

The first thing we read from this Lagrangian is the equality of the following decays

$$\Gamma_{\rho^0 \longrightarrow \pi^+\pi^-} = \Gamma_{\rho^+ \longrightarrow \pi^+\pi^0} = \Gamma_{\rho^- \longrightarrow \pi^0\pi^-} . \tag{2.53}$$

Application of the Feynmann rules to the interaction Lagrangian leads to the amplitude

$$-i\mathcal{M}^a = \varepsilon^a_\mu(\vec{p})(ik_2 - ik_1)i(ig) = -ig\varepsilon^a_\mu(\vec{p})(k_2 - k_1) \equiv ig\varepsilon^a_\mu(\vec{p})k^\mu , \tag{2.54}$$

where $k = k_1 - k_2$ and $k_{1,2} = (\omega_{1,2}, \vec{k}_{1,2})$. Thus, the average of the amplitude squared reads

$$\begin{aligned} \overline{|-i\mathcal{M}|}^2 = \frac{1}{3}\sum_a |-i\mathcal{M}^a|^2 = \frac{g^2}{3}\sum_a \varepsilon^a_\mu(\vec{p})k^\mu \varepsilon^a_\nu(\vec{p})k^\nu &= \frac{g^2}{3}\Big(-g_{\mu\nu} + \frac{p_\mu p_\nu}{m^2_\rho}\Big)k^\mu k^\nu = \\ &= \frac{g^2}{3}\Big(-k^2 + \frac{(p\cdot k)^2}{m^2_\rho}\Big) . \end{aligned} \tag{2.55}$$

[2]For simplicity, I ignore the singlet mesons. The next section is devoted to the proper description of mesons within certain nonets.

Considering the pions as identical particles and in the rest frame of the decaying particle $p = (m_\rho, \vec{0})$, one has:

$$p \cdot k = p \cdot (k_1 - k_2) = m_\rho(\omega_1 - \omega_2) = 0\,, \tag{2.56}$$

thus, we obtain the following kinematical relation:

$$-k^2 = -k_1^2 - k_2^2 + 2k_1 \cdot k_2 = -2m_\pi^2 + 2\Big(\frac{m_\rho^2 - 2m_\pi^2}{2}\Big) = m_\rho^2 - 4m_\pi^2\,. \tag{2.57}$$

The final expression for the decay of $\rho$ meson into two pions has the following form:

$$\Gamma_{\rho\to\pi\pi} = \frac{\sqrt{\frac{m_\rho^2}{4} - m_\pi^2}}{8\pi m_\rho} \cdot \frac{g^2}{3}(m_\rho^2 - 4m_\pi^2) \cdot \Theta(m_\rho - 2m_\pi) = \frac{g^2\Big(\frac{m_\rho^2}{4} - m_\pi^2\Big)^{\frac{3}{2}}}{6\pi m_\rho}\Theta(m_\rho - 2m_\pi)\,. \tag{2.58}$$

From the comparison with experimental data $\Gamma_{\rho\to\pi\pi} \approx 150$ MeV [27], we get $g \approx 6$. In the next three chapters we will use analogous (but more general) Lagrangians to describe the decays of $J = 2, 3$ mesons within effective models.

## 2.4 Conventional mesons within nonets

The classification of observed mesonic resonances into nonets of quark-antiquark ($q\bar{q}$) states using flavor and/or chiral symmetry is one of the most significant achievements of modern high-energy physics.

Let us describe the mesons in Table 1.2 within certain nonets using $3 \times 3$ matrices. For convenience, I use the strange-non-strange basis for the isoscalar members of the nonet entries. The first entries of Table 1.2 are pseudoscalar mesons $\{\pi$, $K, \eta(547), \eta'(958)\}$ (with $J^{PC} = 0^{-+}$) that correspond to:

$$P = \frac{1}{\sqrt{2}}\begin{pmatrix} \frac{\eta_N+\pi^0}{\sqrt{2}} & \pi^+ & K^+ \\ \pi^- & \frac{\eta_N-\pi^0}{\sqrt{2}} & K^0 \\ K^- & \bar{K}^0 & \eta_S \end{pmatrix}, \tag{2.59}$$

where the mesons with purely non-strange quarks $\eta_N$ and pure strange quarks $\eta_S$ are defined below

$$\eta_N \equiv (\bar{u}\,\mathrm{i}\gamma^5\,u + \bar{d}\,\mathrm{i}\gamma^5\,d)/\sqrt{2}\,, \quad \eta_S \equiv \bar{s}\,\mathrm{i}\gamma^5\,s\,. \tag{2.60}$$

The mixing relation between the isospin $I = 0$ mesons is

$$\begin{pmatrix} \eta \\ \eta'(958) \end{pmatrix} = \begin{pmatrix} \cos\beta_p & \sin\beta_p \\ -\sin\beta_p & \cos\beta_p \end{pmatrix} \begin{pmatrix} \eta_N \\ \eta_S \end{pmatrix}\,. \tag{2.61}$$

We will use $\beta_p = -43.4^\circ$ obtained in Ref. [61]. Such a large mixing angle is related to the so-called chiral (or axial) $U_\mathrm{A}(1)$ anomaly [62,63].

Scalar mesons (with $J^{PC} = 0^{++}$) form a chiral multiplet together with the pseudoscalar mesons. We shall not include any decay width regarding this nonet because of their unclear identification. However, it is worth mentioning that $\{a_0(1450)$, $K_0^*(1430)$, $f_0(1370)$, $f_0(1500)/f_0(1710)\}$ has a fair chance of building the scalar nonet with the scalar currents

$$S_{ij} \equiv 2^{-1/2}\bar{q}_j q_i\,. \tag{2.62}$$

The corresponding matrix is

$$S = \frac{1}{\sqrt{2}}\begin{pmatrix} \frac{\sigma_N + a_0^0}{\sqrt{2}} & a_0^+ & K_0^{\star+} \\ a_0^- & \frac{\sigma_N - a_0^0}{\sqrt{2}} & K_0^{\star0} \\ K_0^{\star-} & \bar{K}_0^{\star0} & \sigma_S \end{pmatrix}\,. \tag{2.63}$$

Unlike the pseudoscalar case, consideration of the glueball in the scalar sector is needed, since $\overline{n}n$, $\bar{s}s$, and $\overline{g}g$ are believed to generate the resonances $f_0(1370)$, $f_0(1500)$, and $f_0(1710)$, see e.g. Refs. [22,64–76].

Vector mesons $\{\rho(770)$, $K^*(892)$, $\omega(782)$, $\phi(1020)\}$ (with $J^{PC} = 1^{--}$ ) are given in the third column of Table 1.2 and appear in the matrix form as

$$V_1^\mu = \frac{1}{\sqrt{2}}\begin{pmatrix} \frac{\omega_{1,N}^\mu + \rho_1^{0\mu}}{\sqrt{2}} & \rho_1^{+\mu} & K_1^{*+\mu} \\ \rho_1^{-\mu} & \frac{\omega_{1,N}^\mu - \rho_1^{0\mu}}{\sqrt{2}} & K_1^{*0\mu} \\ K_1^{*-\mu} & \bar{K}_1^{*0\mu} & \omega_{1,S}^\mu \end{pmatrix}\,. \tag{2.64}$$

In this case, the vector currents have the following form:

$$V^{\mu}_{1,ij} = (\bar{q}_j\,\gamma^{\mu}\,q_i)/\sqrt{2}\,, \tag{2.65}$$

The isoscalar states in the strange and non-strange basis are given by

$$\begin{pmatrix} \omega(782) \\ \phi(1020) \end{pmatrix} = \begin{pmatrix} \cos\beta_{v_1} & \sin\beta_{v_1} \\ -\sin\beta_{v_1} & \cos\beta_{v_1} \end{pmatrix} \begin{pmatrix} \omega_{1,N} \\ \omega_{1,S} \end{pmatrix}, \tag{2.66}$$

and the small mixing angle $\beta_{v_1} = -3.9^{\circ}$ is taken from the PDG [27]. It implies that $\omega(782)$ is almost made up of non-strange quarks, while $\phi(1020)$ is nearly composed of strange ones.

The next two meson families of Table 1.2 are the axial-vector mesons $\{a_1(1260)$, $K_1(1270)/K_1(1400)$, $f_1(1420)$, $f_1(1285)\}$ (with $J^{PC} = 1^{++}$)

$$A^{\mu}_{1} = \frac{1}{\sqrt{2}} \begin{pmatrix} \frac{f^{\mu}_{1,N}+a^{0\mu}_{1}}{\sqrt{2}} & a^{+\mu}_{1} & K^{+\mu}_{1,A} \\ a^{-\mu}_{1} & \frac{f^{\mu}_{1,N}-a^{0\mu}_{1}}{\sqrt{2}} & K^{0\mu}_{1,A} \\ K^{-\mu}_{1,A} & \bar{K}^{0\mu}_{1,A} & f^{\mu}_{1,S} \end{pmatrix}, \tag{2.67}$$

and the pseudovector mesons $\{b_1(1235), K_1(1270)/K_1(1400), h_1(1415), h_1(1170)\}$ (with $J^{PC} = 1^{+-}$)

$$B^{\mu}_{1} = \frac{1}{\sqrt{2}} \begin{pmatrix} \frac{h^{\mu}_{1,N}+b^{0\mu}_{1}}{\sqrt{2}} & b^{+\mu}_{1} & K^{+\mu}_{1,B} \\ b^{-\mu}_{1} & \frac{h^{\mu}_{1,N}-b^{0\mu}_{1}}{\sqrt{2}} & K^{0\mu}_{1,B} \\ K^{-\mu}_{1,B} & \bar{K}^{0\mu}_{1,B} & h^{\mu}_{1,S} \end{pmatrix}. \tag{2.68}$$

They are in good agreement with the quark model of Ref. [77] and the Bethe-Salpeter approach of Ref. [37]. The mixing angles in these nonets are still unknown. These two matrices contain the corresponding currents

$$A^{\mu}_{1,ij} = (\bar{q}_j\,\gamma^5\,\gamma^{\mu}\,q_i)/\sqrt{2}\,, \tag{2.69}$$

$$B^{\mu}_{1,ij} = (\bar{q}_j\,\gamma^5\,\overleftrightarrow{\partial}^{\mu}\,q_i)/\sqrt{2} \quad \text{where} \quad \overleftrightarrow{\partial}^{\mu} := \overrightarrow{\partial}^{\mu} - \overleftarrow{\partial}^{\mu}\,. \tag{2.70}$$

The kaonic sector of these nonets displaces an important and different mixing between $K_{1,A}$ and $K_{1,B}$. Throughout this thesis, I focus on the physical resonances $K_1(1270)$ and $K_1(1400)$ that emerge as:

$$\begin{pmatrix} K_1(1270) \\ K_1(1400) \end{pmatrix} = \begin{pmatrix} \cos\varphi_K & -\mathrm{i}\sin\varphi_K \\ -\mathrm{i}\sin\varphi_K & \cos\varphi_K \end{pmatrix} \begin{pmatrix} K_{1,A} \\ K_{1,B} \end{pmatrix} . \tag{2.71}$$

Note, this is a mixing between members of different nonets. Such mixing is sometimes also represented in another basis [78–80]

$$\begin{pmatrix} |K_1^+(1270)\rangle \\ |K_1^+(1400)\rangle \end{pmatrix} = \begin{pmatrix} \sin\theta_K & \cos\theta_K \\ \cos\theta_K & -\sin\theta_K \end{pmatrix} \begin{pmatrix} |K_{1,A}^+\rangle \\ |K_{1,B}^+\rangle \end{pmatrix} . \tag{2.72}$$

The two angles are related via the following relation

$$\theta_K = \varphi_K + 90^\circ . \tag{2.73}$$

Numerically, one obtains

$$\varphi_K = (56.4 \pm 4.3)^\circ , \tag{2.74}$$

which makes us to conclude that $K_1(1270)$ is predominantly $K_{1,B}$ while $K_1(1400)$ is predominantly $K_{1,A}$ ( despite the large mixing angle).

The orbitally excited vector mesons $\{\rho(1700), K^*(1680), \omega(1650), \phi(?)\}$ (with $J^{PC} = 1^{--}$) form a nonet

$$V_E^\mu = \frac{1}{\sqrt{2}} \begin{pmatrix} \frac{\omega_{E,N}^\mu + \rho_E^{0\mu}}{\sqrt{2}} & \rho_E^{+\mu} & K_E^{*+\mu} \\ \rho_E^{-\mu} & \frac{\omega_{E,N}^\mu - \rho_E^{0\mu}}{\sqrt{2}} & K_E^{*0\mu} \\ K_E^{*-\mu} & \bar{K}_E^{*0\mu} & \omega_{E,S}^\mu \end{pmatrix} , \tag{2.75}$$

with the currents

$$V_{E,ij}^\mu = (\bar{q}_j \overleftrightarrow{\partial}^\mu q_i)/\sqrt{2} . \tag{2.76}$$

Recent detailed phenomenological analyses for these mesons were performed in Ref. [81].

Next, I start the discussion about mesons with $J = 2, 3$ which are the last four entries of Table 1.2 and the main objects of this thesis.

The tensor mesons $\{a_2(1320), K_2^*(1430), f_2(1270), f_2'(1525)\}$ (with $J^{PC} = 2^{++}$) correspond to the following tensor currents:

$$T_{ij}^{\mu\nu} = \bar{q}_j\left[(\mathrm{i}\gamma^\mu \overleftrightarrow{\partial}^\nu + \cdots)\right] q_i/\sqrt{2}\,, \tag{2.77}$$

and are grouped in the following nonet matrix

$$T^{\mu\nu} = \frac{1}{\sqrt{2}}\begin{pmatrix} \frac{f_{2,N}^{\mu\nu}+a_2^{0\mu\nu}}{\sqrt{2}} & a_2^{+\mu\nu} & K_2^{*+\mu\nu} \\ a_2^{-\mu\nu} & \frac{f_{2,N}^{\mu\nu}-a_2^{0\mu\nu}}{\sqrt{2}} & K_2^{*0\mu\nu} \\ K_2^{*-\mu\nu} & \bar{K}_2^{*0\mu\nu} & f_{2,S}^{\mu\nu} \end{pmatrix}. \tag{2.78}$$

The mixing relation for the tensor mesons reads

$$\begin{pmatrix} f_2(1270) \\ f_2'(1525) \end{pmatrix} = \begin{pmatrix} \cos\beta_T & \sin\beta_T \\ -\sin\beta_T & \cos\beta_T \end{pmatrix}\begin{pmatrix} f_{2,N} \\ f_{2,S} \end{pmatrix}, \tag{2.79}$$

where the small mixing angle $\beta_T = 5.7^\circ$ is given in the PDG [27][3]. These mesons have been studied previously in various papers (see, e.g., Refs. [71,82,83]).

The pseudotensor mesons $\{\pi_2(1670), K_2(1770)/K_2(1820), \eta_2(1870), \eta_2(1645)\}$ (with $J^{PC} = 2^{-+}$) are described by

$$P_2^{\mu\nu} = \frac{1}{\sqrt{2}}\begin{pmatrix} \frac{\eta_{2N}^{\mu\nu}+\pi_2^{0\,\mu\nu}}{\sqrt{2}} & \pi_2^{+\,\mu\nu} & K_{2P}^{+\,\mu\nu} \\ \pi_2^{-\,\mu\nu} & \frac{\eta_{2N}^{\mu\nu}-\pi_2^{0\,\mu\nu}}{\sqrt{2}} & K_{2P}^{0\,\mu\nu} \\ K_{2P}^{-\,\mu\nu} & \bar{K}_{2P}^{0\,\mu\nu} & \eta_{2S}^{\mu\nu} \end{pmatrix}. \tag{2.80}$$

The form of the pseudotensor current reads Ref. [84]

$$P_{2,ij}^{\mu\nu} = \bar{q}_j\left[(\mathrm{i}\gamma^5\,\partial^\mu\,\partial^\nu + \cdots)\right] q_i/\sqrt{2} \tag{2.81}$$

[3] Later on, we will present an independent determination of this mixing.

The unknown mixing angle $\beta_{pt}$ within isoscalar sector is given below

$$\begin{pmatrix} \eta_2(1645) \\ \eta_2(1870) \end{pmatrix} = \begin{pmatrix} \cos\beta_{pt} & \sin\beta_{pt} \\ -\sin\beta_{pt} & \cos\beta_{pt} \end{pmatrix} \begin{pmatrix} \eta_{2\,N} \\ \eta_{2\,S} \end{pmatrix} . \tag{2.82}$$

A large mixing angle $\beta_{pt} \approx -40^\circ$ is possible (see e.g. Refs. [85–87]).

The next entries in Table 1.2 are axial-tensor mesons (with $J^{PC} = 2^{--}$ ) whose currents are

$$A^{\mu\nu}_{2\,ij} = 2^{-1/2}\,\bar{q}_j \left(\gamma^5\,\gamma^\mu\,\partial^\nu + \cdots\right) q_i\,. \tag{2.83}$$

Apart from the kaonic spin-2 resonances $K_2(1770)/K_2(1820)$ (the second one shall be assigned to $J^{PC} = 2^{--}$, even if the mixing is expected), we know very little about them. One can find recent theoretical works dedicated to the vacuum properties and thermal properties of these resonances in Refs. [88–90] as well as in Refs. [34, 91]. The axial-tensor nonet reads:

$$A_2^{\mu\nu} = \frac{1}{\sqrt{2}} \begin{pmatrix} \frac{\omega^{\mu\nu}_{2,N} + \rho_2^{0\mu\nu}}{\sqrt{2}} & \rho_2^{+\mu\nu} & K_{2A}^{+\mu\nu} \\ \rho_2^{-\mu\nu} & \frac{\omega^{\mu\nu}_{2,N} - \rho_2^{0\mu\nu}}{\sqrt{2}} & K_{2A}^{0\mu\nu} \\ K_{2A}^{-\mu\nu} & \bar{K}_{2A}^{0\mu\nu} & \omega^{\mu\nu}_{2,S} \end{pmatrix} , \tag{2.84}$$

where the mixing in the isoscalar mixing is assumed to be zero

$$\omega_{2,N} \simeq \omega_2\,, \quad \omega_{2,S} \simeq \phi_2\,. \tag{2.85}$$

This assumption is supported because of the so-called homochiral nature of this nonet according to Ref. [87] in which homochiral and heterochiral classification of the multiplets were introduced. Generally, a small mixing angle appears in homochiral multiplets, such as vector and axial-vector mesons.

The last entries of Table 1.2 are spin-3 mesons $\{\rho_3(1690), K_3^*(1780), \phi_3(1850), \omega_3(1670)\}$ (with $J^{PC} = 3^{--}$) which are combined in the following nonet matrix:

$$W_3^{\mu\nu\rho} = \frac{1}{\sqrt{2}} \begin{pmatrix} \frac{\omega^{\mu\nu\rho}_{3,N} + \rho_3^{0\mu\nu\rho}}{\sqrt{2}} & \rho_3^{+\mu\nu\rho} & K_3^{+\mu\nu\rho} \\ \rho_3^{-\mu\nu\rho} & \frac{\omega^{\mu\nu\rho}_{3,N} - \rho_3^{0\mu\nu\rho}}{\sqrt{2}} & K_3^{0\mu\nu\rho} \\ K_3^{-\mu\nu\rho} & \bar{K}_3^{0\mu\nu\rho} & \omega^{\mu\nu\rho}_{3,S} \end{pmatrix} . \tag{2.86}$$

The isoscalar members of this nonet are:

$$\begin{pmatrix} \omega_3(1670) \\ \phi_3(1850) \end{pmatrix} = \begin{pmatrix} \cos\beta_{w_3} & \sin\beta_{w_3} \\ -\sin\beta_{w_3} & \cos\beta_{w_3} \end{pmatrix} \begin{pmatrix} \omega_{3,N} \\ \omega_{3,S} \end{pmatrix}, \tag{2.87}$$

where the small mixing angle is $\beta_{w_3} = 3.5^\circ$ [27]. This small value is also in favor of their homochiral nature. The currents read

$$W_{ij}^{\mu\nu\rho} = \bar{q}_j \left[ (\gamma^\mu \partial^\nu \partial^\rho + \cdots) \right] q_i / \sqrt{2}.$$

Spin-3 mesons were previously studied in Refs. [28, 92–98].
The transformation rules under parity, charge conjugation, and flavor transformations for all the relevant nonets of this thesis ($P, S, V, A_1, B_1, V_E, T, P_2, A_2, W_3$) are summarized in Table 2.1.

| Nonet | Parity $(P)$ | Charge conjugation $(C)$ | Flavour $(U_V(3))$ |
|---|---|---|---|
| $0^{-+} = P$ | $-P(t, -\vec{x})$ | $P^t$ | $UPU^\dagger$ |
| $0^{++} = S$ | $S(t, -\vec{x})$ | $S^t$ | $USU^\dagger$ |
| $1^{--} = V_1^\mu$ | $V_{1\,\mu}(t, -\vec{x})$ | $-(V_1^\mu)^t$ | $UV_1^\mu U^\dagger$ |
| $1^{++} = A_1^\mu$ | $-A_{1\,\mu}(t, -\vec{x})$ | $(A_1^\mu)^t$ | $UA_1^\mu U^\dagger$ |
| $1^{+-} = B_1^\mu$ | $-B_{1,\mu}(t, -\vec{x})$ | $-(B_1^\mu)^t$ | $UB_1^\mu U^\dagger$ |
| $1^{--} = V_E^\mu$ | $V_{E,\mu}(t, -\vec{x})$ | $-(V_E^\mu)^t$ | $UV_E^\mu U^\dagger$ |
| $2^{++} = T^{\mu\nu}$ | $T_{\mu\nu}(t, -\vec{x})$ | $(T^{\mu\nu})^t$ | $UT^{\mu\nu}U^\dagger$ |
| $2^{-+} = P_2^{\mu\nu}$ | $P_{2\,\mu\nu}(t, -\vec{x})$ | $(P_2^{\mu\nu})^t$ | $UP_2^{\mu\nu}U^\dagger$ |
| $2^{--} = A_2^{\mu\nu}$ | $-A_{2\,\mu\nu}(t, -\vec{x})$ | $-(A_2^{\mu\nu})^t$ | $UA_2^{\mu\nu}U^\dagger$ |
| $3^{--} = W_3^{\mu\nu\rho}$ | $W_{3,\mu\nu\rho}(t, -\vec{x})$ | $-(W_3^{\mu\nu\rho})^t$ | $UW_3^{\mu\nu\rho}U^\dagger$ |

**Table 2.1:** Mesonic nonets under P, C and $U_V(3)$ transformations.

I also present the chiral nonets that shall be used in the next chapter for spin-2 eLSM

$$\begin{aligned} \Phi &:= S + iP\,, \quad L^\mu := V_1^\mu + A_1^\mu\,, \quad R^\mu := V_1^\mu - A_1^\mu\,, \\ \mathbf{L}^{\mu\nu} &:= T^{\mu\nu} + A_2^{\mu\nu}\,, \quad \mathbf{R}^{\mu\nu} := T^{\mu\nu} - A_2^{\mu\nu}\,. \end{aligned} \tag{2.88}$$

These objects are useful because they transform in a simple way under chiral transformation and can be used as a starting point to build chiral interaction Lagrangians.

# Chapter 3

# Spin-2 mesons

In this section, we present a phenomenological investigation based on the extended Linear Sigma Model of the masses, strong, and radiative decays of spin-2 ground state mesons (with $J^{PC} = 2^{++}, 2^{--}$), focusing on the missing axial-tensor mesons.

## 3.1 Introduction

According to the Table 1.2, there are three ground state spin-2 meson families:

1. Tensor mesons: $\{a_2(1320), K_2^*(1430), f_2(1270), f_2'(1525)\}$ (with $S = L = 1$, and $J^{PC} = 2^{++}$);

2. Pseudotensor mesons:$\{\pi_2(1670), K_2(1770)/K_2(1820), \eta_2(1870), \eta_2(1645)\}$ (with $S = 0$, $L = 2$, and , $J^{PC} = 2^{-+}$) ;

3. Axial-tensor mesons: $\{\rho_2(?), K_2(1820)/K_2(1770), \phi_2(?), \omega_2(?)\}$ (with $S = 1$, $L = 2$, and , $J^{PC} = 2^{--}$).

For tensor mesons, we know their masses, hadronic and two-photon decay rates. A small mixing angle in the isoscalar sector, obtained from the mass relations, is also confirmed by decay rates (as we shall see in the next sections).

There are two main problems with the pseudotensor mesons:

- A non-trivial mixing in the kaonic sector: Because of the non-diagonal C-parity of the kaons, $2^{--}$ and $2^{-+}$ states may mix in order to form $K_2(1770)$ and $K_2(1820)$ [99].

- Unknown mixing angle between $\eta_2(1870)$ and $\eta_2(1645)$ (see the discussion in Chapter 7).

The resonance $K_2(1820)$, which is supposed to be predominantly an axial-tensor meson, has a larger total decay width compared to $K_2(1770)$

$$\Gamma^{\text{tot}}_{K_2(1820)} = 264 \pm 34\,\text{MeV}\,, \quad \Gamma^{\text{tot}}_{K_2(1770)} = 186 \pm 14\,\text{MeV}\,. \tag{3.1}$$

Other members belonging to axial-tensor meson nonet have not been identified yet. There are additionally, four axial-tensor mesons $\rho_2(1940)$, $\omega_2(1975)$, $\omega_2(2195)$, and $\rho_2(2225)$ indicated as "further states" in PDG [27], but they are too heavy to be identified as ground-state axial-tensor states.

The scalar and the axial-vector mesons, as well as their corresponding chiral partners, the pseudoscalar and vector mesons, introduced in the previous chapter, were successfully described using the eLSM in Ref. [19]. This model is useful in the case of a lack of information on one of the chiral nonet—the same $J$ but opposite G-parity—because it permits the analysis of the chiral partners on an equal basis.

Using the eLSM to describe both tensor and axial-tensor mesons in a chiral framework then seems well-motivated. With sufficient experimental data, we postdict the known results (as well as previously not-yet measured radiative decays) for tensor mesons, fit the model parameter to the experiment, and then evaluate the masses and decay widths of the unknown axial-tensor mesons. The excited (pseudo)scalar mesons [22, 100], pseudovector and orbitally excited vector mesons [101] and hybrid mesons [102] were also studied within this chiral model.

We find that there is generally a good agreement between theory and experiment, which supports the predominant $q\bar{q}$ nature of the ground-state tensor mesons. Next, it turns out that the decay of the axial-tensor mesons into vector-pseudoscalar mesons is dominant and quite large. Our prediction for the mass of $\rho_2$ from the spontaneous breaking of the chiral symmetry agrees with the relativistic quark model [77]. Our findings can be useful for the current experiments at BESIII [103, 104], CLAS12 [105], COMPASS [106], GlueX [107–109], CMD-3 [110] and the future PANDA experiment [111], where axial-tensor mesons can be investigated.

## 3.2 Spin-2 eLSM

In the chiral limit of QCD ($m_{u,d,s} = 0$) the spontaneous breaking of the chiral symmetry

$$SU_{\mathrm{L}}(3) \times SU_{\mathrm{R}}(3) \times U_{\mathrm{V}}(1)\,, \tag{3.2}$$

into $SU_{\mathrm{V}}(3)$ flavor symmetry and $U_{\mathrm{V}}(1)$ symmetry (conservation of baryon number). In 1961, Nambu and Jona-Lasinio presented the first powerful model describing this process using four-fermion Fermi interactions in Refs. [112, 113]. As seen in the previous chapter, quark-antiquark mesons are grouped into nonets at the level of confined light hadrons [114–116].

Effective low-energy hadronic models, such as the previously described LSM, can be used to describe low-energy QCD. Confinement and color neutrality are automatically built in when hadrons are the relevant degrees of freedom. These models typically include chiral symmetry and its spontaneous and explicit breaking in such a way as to mimic properties of QCD.

In order to study spin-2 mesons, we use the chiral nonets introduced in the previous chapter. The transformation laws of these chiral nonets are shown in Table 3.1. As can be seen, these are quite simple transformations that enable the formulation of chirally invariant terms. Moreover, according to the classifications of Ref. [87], (pseudo)scalar mesons form a heterochiral multiplet with a possible large isoscalar mixing angle, while (axial-)vector and (axial-)tensor build homochiral multiplets with expected small isoscalar mixing angles. The (pseudo)scalar and (axial-)vector mesons as well as the

| Nonet | Parity ($P$) | Charge conjugation ($C$) | $U_L(3) \times U_R(3)$ |
|---|---|---|---|
| $\Phi(t,\vec{x})$ | $\Phi^\dagger(t,-\vec{x})$ | $\Phi^t(t,\vec{x})$ | $U_L \Phi U_R^\dagger$ |
| $R^\mu(t,\vec{x})$ | $L_\mu(t,-\vec{x})$ | $-(L^\mu(t,\vec{x}))^t$ | $U_R R^\mu U_R^\dagger$ |
| $L^\mu(t,\vec{x})$ | $R_\mu(t,-\vec{x})$ | $-(R^\mu(t,\vec{x}))^t$ | $U_L L^\mu U_L^\dagger$ |
| $\mathbf{R}^{\mu\nu}(t,\vec{x})$ | $\mathbf{L}_{\mu\nu}(t,-\vec{x})$ | $(\mathbf{L}^{\mu\nu}(t,\vec{x}))^t$ | $U_R \mathbf{R}^{\mu\nu} U_R^\dagger$ |
| $\mathbf{L}^{\mu\nu}(t,\vec{x})$ | $\mathbf{R}_{\mu\nu}(t,-\vec{x})$ | $(\mathbf{R}^{\mu\nu}(t,\vec{x}))^t$ | $U_L \mathbf{L}^{\mu\nu} U_L^\dagger$ |

**Table 3.1:** Chiral nonets under P, C, and $U_L(3) \times U_R(3)$ transformations.

dilaton (or scalar glueball) field $G$ are all ingredients of the standard form of the eLSM

Lagrangian, as shown in Refs. [19,22,102]:

$$\mathcal{L}_{\text{eLSM}} = V_{\text{dil}} + \text{Tr}\Big[\Big(D_\mu\Phi\Big)^\dagger \Big(D_\mu\Phi\Big)\Big] - m_0^2\Big(\frac{G}{G_0}\Big)^2\text{Tr}\Big[\Phi^\dagger\Phi\Big] - \frac{1}{4}\text{Tr}\Big[\Big(L_{\mu\nu}^2 + R_{\mu\nu}^2\Big)\Big]$$
$$+\text{Tr}\Big[\Big(\frac{m_{\text{vec}}^2 G^2}{2G_0^2} + \Delta\Big)\Big(L_\mu^2 + R_\mu^2\Big)\Big] + \Big(\text{higher order terms}\Big)\,, \tag{3.3}$$

where $V_{\text{dil}}$ is the dilaton Lagrangian for the field $G$ [117,118], which has the following form

$$V_{\text{dil}}(G) = \frac{1}{4}\lambda\Big(G^4 \ln\Big|\frac{G}{G_0}\Big| - \frac{G^4}{4}\Big)\,, \tag{3.4}$$

where $G_0 \sim 0.4$ GeV is the corresponding vacuum expectation value (v.e.v.) of a dilaton field $G$. Note, $V_{\text{dil}}(G)$ mimics the trace anomaly feature of the YM part of QCD. The dilaton field $G$ is formally important for dilatation invariance and its breaking even if it will not show up as a decay product. The "covariant derivative" in Eq.(3.3) is defined as

$$D_\mu\Phi := \partial_\mu\Phi - ig_1\left(L_\mu\Phi - \Phi R_\mu\right)\,. \tag{3.5}$$

Information about the explicit breaking of the flavor symmetry, that is the inclusion of strange quark mass larger than of non-strange quarks, is considered in the following matrix $\Delta$

$$\Delta = \begin{pmatrix} \delta_N & 0 & 0 \\ 0 & \delta_N & 0 \\ 0 & 0 & \delta_S \end{pmatrix}\,, \tag{3.6}$$

where $\delta_N \sim m_u^2$ and $\delta_S \sim m_s^2$. Dilatation and chiral invariances, as well as their explicit and spontaneous breaking patterns, are thus properly handled in eLSM.[1] Here I briefly review the results for fitted model parameters from Refs. [19,22] relevant to my thesis. Firstly, as a result of SSB isoscalar members of Eq. (2.63) get a v.e.v. which can be understood as a consequence of $m_0^2 < 0$. This is the generalization of the simple Mexican hat form of effective potential previously studied in [18]. As a consequence,

[1] Examples of additional chiral models that used dilatation invariance can be found, for instance, in Refs. [20,21].

shifts of scalar fields are necessary:

$$S \to S + \Phi_0 \qquad \text{with} \qquad \Phi_0 := \frac{1}{\sqrt{2}} \begin{pmatrix} \frac{\phi_N}{\sqrt{2}} & 0 & 0 \\ 0 & \frac{\phi_N}{\sqrt{2}} & 0 \\ 0 & 0 & \phi_S \end{pmatrix} , \tag{3.7}$$

where the numerical values are given in Table 3.2. In the virtue of the breaking of dilatation symmetry, the dilaton field $G$ condenses to $G_0$ too. The second consequence of SSB is that the mixing between axial-vector and pseudoscalar mesons [19, 100] is proportional to the parameter $g_1$ (see Table 3.2). It requires the following shift:

$$A_{1\,\mu} \to A_{1\,\mu} + \partial_\mu \mathcal{P} \text{ with } \mathcal{P} := \frac{1}{\sqrt{2}} \begin{pmatrix} \frac{Z_\pi w_\pi (\eta_N + \pi^0)}{\sqrt{2}} & Z_\pi w_\pi \pi^+ & Z_K w_K K^+ \\ Z_\pi w_\pi \pi^- & \frac{Z_\pi w_\pi (\eta_N - \pi^0)}{\sqrt{2}} & Z_K w_K K^0 \\ Z_K w_K K^- & Z_K w_K \bar{K}^0 & Z_{\eta_S} w_{\eta_S} \eta_S \end{pmatrix} , \tag{3.8}$$

where the modified version of the pseudoscalar nonet Eq. (2.80) with renormalization factors given in Table 3.2 is introduced. The application of these two shifts related to SSB leads to novel interaction terms in the next sections.

Chiral invariant Lagrangians that consider the spin-2 chiral nonets as well are grouped into two types: mass and interaction terms. We first start with the mass terms

$$\begin{aligned} \mathcal{L}_{\text{mass}} = \text{Tr}\Big[\Big(\frac{m_{\text{ten}}^2\, G^2}{2\, G_0^2} + \Delta^{\text{ten}}\Big)\Big(\mathbf{L}_{\mu\nu}^2 + \mathbf{R}_{\mu\nu}^2\Big)\Big] + \frac{h_1^{\text{ten}}}{2}\text{Tr}\Big[\Phi^\dagger\Phi\Big]\text{Tr}\Big[\mathbf{L}^{\mu\nu}\mathbf{L}_{\mu\nu} + \mathbf{R}^{\mu\nu}\mathbf{R}_{\mu\nu}\Big] \\ + h_2^{\text{ten}}\text{Tr}\Big[\Phi^\dagger\mathbf{L}^{\mu\nu}\mathbf{L}_{\mu\nu}\Phi + \Phi\mathbf{R}^{\mu\nu}\mathbf{R}_{\mu\nu}\Phi^\dagger\Big] + 2h_3^{\text{ten}}\text{Tr}\Big[\Phi\mathbf{R}^{\mu\nu}\Phi^\dagger\mathbf{L}_{\mu\nu}\Big], \end{aligned} \tag{3.9}$$

where the only dimensionful constants are collected within the following matrix:

$$\Delta^{\text{ten}} = \begin{pmatrix} \delta_N^{\text{ten}} & 0 & 0 \\ 0 & \delta_N^{\text{ten}} & 0 \\ 0 & 0 & \delta_S^{\text{ten}} \end{pmatrix} . \tag{3.10}$$

This is analogous to Eq. (3.6) but refers to (axial-)tensor multiplet. Upon setting $\delta_N^{\text{ten}} = 0$ (this is possible because a term proportional to the identity matrix can be subtracted and absorbed in the $m_{\text{ten}}^2$-term) and using the physical masses for the tensor

| Parameters | Expressions | Numerical values [19] |
|---|---|---|
| $Z_\pi = Z_{\eta_N}$ | $\frac{m_{a_1}}{\sqrt{m_{a_1}^2 - g_1^2\phi_S^2}}$ | 1.709 |
| $w_\pi = w_{\eta_N}$ | $\frac{g_1\phi_N}{m_{a_1}^2}$ | $0.683\,\mathrm{GeV}^{-1}$ |
| $Z_K$ | $\frac{2m_{K_{1A}}}{\sqrt{4m_{K_{1A}}^2 - g_1^2(\phi_N+\sqrt{2}\phi_S)^2}}$ | 1.604 |
| $w_K$ | $\frac{g_1(\phi_N+\sqrt{2}\phi_S)}{2m_{K_{1,A}}^2}$ | $0.611\,\mathrm{GeV}^{-1}$ |
| $Z_{\eta_S}$ | $\frac{m_{f_{1,S}}}{\sqrt{m_{f_{1,S}}^2 - 2g_1^2\phi_S^2}}$ | 1.539 |
| $w_{\eta_S}$ | $\frac{\sqrt{2}g_1\phi_S}{m_{f_{1,S}}^2}$ | $0.554\,\mathrm{GeV}^{-1}$ |
| $\phi_N$ | $Z_\pi f_\pi$ | $0.158\,\mathrm{GeV}$ |
| $\phi_S$ | $\frac{2Z_K f_K - \phi_N}{\sqrt{2}}$ | $0.138\,\mathrm{GeV}$ |
| $g_1$ | $\frac{m_{a_1}}{Z_\pi f_\pi}\sqrt{1-\frac{1}{Z_\pi^2}}$ | 5.8 |

**Table 3.2:** Fit results for the parameters of the basic version of the eLSM. They will be used in this and the next chapter (see for details in Ref. [19]). Note, these values are one specific solution of the fit. For instance, in the case of $Z_\pi$, an uncertainty of about 20% was estimated.

mesons from PDG [27], we obtain the following relation

$$\delta_S^{\mathrm{ten}} = m_{K_2}^2 - m_{\mathbf{a}_2}^2 \simeq 0.3\,\mathrm{GeV}^2\,. \tag{3.11}$$

In the large-$N_c$ limit [2] $h_1^{\mathrm{ten}}$ is suppressed because of the following $N_c$ scalings of the parameters

- $m_{\mathrm{ten}}^2/G_0^2 \propto N_c^{-2}$ ;
- $h_1^{\mathrm{ten}} \propto N_c^{-3}$ ;
- $h_2^{\mathrm{ten}} \propto N_c^{-1}$ and $h_3^{\mathrm{ten}} \propto N_c^{-1}$.

[2] Although $N_c = 3$ in nature, studying QCD for large values of $N_c$ is useful since numerous simplifications take place in this limit. In particular, conventional mesons and glueballs become stable [119,120].

The decays of (axial-)tensor mesons are produced by additional interaction terms, which is what we will discuss next. The first and the simplest Lagrangian (that can also be rewritten in a dilation invariant form) reads:

$$\begin{aligned}\mathcal{L}_{g_2^{\text{ten}}} = &\frac{g_2^{\text{ten}}}{2}\Big(\text{Tr}\Big[\mathbf{L}_{\mu\nu}\{L^\mu, L^\nu\}\Big] + \text{Tr}\Big[\mathbf{R}_{\mu\nu}\{R^\mu, R^\nu\}\Big]\Big) \\ &+\frac{g_2'^{\text{ten}}}{6}\text{Tr}\Big[\mathbf{L}_{\mu\nu} + \mathbf{R}_{\mu\nu}\Big]\text{Tr}\Big[\{L^\mu, L^\nu\} + \{R^\mu, R^\nu\}\Big]\ ,\end{aligned} \tag{3.12}$$

where the first term is dominant in the large-$N_c$ limit since the dimensionful coupling constants scale as $g_2^{\text{ten}} \propto N_c^{-1/2}$, while $g_2'^{\text{ten}}$ is suppressed since $g_2'^{\text{ten}} \propto N_c^{-3/2}$. Further suppressed terms are ignored. The Lagrangian in Eq. (3.12) leads to the following two important decay channels, which are the dominant ones, as we shall see in the next sections:

1. Tensor mesons with quantum numbers $J^{PC} = 2^{++}$ decaying into two pseudoscalar mesons (with $J^{PC} = 0^{-+}$);

2. Axial-tensor mesons with quantum numbers $J^{PC} = 2^{--}$ decaying into a pseudoscalar meson (with $J^{PC} = 0^{-+}$) and into a vector meson (with $J^{PC} = 1^{--}$).

Experimental results for the well-established $2^{++}$ nonet will be used to determine the couplings $g_2^{\text{ten}}$ and $g_2'^{\text{ten}}$, that are later on implemented to derive chiral model estimations for the missing $2^{--}$ meson nonet decays. Just as two-pseudoscalar decays are the dominant decay channels for the tensor mesons, the vector and pseudoscalar meson decay channels are the dominant ones for the axial-tensor mesons. Both decay modes are present in the Lagrangian of Eq. (3.12).

The following chiral Lagrangian for the chiral fields

$$L^{\mu\nu} := \partial^\mu L^\nu - \partial^\nu L^\mu\,, \quad R^{\mu\nu} := \partial^\mu R^\nu - \partial^\nu R^\mu\,, \tag{3.13}$$

describes the coupling of (axial-)tensor mesons to two (axial-)vector ones:

$$\begin{aligned}\mathcal{L}_{a^{\text{ten}}} = &\frac{a^{\text{ten}}}{2}\text{Tr}\Big[\mathbf{L}_{\mu\nu}\{L^\mu_\beta, L^{\nu\beta}\} + \mathbf{R}_{\mu\nu}\{R^\mu_\beta, R^{\nu\beta}\}\Big] + \\ &+\frac{a'^{\text{ten}}}{6}\text{Tr}\Big[\mathbf{L}_{\mu\nu} + \mathbf{R}_{\mu\nu}\Big]\text{Tr}\Big[\{L^\mu_\beta, L^{\nu\beta}\} + \{R^\mu_\beta, R^{\nu\beta}\}\Big]\,.\end{aligned} \tag{3.14}$$

This Lagrangian generates radiative decays of the tensor mesons via the vector meson dominance (VDM) assumption [121–123], which features the photon-vector-meson mixing generated by the shift on the massive vector field strength

$$V_{\mu\nu} \mapsto V_{\mu\nu} + \tfrac{eQ}{g_\rho} F_{\mu\nu}\,, \quad F_{\mu\nu} := \partial_\mu A_\nu - \partial_\nu A_\mu\,, \quad V_{\mu\nu} := \partial_\mu V_\nu - \partial_\nu V_\mu\,, \tag{3.15}$$

where the up, down, and strange quark charges are included in the matrix

$$Q = \text{diag}\Big(\frac{2}{3}, -\frac{1}{3}, -\frac{1}{3}\Big)\,. \tag{3.16}$$

Moreover, the electric coupling constant $e$ is expressed in terms of the fine structure constant as $e = \sqrt{4\pi\,\alpha}$ and the transition parameter $g_\rho \simeq 5.5 \pm 0.5$.

The coupling constants of the Lagrangian in Eq. (3.14) carry the dimension of [Energy$^{-1}$] and are distinguished according to their large-$N_c$ behavior, with the dominant $a^{\text{ten}} \propto N_c^{-1/2}$ and the subdominant $a'^{\text{ten}} \propto N_c^{-3/2}$ [3].

Thus, the Lagrangian in Eq. (3.14) will be used to describe the following decays of the tensor mesons:

1. two-photon decay;
2. photon and vector mesons decay;
3. three and four body decays via intermediate two vector mesons.

Since there is experimental data [27] about the vector and pseudoscalar decay channels of the tensor mesons with $J^{PC} = 2^{++}$, we introduce the following (dilatation breaking) chiral Lagrangian that links axial-tensor fields to axial-vector and pseudoscalar ones:

$$\begin{aligned}\mathcal{L}_{c^{\text{ten}}} = c_1^{\text{ten}}\,\text{Tr}\Big[&\partial^\mu \mathbf{L}^{\nu\alpha}\tilde{L}_{\mu\nu}\,\partial_\alpha\Phi\,\Phi^\dagger - \partial^\mu\mathbf{R}^{\nu\alpha}\Phi^\dagger\,\partial_\alpha\Phi\tilde{R}_{\mu\nu}\\ &-\partial^\mu\mathbf{R}^{\nu\alpha}\tilde{R}_{\mu\nu}\partial_\alpha\Phi^\dagger\Phi + \partial^\mu\mathbf{L}^{\nu\alpha}\Phi\partial_\alpha\Phi^\dagger\tilde{L}_{\mu\nu}\Big] +\\ +c_2^{\text{ten}}\,\text{Tr}\Big[&\partial^\mu\mathbf{L}^{\nu\alpha}\partial_\alpha\Phi\tilde{R}_{\mu\nu}\;\Phi^\dagger - \partial^\mu\mathbf{R}^{\nu\alpha}\Phi^\dagger\,\tilde{L}_{\mu\nu}\partial_\alpha\Phi -\\ &-\partial^\mu\mathbf{R}^{\nu\alpha}\partial_\alpha\Phi^\dagger\tilde{L}_{\mu\nu}\Phi + \partial^\mu\mathbf{L}^{\nu\alpha}\Phi\tilde{R}_{\mu\nu}\partial_\alpha\Phi^\dagger\Big]\,,\end{aligned} \tag{3.17}$$

[3]This term can be made dilation invariant upon multiplying $\left(\frac{G}{G_0}\right)^b$ with $b > 0$. Indeed, $b = -1$ would be recovered, but that leads to non-analytic parts. For more details, see [19].

where

$$\tilde{R}_{\mu\nu} := \frac{\varepsilon_{\mu\nu\rho\sigma}}{2}(\partial^\rho R^\sigma - \partial^\sigma R^\rho)\,, \quad \tilde{L}_{\mu\nu} := \frac{\varepsilon_{\mu\nu\rho\sigma}}{2}(\partial^\rho L^\sigma - \partial^\sigma L^\rho)\,. \tag{3.18}$$

The dimensionful constants $c_1^{\text{ten}}$ and $c_2^{\text{ten}}$ scale as $N_c^{-1/2}$ and cannot be constrained independently (but their sum can be fixed, see next section) within the current data for tensor mesons. Upon performing the shift of the axial-vector fields and the scalar fields, the following decay channels will be described using the Lagrangian in Eq. (3.17):

1. $2^{++} \to 0^{-+} + 1^{--}$;
2. $2^{--} \to 0^{-+} + 1^{++}$.

These decay channels are also connected by chiral symmetry.

| Decay Mode | $\frac{1}{5}\lvert\mathcal{M}\rvert^2$ |
|---|---|
| $2^{++} \to 0^{-+} + 0^{-+}$ | $g_2^{\text{ten}\,2} \times \frac{2\lvert\vec{k}_{t,p^{(1)},p^{(2)}}\rvert^4}{15}$ |
| $2^{--} \to 0^{-+} + 1^{--}$ | $g_2^{\text{ten}\,2} \times \frac{\lvert\vec{k}_{t,v,p}\rvert^2}{15}\left(5 + \frac{2\lvert\vec{k}_{t,v,p}\rvert^2}{m_v^2}\right)$ |
| $2^{--} \to 0^{-+} + 2^{++}$ | $\frac{4h_3^{\text{ten}\,2}}{45}\left(45 + \frac{4\lvert\vec{k}_{a_2,t,p}\rvert^4}{m_t^4} + \frac{30\lvert\vec{k}_{a_2,t,p}\rvert^2}{m_t^2}\right)$ |
| $2^{++} \to 0^{-+} + 1^{--}$ | $\frac{(c_1^{\text{ten}}+c_2^{\text{ten}})^2\, m_t^2 \lvert\vec{k}_{t,v,p}\rvert^4}{5}$ |
| $2^{--} \to 0^{-+} + 1^{++}$ | $\frac{(c_1^{\text{ten}}-c_2^{\text{ten}})^2\, m_{a_2}^2 \lvert\vec{k}_{a_2,a_1,p}\rvert^4}{5}$ |

**Table 3.3:** Amplitude squares for the decay of spin-2 resonances.

The decay of a generic massive spin-2 state with mass $m_t$ and decay products with masses $m_a$ and $m_b$ has the form [27,124]:

$$\Gamma_{T\to A+B} = \frac{\lvert\vec{k}_{t,a,b}\rvert}{8\,\pi\, m_t^2} \times \frac{1}{5}\lvert\mathcal{M}\rvert^2 \times \kappa_i \times \Theta(m_t - m_a - m_b)\,, \tag{3.19}$$

where $\Theta(x)$ stands for the Heaviside step-function, the $\kappa_i$ are the numbers (possibly dimensionful) related to the Clebsh-Gordan coefficients, and the modulus of the outgoing particle three-momentum reads

$$|\vec{k}_{t,a,b}| \equiv \frac{1}{2\,m_t}\sqrt{(m_t^2 - m_a^2 - m_b^2)^2 - 4\,m_a^2\,m_b^2}\,. \tag{3.20}$$

The square of the amplitudes $|\mathcal{M}|^2$ in Table 3.3 are derived via Feynman rules (for textbook see e.g., [124]).

# Results

We report the results for the (axial-)tensor mesons, including their masses, hadronic, and radiative decays. Chiral SSB is the basic method to predict the masses of the missing axial-tensor mesons. The model parameters listed in Table 3.4 are obtained via the fitting of the experimental data for the well-established tensor mesons. Sec. 3.3 is devoted to the masses of the spin-2 mesons, while the hadronic and radiative decays of the tensor mesons are presented in Sec. 3.4. Predictions for the missing axial-tensor meson decays are given in Sec. 3.5.

## 3.3 Masses of (axial-)tensor mesons

By extending Eq. (3.9) and considering the v.e.v. and shifts of Eq. (3.7) and Eq. (3.8), we derive the equations for the masses of the missing axial-tensor mesons:

$$m_{\rho_2}^2 = \left(m_{\text{ten}}^2 + 2\delta_N^{\text{ten}}\right) - \frac{h_3^{\text{ten}}\phi_N^2}{2} + \frac{h_2^{\text{ten}}\phi_N^2}{2} + \frac{h_1^{\text{ten}}}{2}(\phi_N^2 + \phi_S^2)\,, \tag{3.21}$$

$$m_{K_{2A}}^2 = \left(m_{\text{ten}}^2 + \delta_N^{\text{ten}} + \delta_S^{\text{ten}}\right) + \frac{h_1^{\text{ten}}}{2}(\phi_N^2 + \phi_S^2) - \frac{1}{\sqrt{2}}h_3^{\text{ten}}\phi_N\phi_S + \frac{h_2^{\text{ten}}}{4}(\phi_N^2 + 2\phi_S^2)\,, \tag{3.22}$$

$$m_{\omega_{2,N}}^2 = \left(m_{\text{ten}}^2 + 2\delta_N^{\text{ten}}\right) - \frac{h_3^{\text{ten}}\phi_N^2}{2} + \frac{h_2^{\text{ten}}\phi_N^2}{2} + \frac{h_1^{\text{ten}}}{2}(\phi_N^2 + \phi_S^2)\,, \tag{3.23}$$

| Parameters | Numerical Values | Fitting Data |
|---|---|---|
| $h_3^{\mathrm{ten}}$ | $-41$ | Table 3.5 |
| $\delta_S^{\mathrm{ten}}$ | $0.3\,\mathrm{GeV}^2$ | Table 3.5 |
| $g_2^{\mathrm{ten}}$ | $(1.392 \pm 0.024) \cdot 10^4(\mathrm{MeV})$ | Table 3.7 |
| $g_2'^{\,\mathrm{ten}}$ | $(0.024 \pm 0.041) \cdot 10^4(\mathrm{MeV})$ | Table 3.7 |
| $g_{2\,\mathrm{lat}}^{\mathrm{ten}}$ | $(0.7 \pm 0.4) \cdot 10^4(\mathrm{MeV})$ | Table 3.13 |
| $c^{\mathrm{ten}}$ | $(4.8 \pm 0.9) \cdot 10^{-7}(\mathrm{MeV})^{-3}$ | Table 3.8 |
| $a^{\mathrm{ten}}$ | $(-2.09 \pm 0.06) \cdot 10^{-2}(\mathrm{MeV})^{-1}$ | Table 3.10 |
| $a'^{\,\mathrm{ten}}$ | $(3.5 \pm 0.4) \cdot 10^{-3}(\mathrm{MeV})^{-1}$ | Table 3.10 |
| $\beta_T$ | $(3.16 \pm 0.81)^{\circ}$ | Table 3.7 |

**Table 3.4:** Fit results to known experimental data of PDG fixes the parameters of the eLSM associated with the (axial-)tensor mesons (the details are given in the upcoming section). It should be noted that the suffix "lat" in $g_{2\,\mathrm{lat}}^{\mathrm{ten}}$ denotes that the data has been obtained through comparison to lattice reports in [31].

$$m^2_{\omega_{2,S}} = \left(m^2_{\mathrm{ten}} + 2\delta_S^{\mathrm{ten}}\right) - h_3^{\mathrm{ten}}\phi_S^2 + h_2^{\mathrm{ten}}\phi_S^2 + \frac{h_1^{\mathrm{ten}}}{2}(\phi_N^2 + \phi_S^2)\,. \tag{3.24}$$

The corresponding formulas for the experimentally well-known tensor mesons are as follows:

$$m^2_{\mathbf{a}_2} = \left(m^2_{\mathrm{ten}} + 2\delta_N^{\mathrm{ten}}\right) + \frac{h_3^{\mathrm{ten}}\phi_N^2}{2} + \frac{h_2^{\mathrm{ten}}\phi_N^2}{2} + \frac{h_1^{\mathrm{ten}}}{2}(\phi_N^2 + \phi_S^2) = m^2_{\rho_2} + h_3^{\mathrm{ten}}\phi_N^2\,, \tag{3.25}$$

$$m^2_{K_2} = \left(m^2_{\mathrm{ten}} + \delta_N^{\mathrm{ten}} + \delta_S^{\mathrm{ten}}\right) + \frac{h_1^{\mathrm{ten}}}{2}(\phi_N^2 + \phi_S^2) + \frac{1}{\sqrt{2}}h_3^{\mathrm{ten}}\phi_N\phi_S + \frac{h_2^{\mathrm{ten}}}{4}(\phi_N^2 + 2\phi_S^2)\,, \tag{3.26}$$

$$m^2_{f_{2,N}} = \left(m^2_{\mathrm{ten}} + 2\delta_N^{\mathrm{ten}}\right) + \frac{h_3^{\mathrm{ten}}\phi_N^2}{2} + \frac{h_1^{\mathrm{ten}}}{2}(\phi_N^2 + \phi_S^2) + \frac{h_2^{\mathrm{ten}}\phi_N^2}{2}\,, \tag{3.27}$$

$$m^2_{f_{2,S}} = \left(m^2_{\text{ten}} + 2\delta^{\text{ten}}_S\right) + \frac{h^{\text{ten}}_1}{2}(\phi^2_N + \phi^2_S) + h^{\text{ten}}_3\phi^2_S + h^{\text{ten}}_2\phi^2_S = m^2_{\omega_{2,S}} + 2h^{\text{ten}}_3\phi^2_S\,. \quad (3.28)$$

We note that for every nonet the masses of the isovector and the nonstrange isoscalar elements are the same:

$$m^2_{\rho_2} = m^2_{\omega_{2,N}}\,, \text{ and } m^2_{\mathbf{a}_2} = m^2_{f_{2,N}}\,. \quad (3.29)$$

However, the isoscalars $\omega_{2,N}$ and $f_{2n}$ are not yet in the physical basis, but we know that the mixing angle for the tensor mesons is small from various experimental and theoretical constraints.

The masses of the kaonic members satisfy the following relation:

$$m^2_{K_{2A}} = m^2_{K_2} - \sqrt{2}h^{\text{ten}}_3\phi_N\phi_S. \quad (3.30)$$

Thus, the parameter $h^{\text{ten}}_3$ is estimated using the PDG masses of $K_2(1820)$ and $K^*_2(1430)$:

$$h^{\text{ten}}_3 = \frac{m^2_{K_{2A}} - m^2_{K_2}}{\sqrt{2}\phi_N\phi_S} \simeq -41\,. \quad (3.31)$$

The second relevant parameter of the mass term Lagrangian is $\delta^{\text{ten}}_S$. We obtain this parameter under the assumptions of $\delta^{\text{ten}}_N \simeq 0$ and $\phi^2_N \simeq 2\phi^2_S$ by the following relation:

$$\delta^{\text{ten}}_S = m^2_{K_2} - m^2_{\mathbf{a}_2} \simeq 0.3\,\text{GeV}^2\,. \quad (3.32)$$

By using these two parameters, we are able to obtain the masses of the missing axial-tensor mesons:

1. The mass of the purely strange resonance $\omega_{2,S} \simeq \phi_2$ reads:

$$m^2_{\phi_2} \simeq m^2_{\omega_{2,S}} = m^2_{\mathbf{a}_2} + 2\delta^{\text{ten}}_S - \frac{3}{2}h^{\text{ten}}_3\phi^2_N\,, \quad (3.33)$$

which means that the not-yet discovered $\bar{s}s$ resonance has the following numerical value:

$$m_{\phi_2} \approx 1971\,\text{MeV}\,. \quad (3.34)$$

2. The mass of the missing $\rho_2$ is obtained via the following relation

$$m^2_{\rho_2} = m^2_{\mathbf{a}_2} - h_3^{\text{ten}} \phi_N^2 \,,$$

which implies:

$$m_{\rho_2} = 1663\,\text{MeV}\,. \tag{3.35}$$

When the isoscalar mixing is disregarded, we also obtain

$$m_{\omega_{2n}} = 1663\,\text{MeV}\,. \tag{3.36}$$

So, it is expected that the new resonances $\rho_2(1663)$, $\omega_2(1663)$ and $\phi_2(1971)$ will show up in the mesonic spectrum between 1.5 and 2.0 GeV.

3. The eLSM estimates the mass of the purely strange isoscalar tensor meson as follows:

$$m^2_{f_{2,S}} = m^2_{\omega_{2,S}} + 2h_3^{\text{ten}} \phi_S^2 \simeq 1520 \;\; \text{MeV}\,, \tag{3.37}$$

which is a bit less than the resonance actual value $m_{f_2'(1525)} = 1538\,\text{MeV}$. The degeneracy relation in this nonet $m^2_{\mathbf{a}_2} = m^2_{f_{2,N}}$, implies $m_{f_{2,N}} = 1317\,\text{MeV}$, which is slightly larger than the experimental mass $m_{f_2(1270)} = 1297$ MeV.

By introducing a small isoscalar mixing angle it is possible to explain the minor deviations of the two isoscalar-tensor states. The new masses were found by including the mixing presented in Table 3.5, which is also the summary table of massed of tensor and axial-tensor mesons.

| Resonances | Masses (in MeV) | Resonances | Masses (in MeV) |
|---|---|---|---|
| $a_2(1320)$ | **1317** | $\rho_2(?)$ | 1663 |
| $K_2^*(1430)$ | **1427** | $K_2(1820)$ | **1819** |
| $f_2(1270)$ | 1297 | $\omega_2$ | 1663 |
| $f_2'(1525)$ | 1538 | $\phi_2$ | 1971 |

**Table 3.5:** Masses of the spin-2 mesons within eLSM where the small mixing angle in the tensor sector and no mixing in the axial-tensor are considered. The PDG values are shown as bold numbers.

## 3.4 Results for $J^{PC} = 2^{++}$ mesons

In this part, we present the strong and radiative decay predictions of tensor mesons and compare them to the PDG data that is currently available.

### 3.4.1 Decay channel $T \to P^{(1)} + P^{(2)}$

The chiral Lagrangian of Eq. (3.12) is followed by the interaction Lagrangian characterizing the decay of spin-2 tensor mesons to two pseudoscalar mesons by applying the shift of Eq. (3.8):

$$\mathcal{L}_{tpp} = g_2^{\text{ten}}\,\text{Tr}\Big[T^{\mu\nu}\{(\partial_\mu\mathcal{P}),(\partial_\nu\mathcal{P})\}\Big] + \frac{g_2^{\prime\,\text{ten}}}{3}\text{Tr}\Big[T^{\mu\nu}\Big]\,\text{Tr}\Big[\{(\partial_\mu\mathcal{P}),(\partial_\nu\mathcal{P})\}\Big]\ . \tag{3.38}$$

In order to achieve the same normalization for the singlet state, the factor $1/3$ in front of the second term was selected in this form [4]. For $\lambda^0 = \sqrt{\frac{2}{3}}\mathbf{1}$ and the eight Gell-Mann matrices $\lambda^a$'s, tensor meson can be decomposed as

$$\sqrt{2}T_{\mu\nu} = \sum_{a=0}^{8} T^a_{\mu\nu}\frac{\lambda^a}{\sqrt{2}} \to T_{\mu\nu} = \frac{1}{\sqrt{6}}T^0_{\mu\nu}\mathbf{1} + \cdots, \tag{3.39}$$

which leads to the Lagrangian for the singlet state as:

$$\mathcal{L}_{tpp} = (g_2^{\text{ten}} + g_2^{\prime\text{ten}})T^{0\,\mu\nu}\,\text{Tr}\Big[\{(\partial_\mu\mathcal{P}),(\partial_\nu\mathcal{P})\}\Big] + \cdots . \tag{3.40}$$

Correspondingly, the tree-level decay rate reads

$$\Gamma_{T\to P^{(1)}+P^{(2)}}(m_t, m_{p^{(1)}}, m_{p^{(2)}}) = \kappa_{tpp,i}\,\frac{|\vec{k}_{t,p^{(1)},p^{(2)}}|^5}{60\,\pi\, m_t^2}\Theta(m_t - m_{p^{(1)}} - m_{p^{(2)}})\ , \tag{3.41}$$

where Table 3.6 lists the coefficients $\kappa_{tpp,i}$ with dimension [Energy$^{-2}$] because of the dimensionful quantities $w$'s.

[4] This is a useful convention, but of course it does not affect the results.

| Decay process | $\kappa_{tpp,i}$ |
|---|---|
| $a_2(1320) \to \bar{K}\,K$ | $2\Big(\frac{g_2^{\text{ten}} Z_k^2 w_k^2}{2}\Big)^2$ |
| $a_2(1320) \to \pi\,\eta$ | $\Big(g_2^{\text{ten}} Z_\pi Z_{\eta_N} w_\pi w_{\eta_N} \cos\beta_P\Big)^2$ |
| $a_2(1320) \to \pi\,\eta'$ | $\Big(-g_2^{\text{ten}} Z_\pi Z_{\eta_N} w_\pi w_{\eta_N} \sin\beta_P\Big)^2$ |
| $K_2^*(1430) \to \pi\,\bar{K}$ | $3\Big(\frac{g_2^{\text{ten}} Z_k w_k Z_\pi w_\pi}{2}\Big)^2$ |
| $f_2(1270) \to \bar{K}\,K$ | $2\Big(\frac{Z_k^2 w_k^2 \big((4g_2'^{\text{ten}}+3g_2^{\text{ten}})\cos\beta_T+\sqrt{2}(2g_2'^{\text{ten}}+3g_2^{\text{ten}})\sin\beta_T\big)}{6}\Big)^2$ |
| $f_2(1270) \to \pi\,\pi$ | $6\Big(\frac{Z_\pi^2 w_\pi^2 \big((2g_2'^{\text{ten}}+3g_2^{\text{ten}})\cos\beta_T+\sqrt{2}g_2'^{\text{ten}}\sin\beta_T\big)}{6}\Big)^2$ |
| $f_2(1270) \to \eta\,\eta$ | $2\Big(\frac{Z_{\eta_N}^2 w_{\eta_N}^2 \cos\beta_P^2 \big((2g_2'^{\text{ten}}+3g_2^{\text{ten}})\cos\beta_T+\sqrt{2}g_2'^{\text{ten}}\sin\beta_T\big)+Z_{\eta_S}^2 w_{\eta_S}^2 \sin\beta_P^2 \big(\sqrt{2}(g_2'^{\text{ten}}+3g_2^{\text{ten}})\sin\beta_T+2g_2'^{\text{ten}}\cos\beta_T\big)}{6}\Big)^2$ |
| $f_2'(1525) \to \bar{K}\,K$ | $2\Big(\frac{Z_k^2 w_k^2 \big(-(4g_2'^{\text{ten}}+3g_2^{\text{ten}})\sin\beta_T+\sqrt{2}(2g_2'^{\text{ten}}+3g_2^{\text{ten}})\cos\beta_T\big)}{6}\Big)^2$ |
| $f_2'(1525) \to \pi\,\pi$ | $6\Big(\frac{Z_\pi^2 w_\pi^2 \big(-(2g_2'^{\text{ten}}+3g_2^{\text{ten}})\sin\beta_T+\sqrt{2}g_2'^{\text{ten}}\cos\beta_T\big)}{6}\Big)^2$ |
| $f_2'(1525) \to \eta\,\eta$ | $2\Big(\frac{Z_{\eta_N}^2 w_{\eta_N}^2 \cos\beta_P^2 \big(-(2g_2'^{\text{ten}}+3g_2^{\text{ten}})\sin\beta_T+\sqrt{2}g_2'^{\text{ten}}\cos\beta_T\big)+Z_{\eta_S}^2 w_{\eta_S}^2 \sin\beta_P^2 \big(\sqrt{2}(g_2'^{\text{ten}}+3g_2^{\text{ten}})\cos\beta_T-2g_2'^{\text{ten}}\sin\beta_T\big)}{6}\Big)^2$ |

**Table 3.6:** Clebsch-Gordan coefficients for $T \to P^{(1)} + P^{(2)}$.

During the fitting procedure, we have considered $g_2^{\text{ten}}$, $g_2'^{\text{ten}}$ and the isoscalar mixing anlge $\beta_T$. The fit results are:

$$\begin{aligned} g_2^{\text{ten}} &= (1.392 \pm 0.024)\cdot 10^4 (\text{MeV})\,, \\ g_2'^{\text{ten}} &= (0.024 \pm 0.041)\cdot 10^4 (\text{MeV})\,, \\ \beta_T &= (3.16 \pm 0.81)^\circ\,. \end{aligned} \tag{3.42}$$

These results support the large-$N_c$ predictions that the coupling constant $g_2'^{\text{ten}}$ is considerably smaller than $g_2^{\text{ten}}$. In this particular case, it is also compatible with zero because of the relatively large error. Yet, in general, the phenomenology of large-$N_c$ suppressed decay (such as the decays of the $J/\psi$; see [125]) reveals that such terms, even though small, do not disappear.

| Decay process (in model) | eLSM (MeV) | PDG (MeV) |
|---|---|---|
| $a_2(1320) \to \bar{K}\,K$ | $4.06 \pm 0.14$ | $7.0^{+2.0}_{-1.5} \leftrightarrow (4.9 \pm 0.8)\%$ |
| $a_2(1320) \to \pi\,\eta$ | $25.37 \pm 0.87$ | $18.5 \pm 3.0 \leftrightarrow (14.5 \pm 1.2)\%$ |
| $a_2(1320) \to \pi\,\eta'(958)$ | $1.01 \pm 0.03$ | $0.58 \pm 0.10 \leftrightarrow (0.55 \pm 0.09)\%$ |
| $K_2^*(1430) \to \pi\,\bar{K}$ | $44.82 \pm 1.54$ | $49.9 \pm 1.9 \leftrightarrow (49.9 \pm 0.6)\%$ |
| $f_2(1270) \to \bar{K}\,K$ | $3.54 \pm 0.29$ | $8.5 \pm 0.8 \leftrightarrow (4.6^{+0.5}_{-0.4})\%$ |
| $f_2(1270) \to \pi\,\pi$ | $168.82 \pm 3.89$ | $157.2^{+4.0}_{-1.1} \leftrightarrow (84.2^{+2.9}_{-0.9})\%$ |
| $f_2(1270) \to \eta\,\eta$ | $0.67 \pm 0.03$ | $0.75 \pm 0.14 \leftrightarrow (0.4 \pm 0.08)\%$ |
| $f_2'(1525) \to \bar{K}\,K$ | $23.72 \pm 0.60$ | $75 \pm 4 \leftrightarrow (87.6 \pm 2.2)\%$ |
| $f_2'(1525) \to \pi\,\pi$ | $0.67 \pm 0.14$ | $0.71 \pm 0.14 \leftrightarrow (0.83 \pm 0.16)\%$ |
| $f_2'(1525) \to \eta\,\eta$ | $1.81 \pm 0.05$ | $9.9 \pm 1.9 \leftrightarrow (11.6 \pm 2.2)\%$ |

**Table 3.7:** Decay rates for $T \to P^{(1)} + P^{(2)}$ compared to PDG [27].

The following prediction is also obtained, but it was not included in the fit because of the large experimental uncertainty.

$$\text{eLSM}: \ \Gamma\left[K_2^*(1430) \to \eta\,\bar{K}\right] = (1.13 \pm 0.03) \cdot 10^{-3}\ \text{MeV}, \quad \text{PDG}: \ \ 0.15^{+0.34}_{-0.10}\ \text{MeV}. \tag{3.43}$$

We observe that our predicted value is in the same range as the PDG [27].

Interestingly, the isospin mixing angle $\beta_T = 3.16^\circ$ obtained from the fit of the decays is not far from the PDG value $\beta_T = 5.7^\circ$ obtained from the mass formula

$$\beta_T = 35.3^\circ - \arctan\left(\sqrt{\frac{4m_{K_2}^2 - m_{a_2}^2 - 3m_{f_2'}^2}{-4m_{K_2}^2 + m_{a_2}^2 + 3m_{f_2}^2}}\right). \tag{3.44}$$

Even if some of the experimental entries conflict with those in Table 3.7, the findings demonstrate a general good qualitative agreement ($\chi^2_{\text{red}} = 37$). It is important to keep in mind that only three parameters were used, and the experimental data are fairly accurate. The main deviations from the experiment are related to the decay rates $f_2(1270) \to \bar{K}K$, $f_2'(1525) \to \bar{K}K$ and $f_2'(1525) \to \eta\eta$. For a more accurate description of the tensor decays, additional large-$N_c$ suppressed terms and flavor-symmetry-breaking contributions would be required.

It is worth mentioning that, considering the following form factor within the decay rate in Eq. (3.41) (see Ref. [67])

$$\exp\Big(-\frac{|\vec{k}_{t,p^{(1)},p^{(2)}}|^2}{8\Lambda^2}\Big)\,, \tag{3.45}$$

does not improve the results. For simplicity, I choose $g_2'^{\text{ten}} = 0$ (it is anyhow subleading) and show the $\chi^2$-$\Lambda$ dependence in Fig. 3.1. This fitting does not improve the overall results since at large values of $\Lambda$ and the $\chi^2$ becomes identical to the previous fit result in Table 3.7. On the other hand, the results get worse for smaller $\Lambda$.

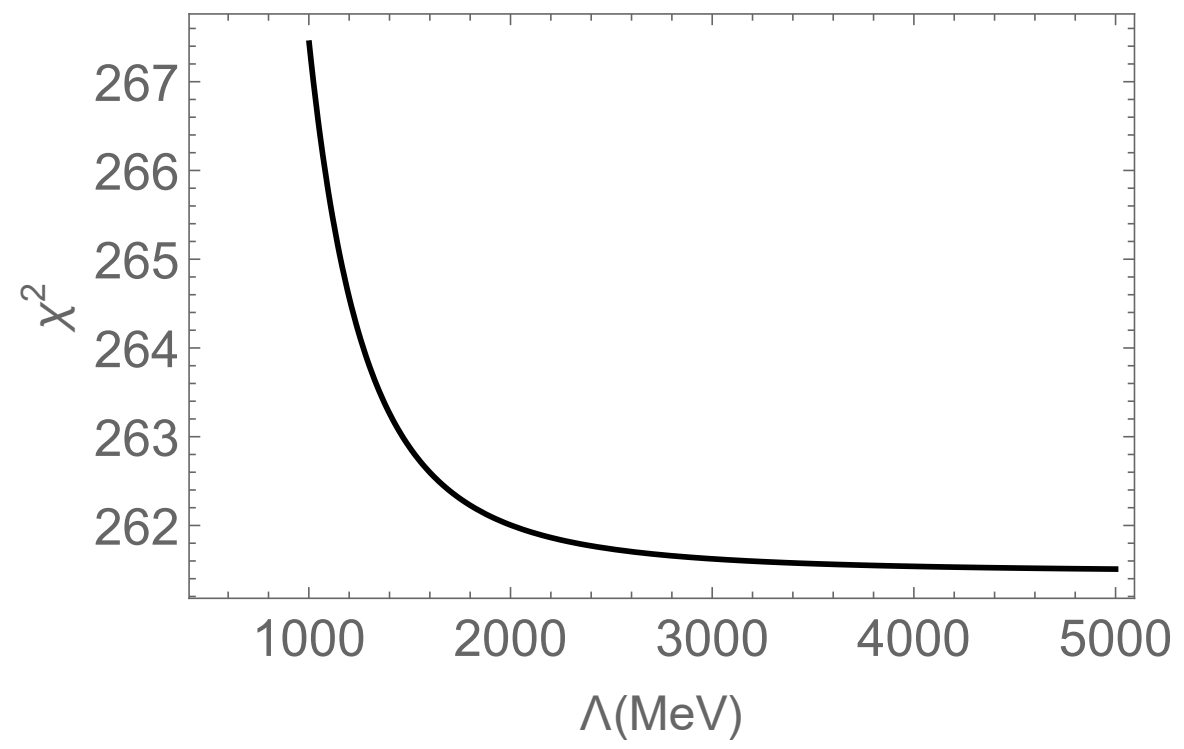

**Figure 3.1:** $\chi^2$ dependence on $\Lambda$ for $g_2'^{\text{ten}} = 0$.

### 3.4.2 Decay channel $T \to V_1 + P$

The decays of the tensor mesons into a vector-pseudoscalar pair are also experimentally measured. They can be described within the following Lagrangian, which is obtained from Eq. (3.17) by taking into account the shift of Eq. (3.7):

$$\mathcal{L}_{tvp} = (c_1^{\text{ten}} + c_2^{\text{ten}})\,\varepsilon_{\mu\nu\rho\sigma}\,\text{Tr}\Big[\partial^\mu T^{\nu\alpha}\,\Big(\partial^\rho V_1^\sigma(\partial_\alpha P)\Phi_0 - \Phi_0(\partial_\alpha P \partial^\rho V_1^\sigma)\Big)\Big]\,. \tag{3.46}$$

The tree-level decay rate formula reads

$$\Gamma_{T\to V+P}(m_t, m_v, m_p) = \frac{c^{\text{ten}\,2}\,|\vec{k}_{t,v,p}|^5}{40\,\pi}\,\kappa_{tvp\,,i}\,\Theta(m_t - m_v - m_p)\,, \tag{3.47}$$

where $c^{\text{ten}} := c_1^{\text{ten}} + c_2^{\text{ten}}$ and the dimensionful factors $\kappa_{tvp,i}$ are given in Table 3.8.

We use a $\chi^2$-fit to the data in Table 3.8 to determine the coupling's value, $c^{\text{ten}}$:

$$c^{\text{ten}} = (4.8 \pm 0.9) \cdot 10^{-7} (\text{MeV})^{-3} , \tag{3.48}$$

that leads to the model estimations of the decay widths in Table 3.8. They are in good agreement with PDG and provide an extra prediction ($f_2'(1525) \to \bar{K}^*(892)\, K + \text{c.c.}$) .

| Decay process (in model) | $\kappa_{tvp,i}$ | eLSM (MeV) | PDG (MeV) |
|---|---|---|---|
| $a_2(1320) \to \rho(770)\, \pi$ | $2\left(\frac{\phi_N}{4}\right)^2$ | $71.0 \pm 2.6$ | $73.61 \pm 3.35 \leftrightarrow (70.1 \pm 2.7)\%$ |
| $K_2^*(1430) \to \bar{K}^*(892)\, \pi$ | $3\left(\frac{\phi_N}{8}\right)^2$ | $27.9 \pm 1.0$ | $26.92 \pm 2.14 \leftrightarrow (24.7 \pm 1.6)\%$ |
| $K_2^*(1430) \to \rho(770)\, K$ | $3\left(\frac{\sqrt{2}\phi_S}{8}\right)^2$ | $10.3 \pm 0.4$ | $9.48 \pm 0.97 \leftrightarrow (8.7 \pm 0.8)\%$ |
| $K_2^*(1430) \to \omega(782)\, \bar{K}$ | $\left(\frac{-\phi_S \cos\beta_V + \phi_N \sin\beta_V}{4\sqrt{2}})\right)^2$ | $3.5 \pm 0.1$ | $3.16 \pm 0.88 \leftrightarrow (2.9 \pm 0.8)\%$ |
| $f_2'(1525) \to \bar{K}^*(892)\, K + \text{c.c.}$ | $4\left(\frac{2\phi_S \cos\beta_T + \phi_N \sin\beta_T}{8})\right)^2$ | $19.89 \pm 0.73$ | |

**Table 3.8:** Clebsch-Gordan coefficients and corresponding decay rates for $T \to P + V$ compared to PDG [27].

### 3.4.3 Decay channel $T \to \gamma + P$

The following Lagrangian is obtained from Eq. (3.17) and VMD shift of Eq. (3.15)

$$\mathcal{L}_{t\gamma p} = \frac{e}{g_\rho} c^{\text{ten}}\, \text{Tr}\Big[\, \partial^\mu T^{\nu\alpha} \left( F_{\mu\nu} Q (\partial_\alpha P) \Phi_0 - \Phi_0 (\partial_\alpha P F_{\mu\nu} Q \right) \Big] , \tag{3.49}$$

which leads to the decay rate

$$\Gamma_{T\to\gamma+P}(m_t, m_p) = \frac{c^{\text{ten}\,2}\, |\vec{k}_{t,\gamma,p}|^5}{40\,\pi}\, \kappa_{t\gamma p,i}\, \Theta(m_t - m_p) . \tag{3.50}$$

We use the couplings $c^{\text{ten}}$, $g_\rho$ (and their errors) in order to evaluate the theoretical errors in the radiative decay rates presented in Table 3.9 via the following expression:

$$\delta\Gamma_{T\to\gamma P}(m_t, m_p) = \sqrt{\Big(\frac{\partial\Gamma_{T\to\gamma P}(m_t, m_p)}{\partial c^{\text{ten}}} \delta c^{\text{ten}}\Big)^2 + \Big(\frac{\partial\Gamma_{T\to\gamma P}(m_t, m_p)}{\partial g_\rho} \delta g_\rho\Big)^2} . \tag{3.51}$$

| Decay process (in model) | $\kappa_{t\gamma p,i}$ | eLSM | PDG |
|---|---|---|---|
| $K_2^{\pm}(1430) \to \gamma K^{\pm}$ | $\frac{e^2}{g_\rho^2}\left(\frac{\phi_N+2\sqrt{2}\phi_S}{12}\right)^2$ | $1.12 \pm 0.47$ MeV | $0.24 \pm 0.05$ MeV |
| $K_2^0(1430) \to \gamma K^0$ | $\frac{e^2}{g_\rho^2}\left(\frac{\phi_N-\sqrt{2}\phi_S}{12}\right)^2$ | $5.1 \pm 2.2$ keV | $< 5.4$ keV |
| $a_2^{\pm}(1320) \to \gamma \pi^{\pm}$ | $\frac{e^2}{g_\rho^2}\left(\frac{\phi_N}{4}\right)^2$ | $1.01 \pm 0.43$ MeV | $0.31 \pm 0.03$ MeV |

**Table 3.9:** Clebsch-Gordan coefficients and corresponding decay rates for $T \to \gamma + P$ compared to PDG [27].

In Table 3.9 our results for the photon-pseudoscalar decays are presented. Despite the large errors, they are somewhat larger but qualitatively consistent with the PDG results.

### 3.4.4 Decay channel $T \to \gamma + V_1$

We move to the next radiative decay of the tensor mesons through the two-photon decays. We first observe that the following vector-vector interaction term is part of the chiral Lagrangian of Eq. (3.14):

$$\mathcal{L}_{tvv} = a^{\text{ten}}\text{Tr}\Big[T_{\mu\nu}\{V^{\mu}_{1\beta}, V_1^{\nu\beta}\}\Big] + \frac{a'^{\,\text{ten}}}{3}\text{Tr}\Big[T_{\mu\nu}\Big]\text{Tr}\big[\{V^{\mu}_{1\beta}, V_1^{\nu\beta}\}\big]\,. \tag{3.52}$$

Applying the VMD shift (3.15) for both vector fields, we obtain the Lagrangian

$$\mathcal{L}_{t\gamma\gamma} = a^{\text{ten}}\frac{e^2}{g_\rho^2}\text{Tr}\Big[T_{\mu\nu}\{QF^{\mu}_{\beta}, QF^{\nu\beta}\}\Big] + \frac{a'^{\,\text{ten}}}{3}\frac{e^2}{g_\rho^2}\text{Tr}\Big[T_{\mu\nu}\Big]\text{Tr}\Big[\{QF^{\mu}_{\beta}, QF^{\nu\beta}\}\Big]\,, \tag{3.53}$$

whose corresponding $\gamma\gamma$ decay rate reads

$$\Gamma_{T\to\gamma\gamma} = \frac{e^4}{g_\rho^4}\frac{2\,|\vec{k}|^5}{5\,\pi\,m_t^2}\kappa_{t\gamma\gamma,i} = \frac{e^4}{g_\rho^4}\frac{m_t^3}{80\pi}\kappa_{t\gamma\gamma,i}\,, \tag{3.54}$$

where we used $m_t = 2|\vec{k}|$ to get the r.h.s. of the above equation.

Using the typical $\chi^2$ technique, employing the three experimental data points in Table 3.10 with two free parameters $a^{\text{ten}}$ and $a'^{\,\text{ten}}$ and keeping the mixing angle

$\beta_T = 3.16^\circ$ unchanged, results in:

$$a^{\text{ten}} = (-2.09 \pm 0.06) \cdot 10^{-2} (\text{MeV})^{-1}, \qquad a'^{\,\text{ten}} = (3.5 \pm 0.4) \cdot 10^{-3} (\text{MeV})^{-1}. \quad (3.55)$$

We present our model results in Table 3.10. The results confirm that the mixing angle $\beta_T = 3.16^\circ$ is consistent with data in this channel as well. In line with the expectations of the large-$N_c$ limit, the parameter $a'^{\,\text{ten}}$ also turns out to be considerably smaller than $a^{\text{ten}}$.

| Decay process (in model) | $\kappa_{t\gamma\gamma,i}$ | eLSM (keV) | PDG (keV) |
|---|---|---|---|
| $a_2(1320) \to \gamma\gamma$ | $\frac{a^{\text{ten}\,2}}{36}$ | $1.01 \pm 0.06$ | $1.00 \pm 0.06$ |
| $f_2(1270) \to \gamma\gamma$ | $\left(\frac{(5a^{\text{ten}}+4a'^{\,\text{ten}})\cos\beta_T+\sqrt{2}(a^{\text{ten}}+2a'^{\,\text{ten}})\sin\beta_T}{18}\right)^2$ | $1.95 \pm 0.10$ | $2.6 \pm 0.5$ |
| $f_2'(1525) \to \gamma\gamma$ | $\left(\frac{-(5a^{\text{ten}}+4a'^{\,\text{ten}})\sin\beta_T+\sqrt{2}(a^{\text{ten}}+2a'^{\,\text{ten}})\cos\beta_T}{18}\right)^2$ | $0.083 \pm 0.009$ | $0.082 \pm 0.009$ |

**Table 3.10:** Clebsch-Gordan coefficients and corresponding decay rates for $T \to \gamma\gamma$ compared to PDG [27].

The following Lagrangian, which describes the decay into a vector-photon pair, is obtained from Eq. (3.52) by VMD shifting only one vector field:

$$\mathcal{L}_{t\gamma V} = 2a^{\text{ten}} \frac{e}{g_\rho} \text{Tr}\Big[T_{\mu\nu}\{QF^{\mu}_{\beta}, V_1^{\nu\beta}\}\Big] + \frac{2a'^{\,\text{ten}}}{3} \frac{e}{g_\rho} \text{Tr}\Big[T_{\mu\nu}\Big] \text{Tr}\Big[\{QF^{\mu}_{\beta}, V_1^{\nu\beta}\}\Big]. \quad (3.56)$$

The decay rate into a vector-photon pair is expressed as

$$\Gamma_{T\to\gamma+V_1}(m_t, m_v) = \frac{\kappa_{t\gamma v,i}\, e^2}{15\,\pi\, m_t^2 m_v^2 g_\rho^2} \Big(3|\vec{k}_{t,v,\gamma}|^7 + 5|\vec{k}_{v,\gamma}|^3 m_v^2 + 5|\vec{k}_{t,v,\gamma}|^4 m_v^2 \sqrt{|\vec{k}_{t,v,\gamma}|^2 + m_v^2}$$
$$+ |\vec{k}_{t,v,\gamma}|^5 \big(10 m_v^2 - 3(|\vec{k}_{t,v,\gamma}|^2 + m_v^2)\big)\Big). \quad (3.57)$$

The related coefficients $\kappa_{t\gamma v,i}$ are included in Table 3.11 and Table 3.12, which display the numerical predictions for the decay rates. Note, while estimating the theoretical errors, we consider $g_\rho = 5.5 \pm 0.5$ together with the ones in Eq. (3.55).

| Decay process (in model) | $\kappa_{t\gamma v,i}$ |
|---|---|
| $a_2(1320) \to \rho(770)\,\gamma$ | $\left(\frac{a^{\text{ten}}}{6}\right)^2$ |
| $a_2^0(1320) \to \omega(782)\,\gamma$ | $\left(\frac{a^{\text{ten}}\cos\beta_V}{2}\right)^2$ |
| $a_2(1320) \to \phi(1020)\,\gamma$ | $\left(\frac{a^{\text{ten}}\sin\beta_V}{2}\right)^2$ |
| $K_2^{*\pm}(1430) \to \bar{K}^{*\pm}(892)\,\gamma$ | $\left(\frac{a^{\text{ten}}}{6}\right)^2$ |
| $K_2^{*0}(1430) \to \bar{K}^{*0}(892)\,\gamma$ | $\left(\frac{a^{\text{ten}}}{3}\right)^2$ |
| $f_2(1270) \to \rho(770)\,\gamma$ | $\left(\frac{(a^{\text{ten}}+2a'^{\text{ten}})\cos\beta_T+\sqrt{2}a'^{\text{ten}}\sin\beta_T}{2}\right)^2$ |
| $f_2(1270) \to \omega(782)\,\gamma$ | $\left(\frac{\cos\beta_V\left((a^{\text{ten}}+2a'^{\text{ten}})\cos\beta_T+\sqrt{2}a'^{\text{ten}}\sin\beta_T\right)-2\sin\beta_V\left((a^{\text{ten}}+a'^{\text{ten}})\sin\beta_T+\sqrt{2}a'^{\text{ten}}\cos\beta_T\right)}{6}\right)^2$ |
| $f_2(1270) \to \phi(1020)\,\gamma$ | $\left(\frac{\sin\beta_V\left((a^{\text{ten}}+2a'^{\text{ten}})\cos\beta_T+\sqrt{2}a'^{\text{ten}}\sin\beta_T\right)+2\cos\beta_V\left((a^{\text{ten}}+a'^{\text{ten}})\sin\beta_T+\sqrt{2}a'^{\text{ten}}\cos\beta_T\right)}{6}\right)^2$ |
| $f_2'(1525) \to \rho(770)\,\gamma$ | $\left(\frac{-(a^{\text{ten}}+2a'^{\text{ten}})\sin\beta_T+\sqrt{2}a'^{\text{ten}}\cos\beta_T}{2}\right)^2$ |
| $f_2'(1525) \to \omega(782)\,\gamma$ | $\left(\frac{\sin\beta_T\left(-(a^{\text{ten}}+2a'^{\text{ten}})\cos\beta_V+2\sqrt{2}a'^{\text{ten}}\sin\beta_V\right)+\cos\beta_T\left(\sqrt{2}a'^{\text{ten}}\cos\beta_V-2(a^{\text{ten}}+a'^{\text{ten}})\sin\beta_V\right)}{6}\right)^2$ |
| $f_2'(1525) \to \phi(1020)\,\gamma$ | $\left(\frac{\cos\beta_T\left(-2(a^{\text{ten}}+a'^{\text{ten}})\cos\beta_V-\sqrt{2}a'^{\text{ten}}\sin\beta_V\right)+\sin\beta_T\left(2\sqrt{2}a'^{\text{ten}}\cos\beta_V+(a^{\text{ten}}+2a'^{\text{ten}})\sin\beta_V\right)}{6}\right)^2$ |

**Table 3.11:** Clebsch-Gordan coefficients for $T \to \gamma + V_1$.

For the sake of simplicity, additional sources of inaccuracy, such as mass uncertainty, were neglected. Because of this, the actual errors can be slightly larger but still in line with those listed in Table 3.12.

### 3.4.5 Decay channel $T \to V_1^{(1)} + V_1^{(2)}$

Our last results for the decay of tensor mesons are related to the Lagrangians in Eq. (3.12) and Eq. (3.14). We have the following data from the PDG:

$$\begin{aligned}
\text{PDG}: \ & \Gamma[f_2(1270) \to \pi^+\pi^-2\pi^0] \approx 19.5\,\text{MeV}; \\
\text{PDG}: \ & \Gamma[a_2(1320) \to \omega(782)\pi\pi] \approx 11.3\,\text{MeV}; \\
\text{PDG}: \ & \Gamma[K_2^\star(1430) \to K^\star(892)\pi\pi] = 13.5 \pm 2.3\,\text{MeV}.
\end{aligned}$$

The vector-vector decay modes $\rho(770)\rho(770)$, $\omega(782)\rho(770)$, and $K^*(892)\rho(770)$, in which the $\rho(770)$ meson further decays into two pions, provide a reasonable interpretation of these decays. In order to describe them, we introduce the properly normalized

| Decay process (in model) | Decay Width (MeV) |
|---|---|
| $a_2(1320) \to \rho(770)\,\gamma$ | $0.22 \pm 0.04$ |
| $a_2^0(1320) \to \omega(782)\,\gamma$ | $1.94 \pm 0.04$ |
| $a_2(1320) \to \phi(1020)\,\gamma$ | $0.0024 \pm 0.0005$ |
| $K_2^{*\,\pm}(1430) \to \bar{K}^{*\,\pm}(892)\,\gamma$ | $0.23 \pm 0.04$ |
| $K_2^{*\,0}(1430) \to \bar{K}^{*\,0}(892)\,\gamma$ | $0.94 \pm 0.18$ |
| $f_2(1270) \to \rho(770)\,\gamma$ | $0.70 \pm 0.17$ |
| $f_2(1270) \to \omega(782)\,\gamma$ | $0.068 \pm 0.017$ |
| $f_2(1270) \to \phi(1020)\,\gamma$ | $0.007 \pm 0.002$ |
| $f_2'(1525) \to \rho(770)\,\gamma$ | $0.32 \pm 0.08$ |
| $f_2'(1525) \to \omega(782)\,\gamma$ | $0.012 \pm 0.005$ |
| $f_2'(1525) \to \phi(1020)\,\gamma$ | $0.61 \pm 0.12$ |

**Table 3.12:** Radiative decay rates for $T \to \gamma + V_1$ .

"Sill" spectral function discussed in Ref. [126] for the $\rho$-meson as

$$d_\rho(y) = \frac{2y}{\pi} \frac{\sqrt{y^2 - 4m_\pi^2}\,\tilde{\Gamma}_\rho}{(y^2 - m_\rho^2)^2 + (\sqrt{y^2 - 4m_\pi^2}\,\tilde{\Gamma}_\rho)^2}\,\Theta(y - 2m_\pi),\ \int_0^\infty dy\, d_\rho(y) = 1\,, \quad (3.58)$$

where the definition of $\tilde{\Gamma}_\rho$ is

$$\tilde{\Gamma}_\rho \equiv \frac{\Gamma_{\rho\to 2\pi}\, m_\rho}{\sqrt{m_\rho^2 - 4m_\pi^2}}\ , \quad (3.59)$$

with the PDG values $\Gamma_{\rho\to 2\pi} = 149.1$ MeV and $m_\rho = 775.11$ MeV.

The Sill spectral function is chosen because of the following advantages: it is normalized even for broad states, it has a vanishing real part contribution of virtual particles, and it takes into account the decay threshold. For further details, see Ref. [126].

We provide the following estimates by taking into account destructive interference between the relevant amplitudes (the constructive one leads to extremely large decay

rates):

$$\Gamma_{f_2\to\rho\rho\to 4\pi} \simeq \int_0^\infty dy_1\, d_\rho(y_1) \int_0^\infty dy_2\, d_\rho(y_2)\, \Gamma_{f_2\to\rho\rho}(m_{f_2}, y_1, y_2) \approx 29.9\,\text{MeV}\,; \quad (3.60)$$

$$\Gamma_{a_2\to\omega\rho\to\omega 2\pi} \simeq \int_0^\infty dy_1\, d_\rho(y_1)\, \Gamma_{a_2\to\rho\omega}(m_{\mathbf{a}_2}, m_{\omega_1}, y_1) \approx 11.1\,\text{MeV}\,; \quad (3.61)$$

$$\Gamma_{K_2\to K^\star\rho\to K2\pi} \simeq \int_0^\infty dy_1\, d_\rho(y_1)\, \Gamma_{K_2\to\rho\omega}(m_{K_2}, m_{K_1}, y_1) \approx 6.6\,\text{MeV}\,. \quad (3.62)$$

Given that no new parameter is required, it is interesting that the results are qualitatively consistent with the experimental data. In fact, this is a complex process that includes the integral over the spectrum function of the $\rho$-meson and the destructive interference of two different interaction terms coming from different Lagrangians.

## 3.5 Results for $J^{PC} = 2^{--}$ mesons

In this subsection, we present the chiral model predictions for the decay rates of the ground-state mesons with $J^{PC} = 2^{--}$. Furthermore, comparisons with existing LQCD and other theoretical results are given.

### 3.5.1 Decay channel $A_2 \to V_1 + P$

The decay of axial-tensor mesons into a vector-pseudoscalar pair is contained in the chiral Lagrangian (3.12):

$$\mathcal{L}_{a_2vp} = g_2^{\text{ten}}\, \text{Tr}\Big[A_2^{\mu\nu}\{(\partial_\mu\mathcal{P}), V_{1\,\nu}\}\Big]\,. \quad (3.63)$$

The corresponding tree-level decay rate reads:

$$\Gamma_{A_2\to V_1+P}(m_{a_2}, m_v, m_p) = \frac{g_2^{\text{ten}\,2}\, |\vec{k}_{a_2,v,p}|^3}{120\,\pi\, m_{a_2}^2}\Big(5 + \frac{2\,|\vec{k}_{a_2,v,p}|^2}{m_v^2}\Big)\, \kappa_{a_2vp\,,i}\, \Theta(m_{a_2} - m_v - m_p)\,, \quad (3.64)$$

where the $\kappa_{a_2vp,i}$ are listed in Table 3.14. Table 3.14 shows rather large results for decay widths. For a number of reasons, we might consider our findings to be qualitative:

1. the data of the tensor mesons alone determines the intensity of the interaction of axial-tensors. Although, in line with chiral symmetry, no modifications have been implemented in the (axial-)tensor sector, chiral symmetry breaking effects can be noticeable between 1-2 GeV;

2. the decays of axial-tensor states depend on variables of the kind $(Zw)^2 \propto g_1^4$ (as seen from the Tables 3.14 and 3.2), therefore even a tiny variation in these parameters of the initial eLSM leads to a significant variation in the outcomes. Sine a 20% uncertainty is estimated in $Z$'s that implies a factor 2 in decay rates;

3. it is assumed that the parameter $g_1$, along with others like $g_2^{\text{ten}}$, is a constant over a wide range of energies; however, it is plausible to suppose that it has a (soft) energy dependence. Also, in this scenario, even a minor variation would have a significant impact on the results of Table 3.14.

In conclusion, it can be said that the chosen parameters, especially $g_1$, represent a significant and challenging source of uncertainty in the decays of axial-tensor states. To be fair, we thus assume an error for the presented values in Table 3.14 of roughly 50%.

For the $\rho_2$, $\omega_{2,N} \simeq \omega_2$, and $\omega_{2,S} \simeq \phi_2$, experimental data is not available, while for the kaonic member of the axial-tensors, it is believed to be a mixture of the resonances $K_2(1820)$ and $K_2(1770)$. This is an analogous mixing between the axial-vector and pseudovector kaonic mesons described in Eq. (2.71).

Assuming that $K_{2A}$ is mostly located in $K_2(1820)$, the decay width for the $K_{2A}$ state described in our model is almost twice as wide as the experimentally observed decay of $K_2(1820)$. This implies that the left sides of Table 3.14 range for the $K_{2A}$ state are preferred.

By comparing our outcomes with the lattice QCD data shown in Table 3.13, this interpretation may be applied to the remaining decay rates. Namely, the outcomes are of the same order of magnitude as the LQCD results but are around 2-3 times greater. This comparison confirms that the lower part of the range of each decay mode of the eLSM should be favored, despite the significant uncertainty in both methods.

| Decay process (in model) | eLSM (MeV) | LQCD (MeV) [28] |
|---|---|---|
| $\rho_2(?) \to \rho(770)\,\eta$ | $\approx 30$ | |
| $\rho_2(?) \to \bar{K}^*(892)\,K + \text{c.c.}$ | $\approx 27$ | 36 |
| $\rho_2(?) \to \omega(782)\,\pi$ | $\approx 122$ | 125 |
| $\rho_2(?) \to \phi(1020)\,\pi$ | $\approx 0.3$ | |
| $K_{2,A} \to \rho(770)\,K$ | $\approx 53$ | |
| $K_{2,A} \to \bar{K}^*(892)\,\pi$ | $\approx 87$ | |
| $K_{2,A} \to \bar{K}^*(892)\,\eta$ | $\approx 0.004$ | |
| $K_{2,A} \to \omega(782)\,\bar{K}$ | $\approx 13.8$ | |
| $K_{2,A} \to \phi(1020)\,\bar{K}$ | $\approx 13.7$ | |
| $\omega_{2,N} \to \rho(770)\,\pi$ | $\approx 363$ | 365 |
| $\omega_{2,N} \to \bar{K}^*(892)\,K + \text{c.c.}$ | $\approx 25$ | 36 |
| $\omega_{2,N} \to \omega(782)\,\eta$ | $\approx 27$ | 17 |
| $\omega_{2,N} \to \phi(1020)\,\eta$ | $\approx 0.02$ | |
| $\omega_{2,S} \to \bar{K}^*(892)\,K + \text{c.c.}$ | $\approx 100$ | 148 |
| $\omega_{2,S} \to \omega(782)\,\eta$ | $\approx 0.2$ | |
| $\omega_{2,S} \to \omega(782)\,\eta'(958)$ | $\approx 0.02$ | |
| $\omega_{2,S} \to \phi(1020)\,\eta$ | $\approx 17$ | 44 |

**Table 3.13:** Decay rates for $A_2 \to V_1 + P$ compared to LQCD results [28].

In order to further check our results, we perform an additional fit to the decay rates obtained in LQCD simulations, which yields

$$g_{2\,\text{lat}}^{\text{ten}} \simeq (0.7 \pm 0.4) \cdot 10^4 (\text{MeV})\,, \tag{3.65}$$

where we assumed of 50 % error in [28]. The result is about two sigmas away from the former result in Eq. (3.42)[5]. The corresponding decays are given in Table 3.13. It's rather intriguing that the different results all match each other, indicating that any ratio of decays generated using lattice data agrees with the appropriate eLSM ratio. Future lattice results could be useful to compare other decay channels too.

[5]It should be noted that the lowest pseudoscalar meson in the $SU(3)$ flavor symmetry-based LQCD analysis has a mass of around 700 MeV.

In conclusion, the result of our theoretical investigation is quite evident: even when assuming the lower limit of our estimations, the axial-tensor states are anticipated to be broad, despite the fact that there are still significant uncertainties. This information may eventually explain why the hypothesized states $\rho_2$, $\omega_2$, and $\phi_2$ have not yet been experimentally seen.

| Decay process (in model) | $\kappa_{a_2vp,i}$ | eLSM (MeV) |
|---|---|---|
| $\rho_2(?) \to \rho(770)\,\eta$ | $\Big(-w_{\eta_N} Z_{\eta_N} \cos\beta_P\Big)^2$ | $\approx 99 \pm 50$ |
| $\rho_2(?) \to \bar{K}^*(892)\,K + \text{c.c.}$ | $4\Big(\frac{Z_K w_K}{2}\Big)^2$ | $\approx 85 \pm 43$ |
| $\rho_2(?) \to \omega(782)\,\pi$ | $\Big(-w_\pi Z_\pi \cos\beta_V\Big)^2$ | $\approx 419 \pm 210$ |
| $\rho_2(?) \to \phi(1020)\,\pi$ | $\Big(w_\pi Z_\pi \sin\beta_V\Big)^2$ | $\approx 0.8$ |
| $K_{2,A} \to \rho(770)\,K$ | $3\Big(\frac{Z_K w_K}{2}\Big)^2$ | $\approx 195 \pm 98$ |
| $K_{2,A} \to \bar{K}^*(892)\,\pi$ | $3\Big(\frac{Z_\pi w_\pi}{2}\Big)^2$ | $\approx 316 \pm 158$ |
| $K_{2,A} \to \bar{K}^*(892)\,\eta$ | $\Big(-\frac{1}{2}(\sqrt{2}Z_{\eta_S} w_{\eta_S} \sin\beta_P + Z_{\eta_N} w_{\eta_N} \cos\beta_P)\Big)^2$ | $\approx 0.01$ |
| $K_{2,A} \to \omega(782)\,\bar{K}$ | $\Big(-\frac{w_K Z_K}{2}(\sqrt{2}\sin\beta_V + \cos\beta_V)\Big)^2$ | $\approx 51 \pm 26$ |
| $K_{2,A} \to \phi(1020)\,\bar{K}$ | $\Big(\frac{w_K Z_K}{2}(-\sqrt{2}\cos\beta_V + \sin\beta_V)\Big)^2$ | $\approx 50 \pm 25$ |
| $\omega_{2,N} \to \rho(770)\,\pi$ | $3\Big(-Z_\pi w_\pi\Big)^2$ | $\approx 1314 \pm 657$ |
| $\omega_{2,N} \to \bar{K}^*(892)\,K + \text{c.c.}$ | $4\Big(-\frac{Z_K w_K}{2}\Big)^2$ | $\approx 85 \pm 43$ |
| $\omega_{2,N} \to \omega(782)\,\eta$ | $\Big(Z_{\eta_N} w_{\eta_N} \cos\beta_P \cos\beta_V\Big)^2$ | $\approx 93 \pm 47$ |
| $\omega_{2,N} \to \phi(1020)\,\eta$ | $\Big(Z_{\eta_N} w_{\eta_N} \cos\beta_P \sin\beta_V\Big)^2$ | $\approx 0.06$ |
| $\omega_{2,S} \to \bar{K}^*(892)\,K + \text{c.c.}$ | $4\Big(-\frac{Z_K w_K}{\sqrt{2}}\Big)^2$ | $\approx 510 \pm 255$ |
| $\omega_{2,S} \to \omega(782)\,\eta$ | $\Big(-\sqrt{2}Z_{\eta_S} w_{\eta_S} \sin\beta_P \sin\beta_V\Big)^2$ | $\approx 1.0 \pm 0.5$ |
| $\omega_{2,S} \to \omega(782)\,\eta'(958)$ | $\Big(-\sqrt{2}Z_{\eta_S} w_{\eta_S} \cos\beta_P \sin\beta_V\Big)^2$ | $\approx 0.3$ |
| $\omega_{2,S} \to \phi(1020)\,\eta$ | $\Big(-\sqrt{2}Z_{\eta_S} w_{\eta_S} \cos\beta_V \sin\beta_P\Big)^2$ | $\approx 101 \pm 51$ |

**Table 3.14:** Clebsch-Gordan coefficients and corresponding decay rates for $A_2 \to V_1 + P$.

### 3.5.2 Decay channel $A_2 \to A_1 + P$

The decay of the axial-tensor mesons into an axial-vector and pseudoscalar pair is contained in the chiral Lagrangian of Eq. (3.17):

$$\mathcal{L}_{a_2 a_1 p} = c'^{\,\text{ten}}\, \varepsilon_{\mu\nu\rho\sigma}\, \text{Tr}\Big[\partial^\mu A_2^{\nu\alpha}\Big((\partial^\rho A_1^\sigma)(\partial_\alpha P)\Phi_0 - \Phi_0(\partial_\alpha P)(\partial^\rho A_1^\sigma)\Big)\Big]\,. \tag{3.66}$$

The tree-level decay rate reads:

$$\Gamma_{A_2\to A_1+P}(m_{a_2}, m_{a_1}, m_p) = \frac{c'^{\,\text{ten}\,2}\,|\vec{k}_{a_2,a_1,p}|^5}{40\,\pi}\,\kappa_{a_2 ap\,,i}\Theta(m_{a_2} - m_{a_1} - m_p)\ , \tag{3.67}$$

where $c'^{\,\text{ten}} := c_1^{\text{ten}} - c_2^{\text{ten}}$ and $\kappa_{a_2 ap\,,i}$ as well as the theoretical predictions proportional to the factor $\frac{c'^{\,\text{ten}}}{c^{\text{ten}}}$ are given in Table 3.15 under the assumption that that $K_{1A} \approx K_1(1400)$.

Note, the presence of two components proportional to $c_1^{\text{ten}}$ and $c_2^{\text{ten}}$ in the main Lagrangian of Eq. (3.17), results in their sum $(c_1^{\text{ten}} + c_2^{\text{ten}})$ and difference $(c_1^{\text{ten}} - c_2^{\text{ten}})$ for tensors and axial-tensors, respectively. In this situation, the decay of the tensor mesons does not fix the coupling for the axial-tensors. In spite of this, it is fair to assume that $c'^{\,\text{ten}}$ is of the same order as $c^{\text{ten}}$, which leads to relatively narrow partial decay widths. Interestingly, under the assumption of $c'^{\,\text{ten}} \simeq c^{\text{ten}}$, our result for the decay channel $\rho_2 \to a_1(1260)\pi$ is remarkably close to the prediction in Ref. [88].

| Decay process (in model) | $\kappa_{a_2 ap\,,i}$ | eLSM (MeV) |
|---|---|---|
| $\rho_2(?) \to a_1(1260)\,\pi$ | $2\left(\frac{\phi_N}{4}\right)^2$ | $\simeq 13\left(\frac{c'^{\,\text{ten}}}{c^{\text{ten}}}\right)^2$ |
| $K_{2,A} \to a_1(1260)\,K$ | $3\left(\frac{\sqrt{2}\phi_S}{8}\right)^2$ | $\simeq 0.1\left(\frac{c'^{\,\text{ten}}}{c^{\text{ten}}}\right)^2$ |
| $K_{2,A} \to \bar{K}_{1,A}\,\pi$ | $3\left(\frac{\phi_N}{8}\right)^2$ | $\simeq 11\left(\frac{c'^{\,\text{ten}}}{c^{\text{ten}}}\right)^2$ |
| $K_{2,A} \to f_1(1285)\,\bar{K}$ | $\left(\frac{\sqrt{2}}{8}(\phi_N \sin\beta_{A_1} - \phi_S\cos\beta_{A_1})\right)^2$ | $\simeq 0.2\left(\frac{c'^{\,\text{ten}}}{c^{\text{ten}}}\right)^2$ |
| $\omega_{2,S} \to \bar{K}_{1,A}\,K + \text{c.c.}$ | $4\left(\frac{\phi_S}{4}\right)^2$ | $\simeq 6\left(\frac{c'^{\,\text{ten}}}{c^{\text{ten}}}\right)^2$ |

**Table 3.15:** Clebsch-Gordan coefficients and corresponding decay rates for $A_2 \to A_1 + P$. We recall that $c^{\text{ten}} := c_1^{\text{ten}} + c_2^{\text{ten}}$ and $c'^{\,\text{ten}} := c_1^{\text{ten}} - c_2^{\text{ten}}$.

| Decay process (in model) | $\kappa_{a_2tp,i}$ | eLSM (MeV) |
|---|---|---|
| $\rho_2(?) \to a_2(1320)\,\pi$ | $2\left(\frac{\phi_N}{4}\right)^2$ | $\approx 88$ |
| $K_{2,A} \to K_2^\star(1430)\,\pi$ | $3\left(\frac{\sqrt{2}\phi_S}{8}\right)^2$ | $\approx 49$ |
| $K_{2,A} \to a_2(1320)\,K$ | $3\left(\frac{\sqrt{2}\phi_N}{8}\right)^2$ | $\approx 84$ |
| $K_{2,A} \to f_2(1270)\,K$ | $\left(\frac{\phi_N \cos\beta_T - 2\phi_S \sin\beta_T}{8}\right)^2$ | $\approx 4$ |
| $\omega_{2,S} \to K_2^\star(1430)\,K + \text{c.c.}$ | $4\left(\frac{\phi_S}{8}\right)^2$ | $\approx 15$ |

**Table 3.16:** Clebsch-Gordan coefficients and corresponding decay rates for $A_2 \to T + P$.

### 3.5.3 Decay channel $A_2 \to T + P$

We estimate how the axial-tensor mesons decay into a tensor-pseudoscalar pair using the last term of the Lagrangian Eq. (3.9), which contains the following interaction Lagrangian:

$$\mathcal{L}_{a_2tp} = 2ih_3^{\text{ten}}\text{Tr}\Big[A_2^{\mu\nu}(PT_{\mu\nu}\Phi_0 - \Phi_0 T_{\mu\nu}P)\Big]\,. \tag{3.68}$$

The tree-level decay rate reads

$$\Gamma_{A_2\to T+P}(m_{a_2}, m_t, m_p) = \frac{(h_3^{\text{ten}})^2|\vec{k}_{a_2,t,p}|}{2\,m_{a_2}^2\pi}\Big(1 + \frac{4|\vec{k}_{a_2,t,p}|^4}{45m_t^4} + \frac{2|\vec{k}_{a_2,t,p}|^2}{3m_t^2}\Big) \kappa_{a_2tp,i}\,\Theta(m_{a_2} - m_t - m_p)\,, \tag{3.69}$$

where $\kappa_{a_2tp,i}$ as well as the theoretical results are presented in Table 3.16. It is visible that rather large decays are produced in certain channels. It should be noted that the decay rate of $\rho_2(?) \to a_2(1320)\,\pi$ is computed in [88] to be roughly 200 MeV. This makes the expected overall decay widths of axial-tensor mesons even larger.

In conclusion, it should be emphasized that even assuming the lower part of the presented ranges, the total decay widths of the axial-tensor mesons are large. Nevertheless, it appears that the missing axial-tensor states could be found in ongoing and future experiments, particularly by examining the decay channels discussed in this subsection, even though it will be challenging to identify such broad states.

## 3.6 Conclusion

In the context of a chiral model for low-energy QCD called the eLSM, we have examined the well-known lightest tensor mesons and their chiral companions, the axial-tensor mesons. Results for the tensor mesons, e.g., masses and decay widths, qualitatively support the idea of a quark-anti-quark picture of this nonet, as indicated in Fig. 3.2.

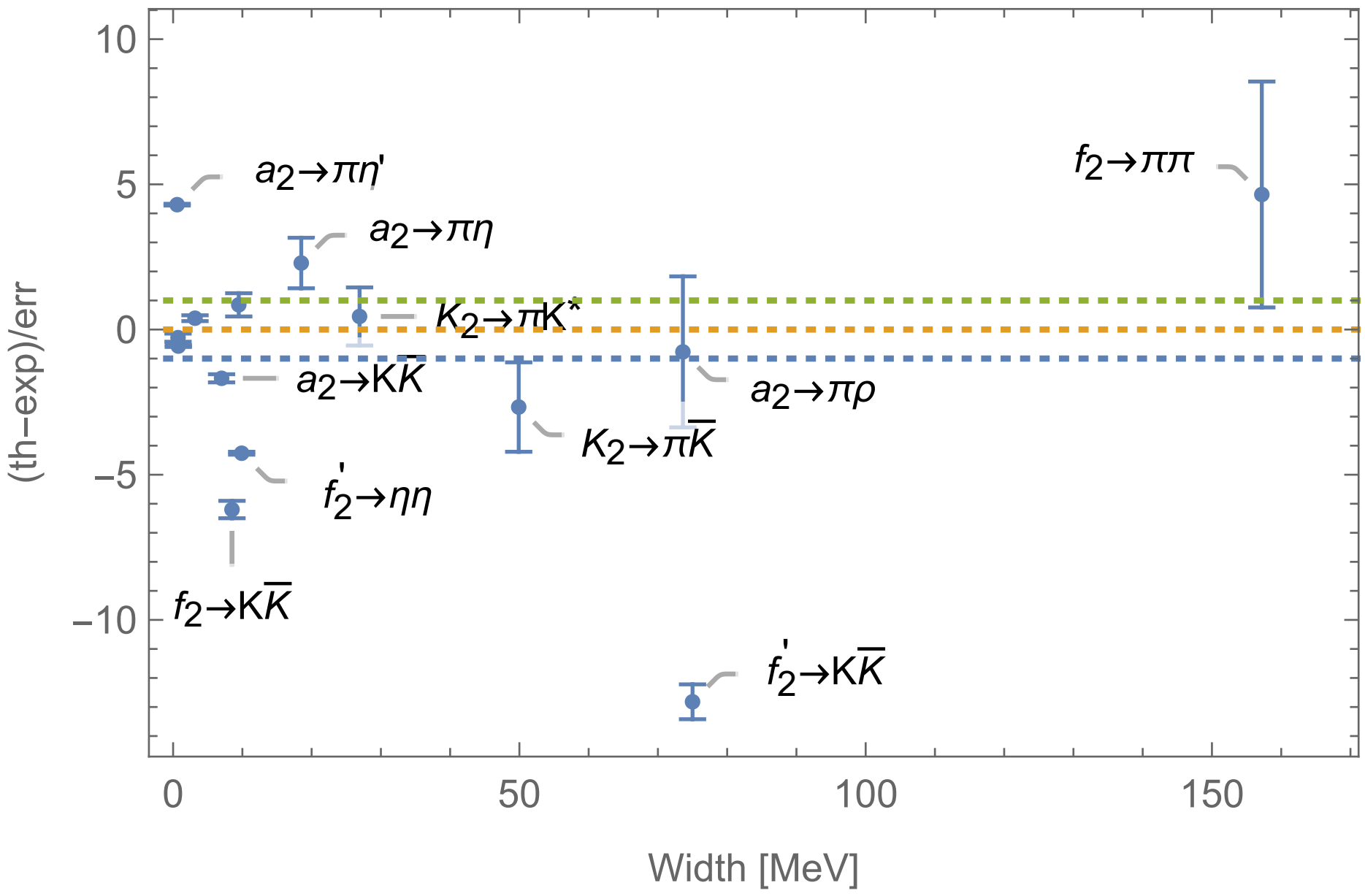

**Figure 3.2:** Comparison of the eLSM results to PDG [27].

Because of chiral symmetry, it is possible to predict masses and decay rates for the missing axial-tensor mesons using the parameters found in the tensor sector. We obtain large and dominant decay widths for the vector-pseudoscalar channel of axial-tensor mesons, yet with large uncertainties. Sizable decay widths for this decay channel are also achieved via LQCD, and the corresponding decay ratios match well with the ones calculated within our model.

# Chapter 4

# Spin-2 Glueball

In this chapter, we include the tensor glueball (with $J^{PC} = 2^{++}$) into the eLSM introduced in the previous chapter. This state is the second-lightest one in the LQCD glueball spectrum after the scalar glueball. The experimental observation of several resonances with $J^{PC} = 2^{++}$, such as $f_2(1430)$, $f_2(1565)$, $f_2(1640)$, $f_2(1810)$, $f_2(1910)$, $f_2(1950)$, $f_2(2010)$, $f_2(2150)$, $f_J(2220)$, poses the question if one of them can be interpreted as a tensor glueball.

## 4.1 Introduction

Mesons known as "glueballs" contain only gluons. Despite the existence of glueballs being one of the earliest predictions of QCD, these objects are still experimentally disputed (some candidates do exist but a final consensus is missing) [64, 127–129]. There are many lattice investigations that confirm the existence of glueballs [29, 130–135]. Yet, it is still difficult to distinguish glueballs in experiments because of their mixing with conventional quark-antiquark mesons. Nevertheless, new data analysis on $J/\psi$ decays appears to be helpful toward identifying the scalar and tensor glueballs [71, 73, 136, 137].

Several masses for the tensor glueball are estimated by LQCD simulations, including:

- 2150(30)(100) MeV [132];
- 2324(42)(32) MeV [133];

- 2376(32) MeV [29];
- 2390(30)(120) MeV [130];
- 2400(25)(120) MeV [131].

The tensor glueball mass is predicted by other theoretical models as well. According to the QCD sum rules, it is predicted to be in the range 2000 MeV-2300 MeV [138, 139], functional methods prediction is around 2610 MeV [36], while the holographic techniques estimate lies in the interval 1900 MeV − 4300 MeV [33, 140–146].

| Resonances | Masses (MeV) | Decay Widths (MeV) | Decay Channels |
|---|---|---|---|
| $f_2(1910)$ | $1900 \pm 9$ | $167 \pm 21$ | $\pi\pi$, $KK$, $\eta\eta$, $\omega\omega$, $\eta\eta'$, $\eta'\eta'$, $\rho\rho$, $a_2(1320)\pi$, $f_2(1270)\eta$ |
| $f_2(1950)$ | $1936 \pm 12$ | $464 \pm 24$ | $\pi\pi$, $K^*K^*$, $KK$, $\eta\eta$ |
| $f_2(2010)$ | $2011^{+60}_{-80}$ | $202 \pm 60$ | $\phi\phi$, $KK$ |
| $f_2(2150)$ | $2157 \pm 12$ | $152 \pm 30$ | $\pi\pi$, $\eta\eta$, $KK$, $a_2(1320)\pi$, $f_2(1270)\eta$ |
| $f_J(2220)$ | $2231.1 \pm 3.5$ | $23^{+8}_{-7}$ | $\eta\eta'$ |
| $f_2(2300)$ | $2297 \pm 28$ | $149 \pm 41$ | $\phi\phi$, $KK$ |
| $f_2(2340)$ | $2345^{+50}_{-40}$ | $322^{+70}_{-60}$ | $\phi\phi$, $\eta\eta$ |

**Table 4.1:** Spin 2 resonances with masses around 2 GeV in PDG [27].

An effective model describing the tensor glueball can be found in Refs. [83, 147]. Recently, the role of the tensor glueball via the glueball scattering within the Glueball Resonance Gas model was studied in [4] (see also chapter 6). Various glueballs were studied within the eLSM, including the scalar glueball [148], the vector glueball [101], the pseudotensor glueball (with $J^{PC} = 2^{-+}$) [85], the spin-3 glueball (with $J^{PC} = 3^{--}$) [1], and glueball molecules [149].

## 4.2 Chiral Lagrangians for $G_2$

Starting from the chiral Lagrangians in Eq. (3.12) and in Eq. (3.14), we perform the following substitution in order to describe the tensor glueball decays:

$$T_{\mu\nu} \longrightarrow \frac{1}{\sqrt{6}} G_{2,\mu\nu} \cdot \mathbb{1}_{3\times 3} \,, \tag{4.1}$$

which implies the flavor blindness of the glueball state. Thus, the Lagrangian for the tensor glueball interaction with (axial-)vector mesons reads

$$\mathcal{L}_{\lambda_2} = \frac{\lambda_2}{\sqrt{6}} G_{2,\mu\nu} \Big( \mathrm{Tr}\Big[ \{L^\mu, L^\nu\} \Big] + \mathrm{Tr}\Big[ \{R^\mu, R^\nu\} \Big] \Big), \tag{4.2}$$

which leads to three kinematically allowed decay channels:

- The vector-vector ($G_2 \to V_1 V_1$) decay rate of the tensor glueball has the following form:

$$\Gamma_{G_2 \to V_1^{(1)} + V_1^{(2)}}(m_{g_2}, m_{v^{(1)}}, m_{v^{(2)}}) = \frac{\kappa_{gvv,i} \lambda_2^2 |\vec{k}_{g_2,v^{(1)},v^{(2)}}|}{120\,\pi\, m_{g_2}^2} \Big( 15 + \frac{5|\vec{k}_{g_2,v^{(1)},v^{(2)}}|^2}{m_{v^{(1)}}^2} + \frac{5|\vec{k}_{g_2,v^{(1)},v^{(2)}}|^2}{m_{v^{(2)}}^2} + \frac{2|\vec{k}_{g_2,v^{(1)},v^{(2)}}|^4}{m_{v^{(1)}}^2 m_{v^{(2)}}^2} \Big) \Theta(m_{g_2} - m_{v^{(1)}} - m_{v^{(2)}}) \,; \tag{4.3}$$

- The decay of $G_2$ into axial-vector and pseudoscalar ($G_2 \to A_1 P$) mesons is obtained via the application of the shift Eq. (3.8) once in Eq. (4.2), leading to:

$$\Gamma_{G_2 \to A_1 + P}(m_{g_2}, m_{a_1}, m_p) = \frac{\kappa_{gap,i}\, \lambda_2^2\, |\vec{k}_{g_2,a_1,p}|^3}{120\,\pi\, m_{g_2}^2} \left( 5 + \frac{2\,|\vec{k}_{g_2,a_1,p}|^2}{m_{a_1}^2} \right) \Theta(m_{g_2} - m_{a_1} - m_p) \,; \tag{4.4}$$

- The decay into two pseudoscalars ($G_2 \to PP$) is obtained considering the SSB-related shift of Eq. (3.8) twice within Eq. (4.2)

$$\Gamma_{G_2 \to P^{(1)} + P^{(2)}}(m_{g_2}, m_{p^{(1)}}, m_{p^{(2)}}) = \frac{\kappa_{gpp,i}\, \lambda_2^2\, |\vec{k}_{g_2,p^{(1)},p^{(2)}}|^5}{60\,\pi\, m_{g_2}^2} \Theta(m_{g_2} - m_{p^{(1)}} - m_{p^{(2)}}) \,. \tag{4.5}$$

Since the coupling constant $\lambda_2$ is not known, we are at first limited to computing decay ratios rather than decay rates. Later on, a guess based on the large-$N_c$ estimate for two pseudoscalar decays will be discussed in Sec. 4.4.

An analogous Lagrangian to the last term of Eq. (3.9) for the tensor glueball decay into a tensor meson and a pseudoscalar meson reads

$$\mathcal{L}_{\alpha_2} = \frac{\alpha_2}{\sqrt{6}} G_{2,\mu\nu} \left( \mathrm{Tr}\left[ \Phi \mathbf{R}^{\mu\nu} \Phi^\dagger \right] + \mathrm{Tr}\left[ \Phi^\dagger \mathbf{L}^{\mu\nu} \Phi \right] \right) . \tag{4.6}$$

The decay rate for this process is given by

$$\Gamma_{G_2 \to T+P}(m_{g_2}, m_t, m_p) = \frac{\alpha_2^2 |\vec{k}_{g_2,t,p}|}{2\, m_{g_2}^2 \pi} \left( 1 + \frac{4|\vec{k}_{g_2,t,p}|^4}{45 m_t^4} + \frac{2|\vec{k}_{g_2,t,p}|^2}{3m_t^2} \right) \times \kappa_{g_2 tp\,,i}\, \Theta(m_{g_2} - m_t - m_p) \,, \tag{4.7}$$

where $\alpha_2$ is a dimensionless unknown parameter.

The last chiral term for $G_2$ is derived from the Lagrangian in Eq. (3.14) and describes the decay of a tensor glueball into a vector and a pseudoscalar meson:

$$\begin{aligned} \mathcal{L}_{c^{\mathrm{glu}}} = &\frac{c_1^{\mathrm{glu}}}{\sqrt{6}} \partial^\mu G_2^{\nu\alpha} \mathrm{Tr}\left[ \tilde{L}_{\mu\nu}\, \partial_\alpha \Phi\, \Phi^\dagger - \Phi^\dagger\, \partial_\alpha \Phi \tilde{R}_{\mu\nu} - \tilde{R}_{\mu\nu} \partial_\alpha \Phi^\dagger \Phi + \Phi \partial_\alpha \Phi^\dagger \tilde{L}_{\mu\nu} \right] \\ &+ \frac{c_2^{\mathrm{glu}}}{\sqrt{6}} \partial^\mu G_2^{\nu\alpha} \mathrm{Tr}\left[ \partial_\alpha \Phi \tilde{R}_{\mu\nu}\ \Phi^\dagger - \Phi^\dagger\, \tilde{L}_{\mu\nu} \partial_\alpha \Phi - \partial_\alpha \Phi^\dagger \tilde{L}_{\mu\nu} \Phi + \Phi \tilde{R}_{\mu\nu} \partial_\alpha \Phi^\dagger \right] . \end{aligned} \tag{4.8}$$

The tree-level decay rate formula reads:

$$\Gamma_{G\longrightarrow VP}(m_{g_2}, m_v, m_p) = \frac{c^{\mathrm{glu}\,2}\, |\vec{k}_{g_2,v,p}|^5}{40\,\pi} \kappa_{gvp\,,i}\, \Theta(m_{g_2} - m_v - m_p) \,, \tag{4.9}$$

where $c^{\mathrm{glu}} := c_1^{\mathrm{glu}} + c_2^{\mathrm{glu}}$. The only non-zero term in this channel is $KK^*$ proportional to $(\phi_N - \sqrt{2}\phi_S)$, which vanishes in the chiral limit. It is then expected to be very small.

## 4.3 Results for the decay ratios

An enhancement in the mass distribution around 2210 MeV was observed in the analyses of BESIII data in Ref. [73]. We chose this value for the mass of the tensor glueball in order to estimate the decay ratios via $\kappa_{goo\,,i}$ in Table 4.2 and Eqs. (4.3), (4.5), (4.4). The corresponding results in Table 4.3 show that the dominant decays are $\rho\rho$ and $K^*K^*$ decay channels.

| Decay process | $\kappa_{goo,i}$ |
|---|---|
| $G_2 \longrightarrow \rho(770)\,\rho(770)$ | 1 |
| $G_2 \longrightarrow \overline{K}^\star(892)\,K^\star(892)$ | $\frac{4}{3}$ |
| $G_2 \longrightarrow \omega(782)\,\omega(782)$ | $\frac{1}{3}$ |
| $G_2 \longrightarrow \omega(782)\,\phi(1020)$ | 0 |
| $G_2 \longrightarrow \phi(1020)\,\phi(1020)$ | $\frac{1}{3}$ |
| $G_2 \longrightarrow \pi\,\pi$ | $(Z_\pi^2 w_\pi^2)^2$ |
| $G_2 \longrightarrow \bar{K}\,K$ | $\frac{4}{3} \times (Z_k^2 w_k^2)^2$ |
| $G_2 \longrightarrow \eta\,\eta$ | $\frac{1}{3} \times (Z_{\eta_N}^2 w_{\eta_N}^2 \cos\beta_P^2 + Z_{\eta_S}^2 w_{\eta_S}^2 \sin\beta_P^2)^2$ |
| $G_2 \longrightarrow \eta\,\eta'(958)$ | $\frac{1}{3} \times ((Z_{\eta_N}^2 w_{\eta_N}^2 - Z_{\eta_S}^2 w_{\eta_S}^2)\cos\beta_P \sin\beta_P)^2$ |
| $G_2 \longrightarrow \eta'(958)\,\eta'(958)$ | $\frac{1}{18} \times (Z_{\eta_S}^2 w_{\eta_S}^2 \cos\beta_P^2 + Z_{\eta_N}^2 w_{\eta_N}^2 \sin\beta_P^2)^2$ |
| $G_2 \longrightarrow a_1(1260)\,\pi$ | $\frac{1}{2} \times (Z_\pi w_\pi)^2$ |
| $G_2 \longrightarrow f_1(1285)\,\eta$ | $\frac{1}{6}(Z_{\eta_S} w_{\eta_S} \sin\beta_{A_1} \sin\beta_P + Z_{\eta_N} w_{\eta_N} \cos\beta_{A_1} \cos\beta_P)^2$ |
| $G_2 \longrightarrow K_{1,A}\,K$ | $\frac{2}{3} \times (Z_k w_k)^2$ |
| $G_2 \longrightarrow f_1(1420)\,\eta'(958)$ | $\frac{1}{6}(Z_{\eta_N} w_{\eta_N} \sin\beta_{A_1} \sin\beta_P + Z_{\eta_S} w_{\eta_S} \cos\beta_{A_1} \cos\beta_P)^2$ |
| $G_2 \longrightarrow f_1(1285)\,\eta'(958)$ | $\frac{1}{6}(Z_{\eta_S} w_{\eta_S} \sin\beta_{A_1} \cos\beta_P - Z_{\eta_N} w_{\eta_N} \cos\beta_{A_1} \sin\beta_P)^2$ |
| $G_2 \longrightarrow f_2(1270)\,\eta$ | $\frac{1}{24}(\phi_N \cos\beta_p \cos\beta_T + \phi_S \sin\beta_p \sin\beta_T)^2$ |
| $G_2 \longrightarrow a_2(1320)\,\pi$ | $\frac{\phi_N^2}{8}$ |
| $G_2 \longrightarrow K^*(892)\,K + \text{c.c.}$ | $\frac{1}{48}\left(\sqrt{2}\phi_N - 2\phi_S\right)^2$ |

**Table 4.2:** Clebsh-Gordan coefficients for the decays of $G_2$.

Quite large errors are expected in the results of Table 4.3 because of SSB-related parameters $Z_i$'s. As we have discussed in the previous chapter, 50% indeterminacy for the decay ratio $\rho\rho$-$\pi\pi$ can be assumed. Nevertheless, this ratio is still expected to be large. Interestingly, dominant decay channels of $G_2$ into $\rho\rho$ and $K^*K^*$ are also obtained in [33] despite being smaller than our estimation. The change in the decay ratios depending on the tensor glueball mass is illustrated in Fig. 4.1. Next, we check the tensor glueball assignment for the resonances listed in Table 4.1.

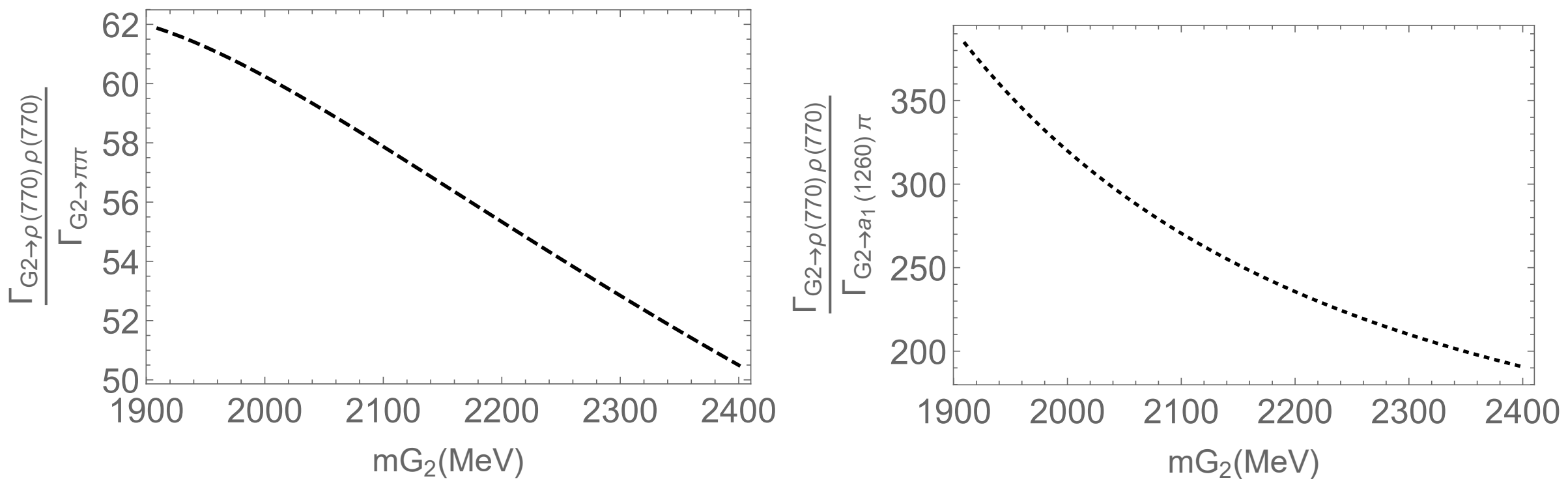

**Figure 4.1:** Decay ratios for different values of tensor glueball mass.

| Decay Ratio | eLSM | Decay Ratio | eLSM | Decay Ratio | eLSM |
|---|---|---|---|---|---|
| $\frac{G_2(2210)\longrightarrow \bar{K}\,K}{G_2(2210)\longrightarrow \pi\,\pi}$ | 0.4 | $\frac{G_2(2210)\longrightarrow \rho(770)\,\rho(770)}{G_2(2210)\longrightarrow \pi\,\pi}$ | 55 | $\frac{G_2(2210)\longrightarrow a_1(1260)\,\pi}{G_2(2210)\longrightarrow \pi\,\pi}$ | 0.24 |
| $\frac{G_2(2210)\longrightarrow \eta\,\eta}{G_2(2210)\longrightarrow \pi\,\pi}$ | 0.1 | $\frac{G_2(2210)\longrightarrow \bar{K}^*(892)\,\bar{K}^*(892)}{G_2(2210)\longrightarrow \pi\,\pi}$ | 46 | $\frac{G_2(2210)\longrightarrow K_{1,A}\,K}{G_2(2210)\longrightarrow \pi\,\pi}$ | 0.08 |
| $\frac{G_2(2210)\longrightarrow \eta\,\eta'}{G_2(2210)\longrightarrow \pi\,\pi}$ | 0.004 | $\frac{G_2(2210)\longrightarrow \omega(782)\,\omega(782)}{G_2(2210)\longrightarrow \pi\,\pi}$ | 18 | $\frac{G_2(2210)\longrightarrow f_1(1285)\,\eta}{G_2(2210)\longrightarrow \pi\,\pi}$ | 0.02 |
| $\frac{G_2(2210)\longrightarrow \eta'\,\eta'}{G_2(2210)\longrightarrow \pi\,\pi}$ | 0.006 | $\frac{G_2(2210)\longrightarrow \phi(1020)\,\phi(1020)}{G_2(2210)\longrightarrow \pi\,\pi}$ | 6 | $\frac{G_2(2210)\longrightarrow f_1(1285)\,\eta}{G_2(2210)\longrightarrow \pi\,\pi}$ | 0.02 |
| | | | | $\frac{G_2(2210)\longrightarrow f_1(1420)\,\eta}{G_2(2210)\longrightarrow \pi\,\pi}$ | 0.01 |

**Table 4.3:** Decay ratios of $G_2$ w.r.t. $\pi\pi$ .

- We start with the resonance $f_2(1910)$ which has a total decay width of $167 \pm 21$ MeV. The following two decay ratios taken from Tables 4.5 and 4.4 agree well with PDG data:

$$\frac{\Gamma^{\text{PDG}}_{f_2(1910)\to\rho(770)\rho(770)}}{\Gamma^{\text{PDG}}_{f_2(1910)\to\omega(782)\omega(782)}} = 2.6 \pm 0.4\,, \qquad \frac{\Gamma^{\text{eLSM}}_{f_2(1910)\to\rho(770)\rho(770)}}{\Gamma^{\text{eLSM}}_{f_2(1910)\to\omega(782)\omega(782)}} = 3.1\,, \tag{4.10}$$

$$\frac{\Gamma^{\text{PDG}}_{f_2(1910)\to f_2(1270)\eta}}{\Gamma^{\text{PDG}}_{f_2(1910)\to a_2(1320)\pi}} = 0.09 \pm 0.05\,, \qquad \frac{\Gamma^{\text{eLSM}}_{f_2(1910)\to f_2(1270)\eta}}{\Gamma^{\text{eLSM}}_{f_2(1910)\to a_2(1320)\pi}} = 0.07\,.$$

| Resonances | Decay Ratios | PDG | eLSM |
|---|---|---|---|
| $f_2(1910)$ | $f_2(1270)\eta/a_2(1320)\pi$ | $0.09\pm0.05$ | 0.07 |
| $f_2(2150)$ | $f_2(1270)\eta/a_2(1320)\pi$ | $0.79\pm0.11$ | 0.1 |

**Table 4.4:** Decay ratios for the tensor-pseudoscalar pair compared to PDG [27].

| Decay Ratios | eLSM |
|---|---|
| $\frac{G_2(1910)\longrightarrow\rho(770)\,\rho(770)}{G_2(1910)\longrightarrow\pi\,\pi}$ | 62 |
| $\frac{G_2(1910)\longrightarrow\omega(782)\,\omega(782)}{G_2(1910)\longrightarrow\pi\,\pi}$ | 20 |
| $\frac{G_2(1910)\longrightarrow\eta\,\eta}{G_2(1910)\longrightarrow\pi\,\pi}$ | 0.077 |
| $\frac{G_2(1910)\longrightarrow\eta\,\eta'(958)}{G_2(1910)\longrightarrow\pi\,\pi}$ | 0.01 |
| $\frac{G_2(1910)\longrightarrow\bar{K}\,K}{G_2(1910)\longrightarrow\pi\,\pi}$ | 0.31 |

**Table 4.5:** Results for the $f_2(1910)$ mass.

| Decay Ratios | eLSM |
|---|---|
| $\frac{G_2(1950)\longrightarrow\bar{K}^\star(892)\,K^\star(892)}{G_2(1950)\longrightarrow\pi\,\pi}$ | 42 |
| $\frac{G_2(1950)\longrightarrow\eta\,\eta}{G_2(1950)\longrightarrow\pi\,\pi}$ | 0.081 |
| $\frac{G_2(1950)\longrightarrow\bar{K}\,K}{G_2(1950)\longrightarrow\pi\,\pi}$ | 0.32 |

**Table 4.6:** Results for the $f_2(1950)$ mass.

However, the interpretation of this resonance as mostly gluonic is problematic because of the following two large disagreements with PDG data:

$$\frac{\Gamma^{\rm PDG}_{f_2(1910)\to\eta\eta}}{\Gamma^{\rm PDG}_{f_2(1910)\to\eta\eta'(958)}} < 0.05\,, \qquad \frac{\Gamma^{\rm eLSM}_{f_2(1910)\to\eta\eta}}{\Gamma^{\rm eLSM}_{f_2(1910)\to\eta\eta'(958)}} \approx 8\,, \tag{4.11}$$
$$\frac{\Gamma^{\rm PDG}_{f_2(1910)\to\omega(782)\omega(782)}}{\Gamma^{\rm PDG}_{f_2(1910)\to\eta\eta'(958)}} = 2.6\pm0.6\,, \qquad \frac{\Gamma^{\rm eLSM}_{f_2(1910)\to\omega(782)\omega(782)}}{\Gamma^{\rm eLSM}_{f_2(1910)\to\eta\eta'(958)}} \approx 2000\,.$$

- The decay ratios of the resonance $f_2(1950)$ as a glueball are listed in Table 4.6. Both the theoretical and experimental ratios, $K\bar{K}/\pi\pi$ and $\eta\eta/\pi\pi$ agree with each other. When we assume that $\rho\rho$ decays into 4 pions, it is possible to estimate the lower bound of the ratio $4\pi/\eta\eta$, which leads to the order of about 100. The experimental total decay width of $f_2(1950)$ is broad, around 460 MeV. At present, $f_2(1950)$ is, according to our results, the best tensor glueball candidate; see the remainder of the chapter. Interestingly, the Witten-Sakai-Sugimoto model [33] also estimates a large total decay width for the tensor glueball. It is noteworthy that the decay $J/\psi \to \gamma K^*(892)\bar{K}^*(892)$ exhibits a branching ratio of $(7.0 \pm 2.2)\,10^{-4}$, which is relatively high. Namely, a tensor glueball is predicted to have a strong coupling to $J/\psi$ meson.

| Decay Ratios | eLSM |
|---|---|
| $\frac{G_2(2150)\longrightarrow\eta\,\eta}{G_2(2150)\longrightarrow\pi\,\pi}$ | 0.1 |
| $\frac{G_2(2150)\longrightarrow\bar{K}\,K}{G_2(2150)\longrightarrow\pi\,\pi}$ | 0.38 |

**Table 4.7:** Results for the $f_2(2150)$ mass.

| Decay Ratios | eLSM |
|---|---|
| $\frac{G_2(2300)\longrightarrow\bar{K}\,K}{G_2(2300)\longrightarrow\phi(1020)\,\phi(1020)}$ | 0.06 |
| $\frac{G_2(2340)\longrightarrow\eta\,\eta}{G_2(2340)\longrightarrow\phi(1020)\,\phi(1020)}$ | 0.02 |

**Table 4.8:** Results for the $f_2(2300)$ and $f_2(2340)$ masses.

- The next resonance in Table 4.1 is $f_2(2010)$ with a total decay width of $202 \pm 60$ MeV. This state is listed as a well-established meson list of PDG [27]. The main reason for its interpretation as a strange-antistrange state and not as a glueball is because the only observed channels are $K\overline{K}$ and $\phi(1020)\phi(1020)$. The same argument for $f_2(2300)$ does not favor it being a tensor glueball state.

- Although the resonance $f_2(2150)$ does not belong to the list of well-established resonances of PDG, it is still interesting to present the comparison between the model and the experiment. The following three PDG data contradict the chiral model outcomes in Table 4.7 and Table 4.4

$$\frac{\Gamma_{f_2(2150)\to K\overline{K}}}{\Gamma_{f_2(2150)\to\eta\eta}} = 1.28 \pm 0.23\,, \tag{4.12}$$

$$\frac{\Gamma_{f_2(2150)\to\pi\pi}}{\Gamma_{f_2(2150)\to\eta\eta}} < 0.33\,, \tag{4.13}$$

$$\frac{\Gamma_{f_2(2150)\to f_2(1270)\eta}}{\Gamma_{f_2(2150)\to a_2(1320)\pi}} = 0.79 \pm 0.11\,. \tag{4.14}$$

  Thus $f_2(2150)$ does not seem to be a good glueball candidate.

- In the literature, it has been thought that the meson $f_J(2220)$ (with $J = 2$ or $J = 4$) could be a good candidate for the tensor glueball. Our results disagree with this view since neither the model-predicted decay widths have been seen nor the ones seen in the experiment, such as $\eta\eta'$ agree with the model result.

- The last entry of the Table 4.1 is $f_2(2340)$, which decays into $\eta\eta$ and $\phi(1020)\phi(1020)$, what also suggests a strong strange-antistrange component. Still, further study of additional decays of this state is needed in the future.

## 4.4 Discussion

### Large-$N_c$ estimate of decays

Besides decay ratios, it is interesting to present an (albeit rough) estimate of the tensor glueball decays based on the large-$N_c$ arguments.

Conventional tensor mesons, which were studied in the previous chapter, have the following quark structure (the small mixing angle between them is ignored)

$$f_2 \equiv f_2(1270) \simeq \sqrt{1/2}(\bar{u}u + \bar{d}d)\,, \qquad f_2' \equiv f_2'(1525) \simeq \bar{s}s\,. \tag{4.15}$$

Let us consider their two pions decay from the PDG [27]

$$\Gamma_{f_2\to\pi\pi} = 157.2\,\text{MeV}\,, \qquad \Gamma_{f_2'\to\pi\pi} = 0.71\,\text{MeV}\,, \tag{4.16}$$

where we do not take into account small errors due to the qualitative nature of analysis. Because of the quark contents, the decay amplitudes have different $N_c$ scalings:

$$A_{f_2'\to\pi\pi} \propto 1/\sqrt{N_c}\,, \qquad A_{f_2'\to\pi\pi} \propto 1/N_c^{3/2}\,. \tag{4.17}$$

The latter process is OZI suppressed [6, 150, 151] compared to the previous one. It can be considered as a result of the chain:

$$\bar{s}s \to gg \to \sqrt{1/2}(\bar{u}u + \bar{d}d) \to \pi\pi\,. \tag{4.18}$$

In order to be more definite, let us introduce a transition Hamiltonian as

$$H_{int} = \lambda_2\Big(|\bar{u}u\rangle\langle gg| + |\bar{d}d\rangle\langle gg| + |\bar{s}s\rangle\langle gg| + \text{h.c.}\Big)\,, \tag{4.19}$$

where $\lambda_2 \propto 1/\sqrt{N_c}$ is controlling $\bar{q}q$-$gg$ the mixing. The value $\lambda_2 \simeq 0.22$ is a consequence of the following ratios between two decays:

$$\frac{A_{f_2'\to\pi\pi}}{A_{f_2\to\pi\pi}} \simeq \sqrt{2}\lambda_2^2 \longrightarrow \frac{\Gamma_{f_2'\to\pi\pi}}{\Gamma_{f_2\to\pi\pi}} \simeq 2\lambda_2^4\,. \tag{4.20}$$

The numerical value of $\lambda_2$ enables us to estimate the two-pion decay of the tensor glueball as well as other decays within eLSM. In fact, the glueball decay is somehow an "intermediate stage" since a glueball is assumed to be a $gg$ bound state. We obtain [1]

$$A_{G_2\to\pi\pi} \simeq \sqrt{2}\lambda_2\, A_{f_2\to\pi\pi} \longrightarrow \Gamma_{G_2\to\pi\pi} \simeq 2\lambda_2^2\,\Gamma_{f_2\to\pi\pi} \simeq 15\,\text{MeV}\,. \tag{4.21}$$

The other two pseudoscalar decay rates are presented in Table 4.9. As we did not consider other contributions such as phase space, form factors etc., this value is obviously just a very rough approximation. Still, it may be useful as a first guess.

| Decay process | $\Gamma_i$(MeV) |
|---|---|
| $G_2 \to \pi\,\pi$ | $\sim 15$ |
| $G_2 \to \bar{K}\,K$ | $\sim 6$ |
| $G_2 \to \eta\,\eta$ | $\sim 1.6$ |
| $G_2 \to \eta\,\eta'(958)$ | $\sim 0.06$ |
| $G_2 \to \eta'(958)\,\eta'(958)$ | $\sim 0.08$ |

**Table 4.9:** Two pseudoscalar decay estimations of the tensor glueball.

Interestingly, our estimation for $\pi\pi$ decay is comparable to the Witten-Sakai-Sugimoto (WSS) model result [33]

$$\frac{\Gamma^{\text{WSS}}_{G_2\to\pi\pi}}{m_{G(2000)}} \simeq 0.014 \longrightarrow \Gamma^{\text{WSS}}_{G_2\to\pi\pi} \simeq 28\,\text{MeV}\,. \tag{4.22}$$

Considering the decay ratio

$$\frac{\Gamma^{\text{eLSM}}_{G_2\to\rho\rho}}{\Gamma^{\text{eLSM}}_{G_2\to\pi\pi}} = 55\,, \tag{4.23}$$

and assuming the decay of each $\rho$ into two pions, a large decay width for the four-pion decay of the tensor glueball is expected based on the large-$N_c$ arguments.

[1] The estimate of the coupling of glueballs to mesons follows a similar method in Refs. [152, 153].

## Conventional excited tensor mesons

It is expected that some (not all) of the isoscalar tensor resonances of Fig. 1.2 fit radially and orbitally excited conventional tensor mesons rather than the tensor glueball (see, e.g., Refs. [154–156] for further details). Namely, we need to distinguish the tensor glueball from the radial excitation (with $2^3P_2$ ) and orbital excitation (with $1^3F_2$). The quark model predicts that [77], orbital excited tensor states are heavier than radial excitations, similar to the vector meson case [81].

Let us look at the radial excitations. The isotriplet $a_2(1700)$ and the isodoublet $K_2^*(1980)$ and the isoscalar (mostly nonstrange) $f_2(1640)$ are believed to belong to this nonet. According to the quark model review of the PDG [27], $f_2(1950)$ is also a radially excited tensor meson (the mostly $\bar{s}s$ ). Nevertheless, its experimental decay ratio does not favor this interpretation. The following decay ratio:

$$\frac{\Gamma^{\text{PDG}}_{f_2(1950)\to\overline{K}K}}{\Gamma^{\text{PDG}}_{f_2(1950)\to\pi\pi}} \sim 0.8\,, \tag{4.24}$$

should be larger for a $\bar{s}s$ state. The second argument for it not being a radially excited tensor is that its mass is too close to the isodoublet $K_2^*(1980)$.

Another possible candidate for the missing radially excited tensor meson would be the state $f_2(1910)$, which is not reported in the summary table of PDG [27]. Its mass difference with $K_2^*(1980)$ is not in favor of this resonance being a $\bar{s}s$ state. From a decay width point of view, the following decay ratio

$$\frac{\Gamma^{\text{PDG}}_{f_2(1910)\to f_2(1270)\eta}}{\Gamma^{\text{PDG}}_{f_2(1910)\to a_2(1320)\pi}} = 0.09 \pm 0.05\,, \tag{4.25}$$

does not support the strange-antistrange interpretation of $f_2(1910)$ as well.

In contrast, there is the well-established $f_2(2010)$ resonance that fits the strange quark structure mesons since it decays into $\phi\phi$ and $\overline{K}K$. Thus, among the three resonances $\{f_2(1910), f_2(1950), f_2(2010)\}$, the last one fits as a radially excited tensor meson. As a consequence, since $f_2(1910)$ is not established yet, the sole $f_2(1950)$ may be regarded as supernumerary. Thus its interpretation as a tensor glueball is feasible.

We have less information about the orbitally excited tensor mesons. There are three resonances $\{ f_2(2150), f_2(2300), f_2(2340)\}$ that could possibly be candidates for isoscalar $1^3F_2$ states. While $f_2(2150)$ is omitted from the mesonic summary table of PDG [27], one may regard $f_2(2300)$ and $f_2(2340)$ as single state because of their similar mass. Their dominant $\phi\phi$ decay channel would support $\bar{s}s$ structure.

Finally, we present a summary of eLSM-based analyses of spin-2 mesons in Table 4.10.

| Resonances | Interpretation status |
|---|---|
| $f_2(1910)$ | Agreement with some data, but excluded by $\eta\eta/\eta\eta'$ and $\omega\omega/\eta\eta\prime$ ratios |
| $f_2(1950)$ | $\eta\eta/\pi\pi$ ok, no contradictions with data, but implies broad tensor glueball<br>Best fit as predominantly glueball among considered resonances |
| $f_2(2010)$ | Likely strange-antistrange content |
| $f_2(2150)$ | Most available data in disagreement with theoretical prediction |
| $f_J(2220)$ | Data on $\pi\pi/K\bar{K}$ disagrees with theory, the largest predicted decay channels are not seen |
| $f_2(2300)$ | Likely strange-antistrange content |
| $f_2(2340)$ | Likely strange-antistrange content |

**Table 4.10:** Status of spin-2 resonances within eLSM-based analyses.

## 4.5 Conclusions

In this chapter we studied the tensor glueball within the eLSM by using a chirally invariant Lagrangian approach. Results for the tree-level decay ratios imply the dominant channel into two vector mesons as decay products. Besides the decay ratios, we present the estimate for the decay rates, e.g., the $\pi\pi$ decay mode, to be of the order of 10 MeV. A small decay width of the tensor glueball into $K^*(892)K$ is another prediction of our model, which would make this mode potentially interesting to rule out future tensor glueball candidates.

Comparison of the eLSM results for various spin-2 resonances of PDG shows, at present, that the best match as a tensor glueball is the resonance $f_2(1950)$. Indeed, the inclusion of the additional meson-meson loop corrections in LQCD computation could potentially diminish the LQCD prediction for the tensor glueball mass from 2.2-2.4 GeV to the nearby value of the mass of $f_2(1950)$. The large decay width of $f_2(1950)$ supports its exotic structure rather than the conventional meson one. In future, mixing between conventional $\bar{q}q$ states should be taken into account, as it is done in the scalar sector.

Our results might be useful for the tensor glueball search in ongoing experiments such as BESIII and LHCb.

# Chapter 5

# Spin-3 mesons

In this chapter, I present the results for spin-3 mesons (with $J^{PC} = 3^{--}$) within an effective model. In this case, the effective model is based on flavor symmetry $SU(3)_V$. This is due to the fact that chiral symmetry can not be used because no information about the ground-state spin-3 axial-tensor mesons (with $J^{PC} = 3^{++}$) is available. Yet, formally, also this model can be seen as a part of the eLSM generalized to include mesons with $J = 3$.

## 5.1 Introduction

In a series of articles, hadronic models have been built for various nonets where only the residual flavor symmetry $SU_{\mathrm{V}}(3)$ is considered [66,81,83,85,157,158]. The expansion in $1/N_{\mathrm{c}}$ dominant and subdominant terms is performed when the interaction terms are written down. In addition, terms that explicitly violate flavor symmetry may be included, either as a result of the chiral anomaly or as a result of the underlying breaking caused by the presence of nonzero and unequal quark masses

$$m_u \approx 2\,\mathrm{MeV}\,, \quad m_d \approx 5\,\mathrm{MeV}\,, \quad m_s \approx 93\,\mathrm{MeV}\,. \tag{5.1}$$

Within such a model approach, we analyze the decays of the ground-state $J^{PC} = 3^{--}$ nonet that contains the resonances $\rho_3(1690)$, $K_3^*(1780)$, $\phi_3(1850)$, $\omega_3(1670)$ existing in the PDG [27]. Because of the current degree of data accuracy, we will only take into account the dominant terms in the large-$N_{\mathrm{c}}$ expansion (see [119,120,159]) and ignore flavor symmetry breaking corrections.

## 5.2 Effective model

Based on $SU(3)_V$ flavor symmetry, we construct effective Lagrangians. In order to construct the interaction Lagrangians in Table 5.1, we use the symmetry transformations of the mesonic nonets listed in Table 2.1. We have derived the decay amplitudes $|\mathcal{M}|^2$ via Feynman rules by using the polarization vectors and tensors as well as their corresponding completeness relations reported in Ref. [1]. The summary Table 5.2 shows the results for $\frac{1}{7}|\mathcal{M}|^2$.

| $J_c^{PC} \to J_a^{PC} + J_b^{PC}$ | Interaction Lagrangians |
|---|---|
| $3^{--} \to 0^{-+} + 0^{-+}$ | $\mathcal{L}_{w_3pp} = g_{w_3pp}\,\mathrm{tr}\big[W_3^{\mu\nu\rho}\,[P\,,(\partial_\mu\partial_\nu\partial_\rho P)]_-\big]$ |
| $3^{--} \to 0^{-+} + 1^{--}$ | $\mathcal{L}_{w_3v_1p} = g_{w_3v_1p}\,\varepsilon^{\mu\nu\rho\sigma}\,\mathrm{tr}\big[W_{3,\mu\alpha\beta}\,\{(\partial_\nu V_{1,\rho})\,,(\partial^\alpha\partial^\beta\partial_\sigma P)\}_+\big]$ |
| $3^{--} \to 0^{-+} + 1^{--}$ | $\mathcal{L}_{w_3\gamma p} = g_{w_3v_1p}\,\frac{e}{g_\rho}\,\varepsilon^{\mu\nu\rho\sigma}\,(\partial_\nu a_\rho)\,\mathrm{tr}\big[W_{3,\mu\alpha\beta}\,\{Q,\,(\partial^\alpha\partial^\beta\partial_\sigma P)\}_+\big]$ |
| $3^{--} \to 0^{-+} + 2^{++}$ | $\mathcal{L}_{w_3tp} = g_{w_3tp}\,\varepsilon_{\mu\nu\rho\sigma}\,\mathrm{tr}\big[W^\mu_{3\alpha\beta}\,[(\partial^\nu T^{\rho\alpha})\,,(\partial^\sigma\partial^\beta P)]_-\big]$ |
| $3^{--} \to 0^{-+} + 1^{+-}$ | $\mathcal{L}_{w_3b_1p} = g_{w_3b_1p}\,\mathrm{tr}\big[W_3^{\mu\nu\rho}\{B_{1,\mu}\,,(\partial_\nu\partial_\rho P)\}_+\big]$ |
| $3^{--} \to 0^{-+} + 1^{++}$ | $\mathcal{L}_{w_3a_1p} = g_{w_3a_1p}\,\mathrm{tr}\big[W_3^{\mu\nu\rho}\,[A_{1,\mu}\,,(\partial_\nu\partial_\rho P)]_-\big]$ |
| $3^{--} \to 1^{--} + 1^{--}$ | $\mathcal{L}_{w_3v_1v_1} = g_{w_3v_1v_1}\,\mathrm{tr}\big[W_3^{\mu\nu\rho}\,[(\partial_\mu V_{1,\nu})\,,V_{1,\rho}]_-\big]$ |

**Table 5.1:** Interaction Lagrangians describing the decay of spin-3 mesons .

Before presenting the results for decays, let us make the following comments for the interaction Lagrangians. The first comment is about the exclusion of the coupling between the spin-3 nonet of Eq. (2.86) to scalar mesons of Eq. (2.63). One kinematically allowed decay product would be a scalar and pseudoscalar meson. Yet, the corresponding CPT- and flavor-invariant interaction term:

$$\varepsilon_{\mu\nu\rho\sigma}\,\mathrm{tr}\big(\partial^\mu W^\nu_{3\,\alpha\beta}\,[(\partial^\rho S),\,(\partial^\sigma\partial^\alpha\partial^\beta P)]_-\big) = 0\,, \tag{5.2}$$

that vanishes because of the contraction of antisymmetric and symmetric tensors. This is why we do not include the scalar mesons in our study in agreement with the absence of such experimental decay channels in the PDG [27]. Besides that, the assignment of the scalar mesons is still an open question in low-energy QCD [65,67,68,160–162].

| $J_c^{PC} \to J_a^{PC} + J_b^{PC}$ | $\frac{1}{7}\,\lvert\mathcal{M}\rvert^2$ |
|---|---|
| $3^{--} \to 0^{-+} + 0^{-+}$ | $g_{w_3pp}^2\,\frac{2}{35}\,\lvert\vec{k}_{w_3,p^{(1)},p^{(2)}}\rvert^6$ |
| $3^{--} \to 0^{-+} + 1^{--}$ | $g_{w_3v_1p}^2\,\frac{8}{105}\,\lvert\vec{k}_{w_3,v,p}\rvert^6\,m_{w_3}^2$ |
| $3^{--} \to 0^{-+} + 1^{--}$ | $g_{w_3v_1p}^2\,\left(\frac{e}{g_\rho}\right)^2\,\frac{8}{105}\,\lvert\vec{k}_{w_3,\gamma,p}\rvert^6\,m_{w_3}^2$ |
| $3^{--} \to 0^{-+} + 2^{++}$ | $g_{w_3tp}^2\,\frac{2}{105}\,\lvert\vec{k}_{w_3,t,p}\rvert^4\,\frac{m_{w_3}^2}{m_t^2}\,(2\,\lvert\vec{k}_{w_3,t,p}\rvert^2 + 7\,m_t^2)$ |
| $3^{--} \to 0^{-+} + 1^{+-}$ | $g_{w_3b_1p}^2\,\frac{2}{105}\,\lvert\vec{k}_{w_3,b_1,p}\rvert^4\left(7 + 3\,\frac{\lvert\vec{k}_{w_3,b_1,p}\rvert^2}{m_{b_1}^2}\right)$ |
| $3^{--} \to 0^{-+} + 1^{++}$ | $g_{w_3a_1p}^2\,\frac{2}{105}\,\lvert\vec{k}_{w_3,a_1,p}\rvert^4\left(7 + 3\,\frac{\lvert\vec{k}_{w_3,a_1,p}\rvert^2}{m_{a_1}^2}\right)$ |
| $3^{--} \to 1^{--} + 1^{--}$ | $\frac{g_{w_3v_1v_1}^2\,\lvert\vec{k}_{w_3,v^{(1)},v^{(2)}}\rvert^2}{105 m_{v^{(1)}}^2\,m_{v^{(2)}}^2}\left[6\,\lvert\vec{k}_{w_3,v^{(1)},v^{(2)}}\rvert^4 + 35\,m_{v^{(1)}}^2\,m_{v^{(2)}}^2 + 14\,\lvert\vec{k}_{w_3,v^{(1)},v^{(2)}}\rvert^2\,(m_{v^{(1)}}^2 + m_{v^{(2)}}^2)\right]$ |

**Table 5.2:** Amplitude squares for the decay of spin-3 resonances.

Furthermore, the Lagrangian terms that are taken into account in Table 5.1 are symmetric under $U_V(3)$ and dominant in large-$N_c$. As we will see later on, such an approximation is adequate at the current precision of the experiment data. For instance, in the case of the $W_3B_1P$ interaction, one can think of the Lagrangian $\mathcal{L}_{w_3b_1p}$ as the first term in the large-$N_c$ expansion and/or symmetry-breaking terms, where further terms look as follows:

$$\begin{aligned}
\mathcal{L}_{w_3b_1p}^{\text{full}} &= g_{w_3b_1p}\,\mathrm{tr}\big[W^{\mu\nu\rho}\{B_{1\mu}\,,(\partial_\nu\partial_\rho P)\}_+\big] + g_{w_3b_1p}^{(2)}\,\mathrm{tr}\big[W^{\mu\nu\rho}\big]\,\mathrm{tr}\big[B_{1\mu}\,(\partial_\nu\partial_\rho P)\big] + \\
&+ g_{w_3b_1p}^{(3)}\,\mathrm{tr}\big[W^{\mu\nu\rho}\big]\,\mathrm{tr}\big[B_{1\mu}\big]\,\mathrm{tr}\big[\partial_\nu\partial_\rho P\big] + g_{w_3b_1p}^{(4)}\,\mathrm{tr}\big[\hat{\delta}\,W^{\mu\nu\rho}\,\{B_{1\mu}\,,(\partial_\nu\partial_\rho P)\}_+\big] + \\
&+ g_{w_3b_1p}^{(5)}\,\mathrm{tr}\big[\hat{\delta}\,W^{\mu\nu\rho}\big]\,\mathrm{tr}\big[B_{1\mu}\,(\partial_\nu\partial_\rho P)\big] + g_{w_3b_1p}^{(6)}\,\mathrm{tr}\big[W^{\mu\nu\rho}\big]\,\mathrm{tr}\big[\hat{\delta}\,B_{1\mu}\,(\partial_\nu\partial_\rho P)\big] + \ldots\,,
\end{aligned}$$

where the first term is reported in Table 5.1. It is a flavor symmetric term and scales as $g_{w_3b_1p} \propto N_c^{-1/2}$, thus it is the dominant term in the large-$N_c$ expansion. The second and third terms are flavor symmetric but are suppressed, since they scale as $g_{w_3b_1p}^{(2)} \propto N_c^{-\frac{3}{2}}$ and $g_{w_3b_1p}^{(3)} \propto N_c^{-\frac{5}{2}}$ (typically they involve the exchange of gluon, see Eq. (4.19)). Starting from the fourth term, flavor symmetry is broken via the matrix

$$\hat{\delta} = \mathrm{diag}\{\delta_u, \delta_d, \delta_s\}\,. \tag{5.3}$$

Flavor violation can be non-negligible (because the $\delta_s \propto m_s - m_u$ Ref. [67,68]), while it is expected to be very small for non-strange quarks. Upon setting $\delta_u = \delta_d = 0$, the coupling is expected to be of the order of

$$g^{(4)}_{w_3 b_1 p} \simeq g_{w_3 b_1 p} \,\mathrm{diag}\{0, 0, 0.1\ -\ 0.2\}\,, \tag{5.4}$$

and the actual value should be determined by performing an independent fit to the data.

Even if all of the interaction terms in Table 5.1 can be subject to the same analysis, it is interesting to note that the second and third terms would disappear anytime the commutator is present. It is expected that the largest next-to-leading-order contribution for those interaction terms will come from flavor-symmetry violation.

The following Lagrangian, in addition to the interaction terms discussed in this section, also includes the typical kinetic and mass terms of the used nonets.

$$\begin{aligned}\mathcal{L}_{W,\text{total}} =& \mathcal{L}_{\text{mass}} + \mathcal{L}_{\text{kin}} + \mathcal{L}_{w_3 pp} + \mathcal{L}_{w_3 v_1 p} + \\ &+ \mathcal{L}_{w_3 tp} + \mathcal{L}_{w_3 a_1 p} + \mathcal{L}_{w_3 b_1 p} + \mathcal{L}_{w_3 v_1 v_1}\,.\end{aligned} \tag{5.5}$$

Only the lowest derivatives for each interaction term were retained in its construction. Two alternative lines of reasoning are used to support this choice. First, if the Lagrangian $\mathcal{L}_{W,\text{total}}$ is understood to be a component of a more complete and general chiral hadronic model in the vacuum, such as the eLSM, our effective model automatically arises. Second, such terms with the lowest derivatives are kept as the main contributions to our effective hadronic model, in accordance with the Functional Renormalization Group approach [46–57].

It is interesting to describe the former point in more detail. To this end, recall that the construction of the eLSM follows the assumption of chiral symmetry:

$$U_{\text{L}}(3) \times U_{\text{R}}(3) = SU_{\text{L}}(3) \times SU_{\text{R}}(3) \times U_{\text{L}}(1) \times U_{\text{R}}(1)\,,$$

and the consideration of the scale invariance of the model. The symmetries are broken dynamically in addition to their small explicit breaking. Spontaneous breaking happens when $SU_{\text{L}}(3) \times SU_{\text{R}}(3) \to SU_{\text{V}}(3)$. Additionally, the *conformal/trace anomaly*, which is also known as the scale invariance breaking, occurs as a result of gluonic quantum fluctuations and the creation of a gluon condensate [124,163,164].The heavy mass of $\eta'(958)$ as well as the large pseudoscalar mixing angle are both explained by a second anomaly, [63], the chiral anomaly, which also breaks $U_{\mathrm{A}}(1)$. Besides anomaly terms, the eLSM interaction terms are scale/dilatation transformation invariant in the chiral limit and when the chiral anomaly is ignored.

All coupling constants of $\mathcal{L}_{w_3,\mathrm{total}}$ in Eq. (5.5) are dimensionful, in contrast to the eLSM [19,22], which uses dimensionless coupling constants for parametrization, with the exception of the last entries of Table 5.1. However, a closer look also reveals the benefits of keeping the lowest number of derivatives. Let us start by remembering that the axial-vector nonet shift $A_1$ of Eq. (3.8) originates from scalar-axial-vector mixing generated by SSB. For our illustrative purposes, it is sufficient to assume that $U_{\mathrm{V}}(3)$ is exact in the simplified scenario [19,102]:

$$A_1^\mu \to A_1^\mu + Z\, w\, \partial^\mu P\,, \tag{5.6}$$

where the pion decay constant $f_\pi = 92.4$ MeV appears in $w \approx \frac{g_1 Z f_\pi}{m_{a_1}^2}$ with $m_{a_1} \simeq 1.4$ GeV, $Z \approx 1.6$ and $g_1 \approx g_\rho \approx 5.5$.

Within this framework, the $W_3PP$-Lagrangian,

$$\mathcal{L}_{w_3pp} = g_{w_3pp}\,\mathrm{tr}\big[W^{\mu\nu\rho}\,[P,\,(\partial_\mu\partial_\nu\partial_\rho P)]_-\big]\,, \tag{5.7}$$

with the dimensionful coupling $g_{w_3pp} = [\mathrm{Energy}^{-2}]$ and the $W_3A_1P$-Lagrangian,

$$\mathcal{L}_{w_3a_1p} = g_{w_3a_1p}\,\mathrm{tr}\big[W^{\mu\nu\rho}\,[A_{1,\mu},\,(\partial_\nu\partial_\rho P)]_-\big]\,, \tag{5.8}$$

where $g_{w_3pp} = [\mathrm{Energy}^{-1}]$ can be seen as the outcome of the interaction Lagrangian

$$\mathcal{L}_{w_3a_1a_1} = g_{w_3a_1a_1}\,\mathrm{tr}\big(W^{\mu\nu\rho}\,[A_{1,\mu},\,\partial_\nu A_{1,\rho}]_-\big)\,, \tag{5.9}$$

using only the dimensionless coupling $g_{w_3a_1a_1}$ obtained by shifting Eq. (5.6), applied once for $W_3A_1P$ and twice for $W_3PP$. In other words, only the Lagrangian $\mathcal{L}_{w_3a_1a_1}$ is included in an eLSM chiral Lagrangian generalized to include $J = 3$ mesons:

$$\mathcal{L}_{g_3} = g_3\mathrm{tr}\big(L_3^{\mu\nu\rho}\,L_{1,\mu}\,\partial_\nu L_{1,\rho} + R_3^{\mu\nu\rho}\,R_{1,\mu}\,\partial_\nu R_{1,\rho}\big)\,. \tag{5.10}$$

The Lagrangians $\mathcal{L}_{w_3 a_1 p}$ and $\mathcal{L}_{w_3 pp}$ are then a consequence of $\mathcal{L}_{w_3 a_1 a_1}$ via SSB, explaining how dimensional couplings occur in the model even if one starts just with dimensionless ones, although the decay channel $W_3 \to A_1\, A_1$ is not kinematically allowed. The following two relations between the coupling constants

$$g_{w_3 pp} \approx Z^2\, w^2\, g_{w_3 a_1 a_1}\,, \quad \text{and} \quad g_{w_3 a_1 p} \approx Z\, w\, g_{w_3 a_1 a_1}\,, \tag{5.11}$$

leads to

$$g_{w_3 a_1 p}/g_{w_3 pp} \approx (Z\, w)^{-1}\,. \tag{5.12}$$

Based on this analysis, we shall estimate the decay rates of spin-3 mesons into an axial vector and a pseudoscalar meson pair in Sec. 5.3.6. However, it must be emphasized that the current heuristic discussion cannot give exact ratios of couplings; rather, it just serves as a guide to understanding the model's origin.

Similar arguments about understanding the chiral origin of the model can be applied to the following Lagrangian,

$$\mathcal{L}_{w_3 b_1 p} = g_{w_3 b_1 p}\, \mathrm{tr}\big(W^{\mu\nu\rho}\, \{B_{1,\mu},\, (\partial_\nu \partial_\rho P)\}_+\big)\,, \tag{5.13}$$

where the dimensionful constant $g_{w_3 b_1 p} = [\text{Energy}^{-1}]$, which can be seen as emerging from the following Lagrangian with a dimensionless coupling $g_{w_3 b_1 a_1}$

$$\mathcal{L}_{w_3 b_1 a_1} = g_{w_3 b_1 a_1}\, \mathrm{tr}\big[W^{\mu\nu\rho}\, \{B_{1,\mu},\, (\partial_\nu A_{1,\rho})\}_+\big]\,. \tag{5.14}$$

The following ratio

$$\frac{g_{w_3 b_1 p}}{g_{w_3 b_1 a_1}} \approx Z\, w\,, \tag{5.15}$$

shows the magnitude of the coupling constant $g_{w_3 b_1 a_1}$, which cannot be used since this decay channel is kinematically forbidden.

Unlike the previous Lagrangians, the two other Lagrangians $\mathcal{L}_{w_3 v_1 p}$ and $\mathcal{L}_{w_3 tp}$ in Table 5.1 are not dilatation invariant, and one cannot obtain the dimensionful couplings $g_{w_3 v_1 p} = [\text{Energy}^{-3}]$ and $g_{w_3 tp} = [\text{Energy}^{-2}]$ via SSB of $\mathcal{L}_{w_3 v_1 a_1}$ and $\mathcal{L}_{w_3 t a_1}$ since their coupling constants $g_{w_3 v_1 a_1}$ and $g_{w_3 t a_1}$ still have dimension $[\text{Energy}^{-2}]$ and [Energy$^{-1}$]. Nevertheless, this makes sense because these terms involve the Levi-Civita pseudotensor $\varepsilon_{\mu\nu\rho\sigma}$ and can be seen as a manifestation of the chiral anomaly.

Summarizing the discussion, terms that are components of the model are either (i) dilatation invariant, (ii) may be generated from dilatation invariant terms of a more fundamental underlying chiral model by shifting the axial-vector nonet, or (iii) they are connected to the chiral anomaly.

## 5.3 Results for $J^{PC} = 3^{--}$ mesons

This section is about our results for decay rates and branching ratios of the $J^{PC} = 3^{--}$ mesons and their comparison to available experimental data. Moreover, we confirm that our theoretical outcomes for the sum of all single decay channels never exceed the total decay widths of the $J^{PC} = 3^{--}$listed in Table 5.3.

| Resonances | Masses (MeV) | Total Decay Widths (MeV) |
|---|---|---|
| $\rho_3(1690)$ | 1690 | $161 \pm 10$ |
| $K_3^*(1780)$ | 1780 | $159 \pm 21$ |
| $\omega_3(1670)$ | 1670 | $168 \pm 10$ |
| $\phi_3(1850)$ | 1850 | $87^{+28}_{-23}$ |

**Table 5.3:** Total decay widths and masses of $J^{PC} = 3^{--}$ mesons [27] .

Wherever possible, we calculate the coupling constants and their error using a fit to experimental decay widths and ratios. Because the experimental errors of the masses are typically small and have little effect on the overall errors, we disregard the experimental uncertainties. This approximation is reasonable because the systematic errors in the model, as well as the experimental errors for branching ratios and decay widths, are substantially larger. The quoted errors, however, only represent a lower bound of the actual errors, as other indeterminacies are present. Moreover, for $J = 3$ mesons, the contributions from loops are believed to be negligible due to the relatively small width/mass ratio for the decaying resonances [165,166]. In any case, calculations at tree-level are required if the interaction terms are taken to be effective couplings in a full quantum-effective IR action [44].

### 5.3.1 Decay channel $W_3 \to P^{(1)} + P^{(2)}$

The decay of spin-3 tensor mesons into two pseudoscalar mesons is described by the following effective interaction term:

$$\mathcal{L}_{w_3pp} = g_{w_3pp}\,\mathrm{tr}\big[W_3^{\mu\nu\rho}\,[P,\,(\partial_\mu\partial_\nu\partial_\rho P)]_-\big]\,. \tag{5.16}$$

The tree-level decay rate is

$$\Gamma_{W_3\to P^{(1)}+P^{(2)}}(m_{w_3}, m_{p^{(1)}}, m_{p^{(2)}}) = g^2_{w_3pp}\,\frac{|\vec{k}_{w_3,p^{(1)},p^{(2)}}|^7}{140\,m^2_{w_3}} \times$$
$$\kappa_i\,\Theta(m_{w_3} - m_{p^{(1)}} - m_{p^{(2)}})\,, \tag{5.17}$$

where the amplitude square is given in Table 5.2 and $\kappa_i$ are reported in Table 5.4. Using the above formula as well as the experimental results, which are listed in the fourth row of Table 5.4, we can calculate the coupling constant $g_{w_3pp}$ by using a standard $\chi^2$-approach. We consider $\chi^2$ as a function of the theoretical coupling square $\tilde{g}$

$$\chi^2 \equiv \sum_{i=1}^{N} \frac{(\tilde{g}-\tilde{g}_i)^2}{\Delta\tilde{g}_i}\,, \tag{5.18}$$

where $\tilde{g}_i$ are the squared coupling square and $\Delta\tilde{g}_i$ the experimental errors on them defined from the experiment. Minimizing the $\chi^2$ with respect to coupling $\frac{d\chi^2}{d\tilde{g}} = 0$ leads to the following relation for $N = 4$:

$$\tilde{g} \equiv g^2_{w_3pp} = \frac{\sum_{i=1}^4 \frac{\tilde{g}_i}{\Delta\tilde{g}_i}}{\sum_{j=1}^4 \frac{1}{\Delta\tilde{g}_j}}\,, \quad \text{and} \quad \Delta\tilde{g} \equiv \Delta g^2_{w_3pp} = \frac{1}{\sum_{j=1}^4 \frac{1}{\Delta\tilde{g}_j}}\,. \tag{5.19}$$

Thus, we obtain

$$g^2_{w_3pp} = (1.5 \pm 0.1)\cdot 10^{-10}\ \mathrm{MeV}^{-4}\,. \tag{5.20}$$

Table 5.4 shows the comparison of theoretical and experimental results obtained by using this value for the coupling constant. Although there is a good overall agreement, there is also a significant mismatch, as the experimental value for $K_3^*(1780) \to \bar{K}\,\eta$ is much larger than our theoretical prediction. However, the experimental error is substantial, and more accurate experimental results would be needed to clarify this point. Furthermore, a notable prediction for $\phi_3(1850) \to \bar{K}\,K$ is made for this channel: an experimental determination ought to be possible in the future.

| decay process | $\kappa_i$ | theory $\Gamma/\text{MeV}$ | experiment $\Gamma/\text{MeV}$ |
|---|---|---|---|
| $\rho_3(1690) \to \pi\,\pi$ | $1$ | $32.7 \pm 2.3$ | $38.0 \pm 3.2$ |
| $\rho_3(1690) \to \bar{K}\,K$ | $2\,(\frac{1}{2})^2$ | $4.0 \pm 0.3$ | $2.54 \pm 0.45$ |
| $K_3^*(1780) \to \pi\,\bar{K}$ | $(\frac{1}{2})^2 + (\frac{\sqrt{2}}{2})^2$ | $18.5 \pm 1.3$ | $29.9 \pm 4.3$ |
| $K_3^*(1780) \to \bar{K}\,\eta$ | $\left[\frac{1}{2}\,(-\cos\beta_p + \sqrt{2}\sin\beta_p)\right]^2$ | $7.4 \pm 0.5$ | $48 \pm 22$ |
| $K_3^*(1780) \to \bar{K}\,\eta'(958)$ | $\left[\frac{1}{2}\,(\sqrt{2}\cos\beta_p + \sin\beta_p)\right]^2$ | $0.021 \pm 0.001$ | |
| $\omega_3(1670) \to \bar{K}\,K$ | $2\left[\frac{1}{2}\,(-\cos\beta_{w_3} + \sqrt{2}\sin\beta_{w_3})\right]^2$ | $3.0 \pm 0.2$ | |
| $\phi_3(1850) \to \bar{K}\,K$ | $2\left[\frac{1}{2}\,(\sqrt{2}\cos\beta_{w_3} + \sin\beta_{w_3})\right]^2$ | $18.8 \pm 1.3$ | seen |

**Table 5.4:** Clebsch-Gordan coefficients and corresponding decay rates for $W_3 \to P^{(1)} + P^{(1)}$ compared to PDG [27].

## 5.3.2 Decay channel $W_3 \to V_1 + P$

For the vector and pseudoscalar decay modes, the interaction Lagrangian is as follows:

$$\mathcal{L}_{w_3 v_1 p} = g_{w_3 v_1 p}\,\varepsilon^{\mu\nu\rho\sigma}\,\mathrm{tr}\left[W_{3,\mu\alpha\beta}\left\{(\partial_\nu V_{1,\rho}),\,(\partial^\alpha\partial^\beta\partial_\sigma P)\right\}_+\right]. \tag{5.21}$$

The tree-level decay rate formula in this scenario takes the form

$$\Gamma_{W_3 \to V_1 + P}(m_{w_3}, m_v, m_p) = g^2_{w_3 v_1 p}\,\frac{|\vec{k}_{w_3,v,p}|^7}{105}\,\kappa_i\,\Theta(m_{w_3} - m_v - m_p)\,, \tag{5.22}$$

where the factors $\kappa_i$ are reported in Table 5.5. We follow the same steps as in the pseudoscalar-pseudoscalar case to define the coupling constant. We use the three experimental data points listed in Table 5.5, leading to:

$$g^2_{w_3 v_1 p} = (9.2 \pm 1.9) \cdot 10^{-16}\ \text{MeV}^{-6}\,, \tag{5.23}$$

out of which we get the theoretical estimates listed in Table 5.5. Although the $\Gamma_{K_3^*(1780)\to\rho(770)\,K}$ mode is theoretically underestimated (the experimental error is nevertheless large), we notice that a satisfactory agreement is achieved.

Interestingly, although they could not be quantified, the two theoretically large decays $\omega_3(1670) \to \rho(770)K$ and $\phi_3(1850) \to K^*(892)K$ have both been observed in experiments. Their determination in the future would serve as a test of our method.

| decay process | $\kappa_i$ | theory $\Gamma/\text{MeV}$ | experiment $\Gamma/\text{MeV}$ |
|---|---|---|---|
| $\rho_3(1690) \to \rho(770)\,\eta$ | $\left(\frac{1}{2}\cos\beta_p\right)^2$ | $3.8 \pm 0.8$ | seen |
| $\rho_3(1690) \to \bar{K}^*(892)\,K$ | $4\left(\frac{1}{4}\right)^2$ | $3.4 \pm 0.7$ | |
| $\rho_3(1690) \to \omega(782)\,\pi$ | $\left(\frac{1}{2}\cos\beta_{v_1}\right)^2$ | $35.8 \pm 7.4$ | $25.8 \pm 9.8$ |
| $\rho_3(1690) \to \phi(1020)\,\pi$ | $\left(\frac{1}{2}\sin\beta_{v_1}\right)^2$ | $0.036 \pm 0.007$ | |
| $K_3^*(1780) \to \rho(770)\,K$ | $\left(\frac{1}{4}\right)^2 + \left(\frac{\sqrt{2}}{4}\right)^2$ | $16.8 \pm 3.5$ | $49.3 \pm 15.7$ |
| $K_3^*(1780) \to \bar{K}^*(892)\,\pi$ | $\left(\frac{1}{4}\right)^2 + \left(\frac{\sqrt{2}}{4}\right)^2$ | $27.2 \pm 5.6$ | $31.8 \pm 9.0$ |
| $K_3^*(1780) \to \bar{K}^*(892)\,\eta$ | $\left[\frac{1}{4}\left(\cos\beta_p + \sqrt{2}\sin\beta_p\right)\right]^2$ | $0.09 \pm 0.02$ | |
| $K_3^*(1780) \to \omega(782)\,\bar{K}$ | $\left[\frac{1}{4}\left(\cos\beta_{v_1} + \sqrt{2}\sin\beta_{v_1}\right)\right]^2$ | $4.3 \pm 0.9$ | |
| $K_3^*(1780) \to \phi(1020)\,\bar{K}$ | $\left[\frac{1}{4}\left(\sqrt{2}\cos\beta_{v_1} - \sin\beta_{v_1}\right)\right]^2$ | $1.2 \pm 0.3$ | |
| $\omega_3(1670) \to \rho(770)\,\pi$ | $3\left(\frac{1}{2}\cos\beta_{w_3}\right)^2$ | $97 \pm 20$ | seen |
| $\omega_3(1670) \to \bar{K}^*(892)\,K$ | $4\left[\frac{1}{4}\left(\cos\beta_{w_3} + \sqrt{2}\sin\beta_{w_3}\right)\right]^2$ | $2.9 \pm 0.6$ | |
| $\omega_3(1670) \to \omega(782)\,\eta$ | $\left[\frac{1}{2}\left(\cos\beta_{w_3}\cos\beta_{v_1}\cos\beta_p + \sqrt{2}\sin\beta_{w_3}\sin\beta_{v_1}\sin\beta_p\right)\right]^2$ | $2.8 \pm 0.6$ | |
| $\omega_3(1670) \to \phi(1020)\,\eta$ | $\left[\frac{1}{2}\left(-\cos\beta_{w_3}\sin\beta_{v_1}\cos\beta_p + \sqrt{2}\sin\beta_{w_3}\cos\beta_{v_1}\sin\beta_p\right)\right]^2$ | $(7.6 \pm 1.6) \cdot 10^{-6}$ | |
| $\phi_3(1850) \to \rho(770)\,\pi$ | $3\left(\frac{1}{2}\sin\beta_{w_3}\right)^2$ | $1.1 \pm 0.2$ | |
| $\phi_3(1850) \to \bar{K}^*(892)\,K$ | $4\left[\frac{1}{4}\left(\sqrt{2}\cos\beta_{w_3} - \sin\beta_{w_3}\right)\right]^2$ | $35.5 \pm 7.3$ | seen |
| $\phi_3(1850) \to \omega(782)\,\eta$ | $\left[\frac{1}{2}\left(-\sin\beta_{w_3}\cos\beta_{v_1}\cos\beta_p + \sqrt{2}\cos\beta_{w_3}\sin\beta_{v_1}\sin\beta_p\right)\right]^2$ | $0.015 \pm 0.003$ | |
| $\phi_3(1850) \to \omega(782)\,\eta'(958)$ | $\left[\frac{1}{2}\left(\sin\beta_{w_3}\cos\beta_{v_1}\sin\beta_p + \sqrt{2}\cos\beta_{w_3}\sin\beta_{v_1}\cos\beta_p\right)\right]^2$ | $0.003 \pm 0.001$ | |
| $\phi_3(1850) \to \phi(1020)\,\eta$ | $\left[\frac{1}{2}\left(\sin\beta_{w_3}\sin\beta_{v_1}\cos\beta_p + \sqrt{2}\cos\beta_{w_3}\cos\beta_{v_1}\sin\beta_p\right)\right]^2$ | $3.8 \pm 0.8$ | |

**Table 5.5:** Clebsch-Gordan coefficients and corresponding decay rates for $W_3 \to V + P$ compared to PDG [27].

We also provide the following ratio

$$\frac{\Gamma_{\phi_3(1850)\to\bar{K}^*(892)\,K}}{\Gamma_{\phi_3(1850)\to\bar{K}\,K}} = 1.9 \pm 0.4\,, \tag{5.24}$$

which is roughly consistent with the PDG quote [27] average obtained from Ref. [167]

$$\frac{\Gamma^{\text{exp}}_{\phi_3(1850)\to\bar{K}^*(892)\,K}}{\Gamma^{\text{exp}}_{\phi_3(1850)\to\bar{K}\,K}} = 0.55^{+0.85}_{-0.45}\,. \tag{5.25}$$

Interestingly, Ref. [168] reports a slightly higher ratio of

$$\frac{\Gamma^{\text{exp}}_{\phi_3(1850)\to\bar{K}^*(892)\,K}}{\Gamma^{\text{exp}}_{\phi_3(1850)\to\bar{K}\,K}} = 0.8 \pm 0.4\,. \tag{5.26}$$

Additionally, we would like to draw attention to a current LQCD analysis [28], which supports both results that we have predicted. Table 5.6 shows up the comparison to LQCD results [28] (whose errors are not provided due to large uncertainties). It is rather remarkable how well our (quite simple) model outcomes agree with the LQCD study that makes contact with the underlying QCD. This leads us to the conclusion that even for conventional mesons with $J^{PC} = 3^{--}$ and with masses exceeding 1 GeV, symmetry-based predictions are useful.

| decay process (in model) | Theory (MeV) | LQCD (MeV) |
|---|---|---|
| $\rho_3(1690) \longrightarrow \bar{K}^*(892)\,K + \text{c.c.}$ | 3 | 2 |
| $\rho_3(1690) \longrightarrow \omega(782)\,\pi$ | 36 | 22 |
| $\omega_3(1670) \longrightarrow \rho(770)\,\pi$ | 97 | 62 |
| $\omega_3(1670) \longrightarrow \bar{K}^*(892)\,K + \text{c.c.}$ | 2.9 | 2 |
| $\omega_3(1670) \longrightarrow \omega(782)\,\eta$ | 2.8 | 1 |
| $\phi_3(1850) \longrightarrow \bar{K}^*(892)\,K + \text{c.c.}$ | 36 | 20 |
| $\phi_3(1850) \longrightarrow \phi(1020)\,\eta$ | 4 | 3 |

**Table 5.6:** Decay rates for $W_3 \to V_1 + P$ compared to LQCD [28].

Next, we study three body decays using the Sill distribution described in Chapter 3.

The decay process $\Gamma_{\rho_3(1690)\to K\bar{K}\pi} = 6.12 \pm 2.00$ MeV in [27] can be estimated as follows:

$$\begin{aligned}\Gamma_{\rho_3(1690)\to\bar{K}^*(892)\,K\to K\bar{K}\pi} &\simeq \int_0^\infty dx\, d_{K^\star}(x)\, \Gamma_{\rho_3(1690)\to\bar{K}^*(892)\,K}(m_{K_3}, x, m_K) \\ &\simeq (3.4 \pm 0.7)\,\text{MeV}\,, \end{aligned} \tag{5.27}$$

where

$$\begin{aligned} d_{K^\star}(y) &= \frac{2y}{\pi} \frac{\sqrt{y^2 - m_K^2 - m_\pi^2}\,\tilde{\Gamma}_{K^\star}}{(y^2 - m_{K^\star}^2)^2 + (\sqrt{y^2 - m_K^2 - m_\pi^2}\,\tilde{\Gamma}_{K^\star})^2}\,\Theta(y - m_K - m_\pi), \\ \text{with} \quad & \int_0^\infty dy\, d_{K^\star}(y) = 1\,, \end{aligned} \tag{5.28}$$

in which $\tilde{\Gamma}_{K^\star}$ is linked to the PDG values $\Gamma_{K^\star\to K\pi} = 51.4$ MeV and $m_{K^\star} = 890$ MeV according to:

$$\tilde{\Gamma}_{K^\star} \equiv \frac{\Gamma_{K^\star\to K\pi}\, m_{K^\star}}{\sqrt{m_{K^\star}^2 - m_{K^2} - m_\pi^2}}\ . \tag{5.29}$$

The other decay channel $\Gamma_{\rho_3\to\eta\pi\pi}$ that has been seen experimentally is estimated as:

$$\Gamma_{\rho_3\to\rho\eta\to\eta\pi\pi} \simeq \int_0^\infty dx\, d_\rho(x)\, \Gamma_{\rho_3\to\rho\eta}(m_{\rho_3}, x, m_\eta) \simeq (3.95 \pm 0.81)\,\text{MeV}\,. \tag{5.30}$$

### 5.3.3 Decay channel $W_3 \to \gamma + P$

In this subsection, we present the results of the radiative decay into a photon and a pseudoscalar meson. As a consequence of the VMD shift in Eq. (3.15), we obtain the following Lagrangian

$$\mathcal{L}_{w_3\gamma p} = g_{w_3 v_1 p}\, \tfrac{e}{g_\rho}\, \varepsilon^{\mu\nu\rho\sigma}\, (\partial_\nu a_\rho)\, \mathrm{tr}\big[W_{3,\mu\alpha\beta}\, \{Q,\, (\partial^\alpha\partial^\beta\partial_\sigma P)\}_+\big]\,. \tag{5.31}$$

Using the amplitude squared in Table 5.2, we get the tree-level decay rate formula:

$$\Gamma_{W_3\to\gamma+P}(m_{w_3}, m_p) = g_{w_3 v_1 p}^2\, \big(\tfrac{e}{g_\rho}\big)^2\, \frac{|\vec{k}_{w_3,\gamma,p}|^7}{105\pi}\, \kappa_i^\gamma\,. \tag{5.32}$$

All $\kappa_i^\gamma$ are presented in Table 5.7 [1].

| decay process | $\kappa_i$ | theory $\Gamma/\mathrm{keV}$ |
|---|---|---|
| $\rho_3^{\pm/0}(1690) \to \gamma\,\pi^{\pm/0}$ | $\left(\frac{1}{6}\right)^2$ | $69 \pm 14$ |
| $\rho_3^0(1690) \to \gamma\,\eta$ | $\left(\frac{1}{2}\cos\beta_p\right)^2$ | $157 \pm 32$ |
| $\rho_3^0(1690) \to \gamma\,\eta'(958)$ | $\left(\frac{1}{2}\sin\beta_p\right)^2$ | $20 \pm 4$ |
| $K_3^{\pm}(1780) \to \gamma\,K^{\pm}$ | $\left(\frac{1}{6}\right)^2$ | $58 \pm 12$ |
| $K_3^0(1780) \to \gamma\,K^0$ | $\left(\frac{1}{3}\right)^2$ | $231 \pm 48$ |
| $\omega_3(1670) \to \gamma\,\pi^0$ | $\left(\frac{1}{2}\cos\beta_{w_3}\right)^2$ | $560 \pm 120$ |
| $\omega_3(1670) \to \gamma\,\eta$ | $\left[\frac{1}{6}\left(\cos\beta_{w_3}\cos\beta_p - 2\sin\beta_{w_3}\sin\beta_p\right)\right]^2$ | $19 \pm 4$ |
| $\omega_3(1670) \to \gamma\,\eta'(958)$ | $\left[\frac{1}{6}\left(\cos\beta_{w_3}\sin\beta_p + 2\sin\beta_{w_3}\cos\beta_p\right)\right]^2$ | $1.4 \pm 0.3$ |
| $\phi_3(1850) \to \gamma\,\pi^0$ | $\left(\frac{1}{2}\sin\beta_{w_3}\right)^2$ | $4 \pm 1$ |
| $\phi_3(1850) \to \gamma\,\eta$ | $\left[\frac{1}{6}\left(\sin\beta_{w_3}\cos\beta_p + 2\cos\beta_{w_3}\sin\beta_p\right)\right]^2$ | $129 \pm 26$ |
| $\phi_3(1850) \to \gamma\,\eta'(958)$ | $\left[\frac{1}{6}\left(\sin\beta_{w_3}\sin\beta_p - 2\cos\beta_{w_3}\cos\beta_p\right)\right]^2$ | $35 \pm 7$ |

**Table 5.7:** Clebsch-Gordan coefficients and corresponding decay rates for $W_3 \to \gamma + P$.

In Table 5.7, predictions for the radiative decays $W_3 \to \gamma\,P$ are provided. The decay channels $\omega_3(1670) \to \gamma\,\pi^0$ and $K_3^0(1780) \to \gamma\,K^0$ are quite large. In fact, these processes imply that the photoproduction of mesons with spin $J^{PC} = 3^{--}$ is possible at the ongoing *GlueX* [107–109] and *CLAS12* [105] experiments at *Jefferson Lab*.

[1] Due to the fact that all pseudoscalars are lighter than spin-3 mesons and that photons are massless, the Heaviside function is not required in the decay rate formula.

### 5.3.4 Decay channel $W_3 \to T + P$

The decay of spin-3 tensor mesons into spin-2 tensor and pseudoscalar mesons is described by the interaction Lagrangian

$$\mathcal{L}_{w_3tp} = g_{w_3tp}\,\varepsilon_{\mu\nu\rho\sigma}\,\mathrm{tr}\big[W^{\mu}_{3\;\alpha\beta}\,\big[(\partial^{\nu}T^{\rho\alpha}),\,(\partial^{\sigma}\partial^{\beta}P)\big]_{-}\big]\,. \tag{5.33}$$

Using the amplitude square in Table 5.2 yields the relevant decay equation as follows:

$$\Gamma_{W_3\to T+P}(m_{w_3}, m_t, m_p) = g^2_{w_3tp}\,\frac{|\vec{k}_{w_3,t,p}|^5}{420\,m_t^2}\,\big(2\,|\vec{k}_{w_3,t,p}|^2 + 7\,m_t^2\big)\times \kappa_i\Theta(m_{w_3} - m_t - m_p)\,, \tag{5.34}$$

where Table 5.8 contains a list of the $\kappa_i$. At present, there are no experimentally known branching ratios for this channel that make it possible to calculate the coupling constant.

Yet, the experimental ratio given by the PDG [27] and extracted from Ref. [169] can be used

$$\frac{\Gamma^{\mathrm{exp}}_{\rho_3(1690)\to a_2(1320)\,\pi}}{\Gamma^{\mathrm{exp}}_{\rho_3(1690)\to\rho(770)\,\eta}} = 5.5 \pm 2.0\,. \tag{5.35}$$

Using the theoretical prediction $\Gamma_{\rho_3(1690)\to\rho(770)\,\eta} = (3.8 \pm 0.8)$ MeV in Table 5.5 leads to the coupling constant

$$g^2_{w_3tp} = (2.8 \pm 1.2)\cdot 10^{-9}\ \mathrm{MeV}^{-4}\,. \tag{5.36}$$

After fixing the coupling constant in this way, we obtain the results for the decay rates shown in Table 5.8. Our model prediction for the decay channel $\Gamma_{K_3^*(1780)\to\bar{K}_2^*(1430)\,\pi}$ is safely smaller than the experimental upper limit [27,170]. In addition, the rather large mode $\rho_3(1690)\to a_2(1320)\,\pi$ is is quoted as "seen" in Ref. [27].

| decay process | $\kappa_i$ | theory $\Gamma/\text{MeV}$ | experiment $\Gamma/\text{MeV}$ |
|---|---|---|---|
| $\rho_3(1690) \to a_2(1320)\,\pi$ | $2\left(\frac{1}{2}\right)^2$ | $20.9 \pm 8.7$ | seen |
| $K_3^*(1780) \to \bar{K}_2^*(1430)\,\pi$ | $\left(\frac{1}{4}\right)^2 + \left(\frac{\sqrt{2}}{4}\right)^2$ | $5.8 \pm 2.4$ | $< 25.4 \pm 3.4$ |
| $K_3^*(1780) \to f_2(1270)\,\bar{K}$ | $\left[\frac{1}{4}\left(\cos\beta_t - \sqrt{2}\sin\beta_t\right)\right]^2$ | $(5.4 \pm 2.2) \cdot 10^{-5}$ | |

**Table 5.8:** Clebsch-Gordan coefficients and corresponding decay rates for $W_3 \to T + P$ compared to PDG [27].

### 5.3.5 Decay channel $W_3 \to B_1 + P$

The interaction Lagrangian given below describes the decay of the spin-3 mesons into a pseudovector and a pseudoscalar meson

$$\mathcal{L}_{w_3 b_1 p} = g_{w_3 b_1 p}\,\text{tr}\left[W_3^{\mu\nu\rho}\left\{B_{1,\mu},\,(\partial_\nu\partial_\rho P)\right\}_+\right]. \tag{5.37}$$

Using the results in Table 5.2, the tree-level decay rate reads:

$$\Gamma_{W_3\to B_1+P}(m_{w_3}, m_{b_1}, m_p) = g_{w_3 b_1 p}^2\,\frac{|\vec{k}_{w_3,b_1,p}|^5}{420\,m_{w_3}^2}\left(7 + 3\,\frac{|\vec{k}_{w_3,b_1,p}|^2}{m_{b_1}^2}\right)$$
$$\kappa_i\Theta(m_{w_3} - m_{b_1} - m_p)\,, \tag{5.38}$$

where the $\kappa_i$ are listed in Table 5.9. Also, in this case, the lack of experimental data makes it impossible to directly determine the coupling constant $g_{w_3 b_1 p}$. Nevertheless, we proceed as shown below in order to estimate it in an indirect way:

1. The following lower bound in PDG [27] is taken from Ref. [171] (where it is not included as a PDG fit of average)

$$\frac{\Gamma_{\omega_3(1670)\to b_1(1235)\,\pi}}{\Gamma_{\omega_3(1670)\to\omega(782)\,\pi\,\pi}} > 0.75\ [27]\,, \qquad \frac{\Gamma_{\omega_3(1670)\to b_1(1235)\,\pi}}{\Gamma_{\omega_3(1670)\to\omega(782)\,\pi\,\pi}} = 1.0^{+0.0}_{-0.25}\ [171]\,. \tag{5.39}$$

According to Ref. [171] the decay mode $\omega_3(1670) \to b_1(1235)\,\pi$ is the dominant contribution to the $\omega(782)\,\pi\,\pi$ final state via the subsequent decay $b_1(1235) \to$

$\omega(782)\,\pi$. Thus one can assume the following relation

$$\Gamma_{\omega_3(1670)\to b_1(1235)\,\pi} \approx \Gamma_{\omega_3(1670)\to\omega(782)\,\pi\,\pi}\,. \tag{5.40}$$

2. On the basis of this assumption, we estimate the following:

$$\frac{\Gamma_{\omega_3(1670)\to b_1(1235)\,\pi}}{\Gamma_{\omega_3(1670)\to\rho(770)\,\pi}} \overset{\text{Eq. (5.40)}}{\approx} \frac{\Gamma_{\omega_3(1670)\to\omega\,\pi\,\pi}}{\Gamma_{\omega_3(1670)\to\rho(770)\,\pi}} \overset{\text{PDG}}{=} 0.71 \pm 0.27\,. \tag{5.41}$$

It should be noted that the PDG refers to Ref. [172] for the following ratios but still does not include them in their fitting analyses

$$\frac{\Gamma_{\omega_3(1670)\to\omega\,\pi\,\pi}}{\Gamma_{\omega_3(1670)\to\rho(770)\,\pi}} = 0.47 \pm 0.18\,, \quad \frac{\Gamma_{\omega_3(1670)\to b_1(1235)\,\pi}}{\Gamma_{\omega_3(1670)\to\rho(770)\,\pi}} = 0.32 \pm 0.16\,. \tag{5.42}$$

It may be important to point out that all experimental data on the $\omega_3(1670)$ are quite old and, at the time Refs. [171, 172] were published, its mass and total decay width could not be determined with high accuracy. A novel experimental determination would be welcome.

| decay process | $\kappa_i$ | theory for $\beta_{b_1} = 0$ $\Gamma/\text{MeV}$ | theory for $\beta_{b_1} = -40^\circ$ $\Gamma/\text{MeV}$ |
|---|---|---|---|
| $\rho_3(1690) \to h_1(1170)\,\pi$ | $(\frac{1}{2}\cos\beta_{b_1})^2$ | $53 \pm 23$ | $31 \pm 13$ |
| $\rho_3(1690) \to h_1(1415)\,\pi$ | $(\frac{1}{2}\sin\beta_{b_1})^2$ | $0$ | $0.73 \pm 0.31$ |
| $K_3^*(1780) \to b_1(1235)\,K$ | $(\frac{1}{4})^2 + (\frac{\sqrt{2}}{4})^2$ | $0.55 \pm 0.24$ | $0.55 \pm 0.24$ |
| $K_3^*(1780) \to \bar{K}_{1,B}\,\pi$ | $(\frac{1}{4})^2 + (\frac{\sqrt{2}}{4})^2$ | $37 \pm 16$ | $37 \pm 16$ |
| $K_3^*(1780) \to h_1(1170)\,\bar{K}$ | $[\frac{1}{4}\,(\cos\beta_{b_1} + \sqrt{2}\sin\beta_{b_1})]^2$ | $1.61 \pm 0.70$ | $0.03 \pm 0.01$ |
| $\omega_3(1670) \to b_1(1235)\,\pi$ | $3\,(\frac{1}{2}\cos\beta_{w_3})^2$ | $69 \pm 30$ | $69 \pm 30$ |
| $\phi_3(1850) \to b_1(1235)\,\pi$ | $3\,(\frac{1}{2}\sin\beta_{w_3})^2$ | $1.17 \pm 0.51$ | $1.17 \pm 0.51$ |
| $\phi_3(1850) \to \bar{K}_{1,B}\,K + \text{c.c.}$ | $4\,[\frac{1}{4}\,(\sqrt{2}\cos\beta_{w_3} - \sin\beta_{w_3})]^2$ | $9.1 \pm 3.9$ | $9.1 \pm 3.9$ |
| $\phi_3(1850) \to h_1(1170)\,\eta$ | $[\frac{1}{2}\,(-\sin\beta_{w_3}\cos\beta_{b_1}\cos\beta_p + \sqrt{2}\cos\beta_{w_3}\sin\beta_{b_1}\sin\beta_p)]^2$ | $0.006 \pm 0.003$ | $1.1 \pm 0.5$ |

**Table 5.9:** Clebsch-Gordan coefficients and corresponding decay rates for $W_3 \to B_1 + P$ by using the mixing angle $\beta_{b_1} = 0$ and $\beta_{b_1} = -40^\circ$, respectively. No experimental data is given for this channel in PDG [27] apart from the "possibly seen" decay channel of $\omega_3(1670) \to b_1(1235)\,\pi$.

For the purposes of our study, we make use of the theoretical prediction for $\omega_3(1670) \to \rho(770)\,\pi$ and the ratio in Eq. (5.41) in order to estimate the coupling constant

$$g^2_{w_3 b_1 p} \approx (0.008 \pm 0.003)\ \mathrm{MeV}^{-2}\,. \tag{5.43}$$

Even though they are only the first approximate estimations, the coupling and the results in Table 5.9 could still be useful in identifying dominating decay channels. The following factors are also taken into account while determining and interpreting our results:

1. The decay rate $\Gamma_{\phi_3(1850)\to \bar{K}_{1,B}\,K}$ is rather small when we take into account $K_{1,B} \approx K_1(1270)$. Yet, the results involving the identification $K_{1,B} \approx K_1(1270)$ may only be treated as a first approximation due to the large mixing between $K_{1,A}$ and $K_{1,B}$. By taking into account the full mixing in Eq. (2.71), $K_{1,B}$ should be written as a superposition of $K_1(1270)$ and $K_1(1400)$ which renders the interaction Lagrangian in Eq. (5.37) insufficient on its own and needs the consideration of the interaction terms for the decays $W_3 \to A_1\,P$ in Eq. (5.44), which includes $K_{1,A}$. For the sake of simplicity, we disregard any potential interference between the two Lagrangians, Eq. (5.37) and Eq. (5.44), via the mixing Eq. (2.71). The dominant assignment $K_{1,B} \approx K_1(1270)$ in this subsection is, although a crude approximation, enough for our goals.

2. We present the results in Table 5.9 for two values of the unknown mixing angle $\beta_{b_1} \approx 0$ and $\beta_{b_1} \approx -40^\circ$, the latter being similar to the mixing angle for the pseudoscalar-isoscalars. According to Table 5.9, the mixing angle $\beta_{b_1}$ appears in the decay channels $\rho_3(1690) \to h_1(1170)\,\pi$, $\rho_3(1690) \to h_1(1415)\,\pi$, $K_3^*(1780) \to h_1(1170)\,\bar{K}$, and $\phi_3(1850) \to h_1(1170)\,\eta$. The decay channel $\omega_3(1670) \to b_1(1235)\,\pi$ is the only experimentally "possibly seen" [172], and corresponds to a rather large theoretical partial decay width.

3. The decay channel $\rho_3(1690) \to h_1(1170)\,\pi$ is also independent on the mixing $\beta_{b_1}$. This decay is the dominant one and thus quite interesting for future investigations. The fact that the remaining decay channels are rather small may be the reason why their experimental observation was not yet possible.

### 5.3.6 Decay channel $W_3 \to A_1 + P$

For the decay into an axial-vector and a pseudoscalar meson, the interaction Lagrangian and the corresponding decay formula are given by

$$\mathcal{L}_{w_3 a_1 p} = g_{w_3 a_1 p}\, \mathrm{tr}\big[W_3^{\mu\nu\rho}\, \big[A_{1,\mu},\, (\partial_\nu \partial_\rho P)\big]_-\big]\,, \tag{5.44}$$

$$\Gamma_{W_3 \to A_1 + P}(m_{w_3}, m_{a_1}, m_p) = g^2_{w_3 a_1 p}\, \frac{|\vec{k}_{w_3,a_1,p}|^5}{420\, m^2_{w_3}} \left(7 + 3\, \frac{|\vec{k}_{w_3,a_1,p}|^2}{m^2_{a_1}}\right) \times \kappa_i \Theta(m_{w_3} - m_{a_1} - m_p)\,, \tag{5.45}$$

where $\kappa_i$'s are given in Table 5.10. As previously stated, we are assuming that $K_{1,A} \approx K_1(1400)$ in this estimate. Only some predictions for ratios of various decay channels can be made because we lack sufficient experimental data to determine the coupling constant $g_{w_3 a_1 p}$. These predictions are listed in Table 5.10 and are independent of the mixing angle $\beta_{a_1}$ as can be seen from Table 5.10. In any case, a small mixing is expected [87].

| decay ratio | $\frac{\kappa_i}{\kappa_{w_3 a_1 p}}$ | theory |
|---|---|---|
| $\frac{\Gamma_{K_3^*(1780) \to K_{1,A}\, \pi}}{\Gamma_{\rho_3(1690) \to a_1(1260)\, \pi}}$ | $\frac{\left(\frac{1}{4}\right)^2 + \left(\frac{\sqrt{2}}{4}\right)^2}{2\left(\frac{1}{2}\right)^2}$ | 0.12 |
| $\frac{\Gamma_{K_3^*(1780) \to a_1(1260)\, K}}{\Gamma_{\rho_3(1690) \to a_1(1260)\, \pi}}$ | $\frac{\left(\frac{1}{4}\right)^2 + \left(\frac{\sqrt{2}}{4}\right)^2}{2\left(\frac{1}{2}\right)^2}$ | 0.01 |

**Table 5.10:** Clebsch-Gordan coefficients and corresponding decay ratios for $W_3 \to A_1 + P$ w.r.t. $\rho_3(1690) \to a_1(1260)\, \pi$ decay mode.

Finally, it is intriguing to note that connecting the current model to an underlying chiral model, such as the eLSM, makes it possible to estimate the coupling constant $g_{w_3 a_1 p}$ using Eq. (5.12). This rough estimation leads to

$$g^2_{w_3 a_1 p} \approx 3 \cdot 10^{-4}\ \mathrm{MeV}^{-2} \longrightarrow \Gamma_{\rho_3(1690) \to a_1(1260)\, \pi} = 2\, \mathrm{MeV}\,, \tag{5.46}$$

which implies the suppression of the decays $W_3 \to A_1 + P$ type.

### 5.3.7 Decay channel $W_3 \longrightarrow V_1^{(1)} + V_1^{(2)}$

The decaying process into two vector mesons is described by

$$\mathcal{L}_{w_3 v_1 v_1} = g_{w_3 v_1 v_1} \,\mathrm{tr}\big[W_3^{\mu\nu\rho}\, \big[(\partial_\mu V_{1,\nu}),\, V_{1,\rho}\big]_-\big]\,, \tag{5.47}$$

with the corresponding decay rate

$$\Gamma_{W_3 \to V_1^{(1)} + V_1^{(2)}}(m_{w_3}, m_{v^{(1)}}, m_{v^{(2)}}) = g^2_{w_3 v_1 v_1} \frac{|\vec{k}_{w_3, v^{(1)}, v^{(2)}}|^3}{840\, m^2_{v^{(1)}}\, m^2_{v^{(2)}}\, m^2_{w_3}} \Big[6\, |\vec{k}_{w_3, v^{(1)}, v^{(2)}}|^4 +$$

$$+ 35\, m^2_{v^{(1)}}\, m^2_{v^{(2)}} + 14\, |\vec{k}_{w_3, v^{(1)}, v^{(2)}}|^2 \left(m^2_{v^{(1)}} + m^2_{v^{(2)}}\right)\Big] \kappa_i\, \Theta(m_{w_3} - m_{v^{(1)}} - m_{v^{(2)}})\,, \tag{5.48}$$

where the $\kappa_i$ are given in Table 5.11. Also here, we are unable to estimate the coupling

| decay process | $\kappa_i$ | theory using (5.49) & (5.50) $\Gamma$/MeV | theory using (5.52) & (5.54) $\Gamma$/MeV |
|---|---|---|---|
| $\rho_3(1690) \to \rho(770)\,\rho(770)$ | 1 | $\lesssim 108.6 \pm 35.3$ | $\lesssim 30$ |
| $K_3^*(1780) \to \rho(770)\,\bar{K}^*(892)$ | $(\frac{1}{2})^2 + (\frac{\sqrt{2}}{2})^2$ | $\lesssim 53.6 \pm 17.4$ | $\lesssim 15$ |
| $K_3^*(1780) \to \bar{K}^*(892)\,\omega(782)$ | $[\frac{1}{2}\,(\cos\beta_{v_1} - \sqrt{2}\sin\beta_{v_1})]^2$ | $\lesssim 19.0 \pm 6.2$ | $\lesssim 5$ |
| $\phi_3(1850) \to \bar{K}^*(892)\,K^*(892)$ | $2\,[\frac{1}{2}\,(\sqrt{2}\cos\beta_{w_3} + \sin\beta_{w_3})]^2$ | $\lesssim 38.3 \pm 12.5$ | $\lesssim 11$ |

**Table 5.11:** Estimates of the upper limits for two-vector decay channels.

constant, but we present the decay ratios listed in Table 5.12.

| decay ratio | theory |
|---|---|
| $\frac{\Gamma_{K_3^*(1780)\to\rho(770)\,K^*(892)}}{\Gamma_{\rho_3(1690)\to\rho(770)\,\rho(770)}}$ | 0.49 |
| $\frac{\Gamma_{K_3^*(1780)\to K^*(892)\,\omega(782)}}{\Gamma_{\rho_3(1690)\to\rho(770)\,\rho(770)}}$ | 0.17 |
| $\frac{\Gamma_{\phi_3(1850)\to K^*(892)\,K^*(892)}}{\Gamma_{\rho_3(1690)\to\rho(770)\,\rho(770)}}$ | 0.35 |

**Table 5.12:** Decay ratios for $W_3 \to V_1^{(1)} + V_1^{(2)}$ w.r.t. $\rho_3(1690) \to \rho\,\rho$ decay mode.

We also determine an upper bound of the coupling constant via

$$\frac{\Gamma_{\rho_3(1690)\to\pi\,\pi}}{\Gamma_{\rho_3(1690)\to\rho(770)\,\rho(770)}} \gtrapprox \frac{\Gamma_{\rho_3(1690)\to\pi\,\pi}}{\Gamma_{\rho_3(1690)\to\pi^{\pm}\,\pi^{+}\,\pi^{-}\,\pi^{0}}} \overset{\text{PDG}}{=} 0.35 \pm 0.11\,, \tag{5.49}$$

which is inferred from the fact that the $\rho_3(1690) \to \rho(770)\,\rho(770)$ decay channel is part of the process $\rho_3(1690) \to \pi^{\pm}\,\pi^{+}\,\pi^{-}\,\pi^{0}$, which is one of the dominating channels.

Using the experimental result $\Gamma_{\rho_3(1690)\to\pi\,\pi} = (38.0 \pm 3.2)$ MeV we obtain

$$g^2_{w_3 v_1 v_1} \lesssim 535 \pm 174\,, \tag{5.50}$$

and calculate upper bounds for the decay rates in Table 5.11.

However, using a value near this upper limit of Eq. (5.50) would suggest that the sum of all the decays of the state $\rho_3(1690)$ would exceed the experimental total width of $(161 \pm 10)$ MeV, indicating that the decay width of roughly 108 MeV is undoubtedly too large. This is consistent with the ratio given in PDG [27] where the experimental measurements [173–176],

$$\frac{\Gamma_{\rho_3(1690)\to\rho(770)\,\rho(770)}}{\Gamma_{\rho_3(1690)\to\pi^{\pm}\,\pi^{+}\,\pi^{-}\,\pi^{0}}} < 1\,. \tag{5.51}$$

To get a second, more accurate estimate of these decay channels, we first add up the biggest decay channels:

$$\Gamma^{\text{tot}}_{\rho_3} = 161 \text{ MeV} \approx \tag{5.52}$$

$$\begin{aligned} &= \Gamma_{\rho_3(1690)\to\pi\,\pi} + \Gamma_{\rho_3(1690)\to\bar{K}\,K} + \Gamma_{\rho_3(1690)\to\rho(770)\,\eta} + \\ &+ \Gamma_{\rho_3(1690)\to\bar{K}^*(892)\,K} + \Gamma_{\rho_3(1690)\to\omega(782)\,\pi} + \Gamma_{\rho_3(1690)\to\phi(1020)\,\pi} + \\ &+ \Gamma_{\rho_3(1690)\to a_2(1320)\,\pi} + \Gamma_{\rho_3(1690)\to a_1(1260)\,\pi} + \Gamma_{\rho_3(1690)\to h_1(1170)\,\pi} + \\ &+ \Gamma_{\rho_3(1690)\to h_1(1415)\,\pi} + \Gamma_{\rho_3(1690)\to\rho(770)\,\rho(770)} + \Gamma_{\rho_3(1690)\to\text{other}}\,, \end{aligned}$$

where $\Gamma_{\rho_3\to\text{other}}$ is an abbreviation for any other decay channels that are not mentioned above and are expected to be small. We use the smaller value shown in Table 5.9 for $\Gamma_{\rho_3(1690)\to h_1(1170)\,\pi}$ while assuming a large mixing angle of $\beta_{b_1} = -40^{\circ}$. This method allows us to calculate a reasonable upper limit

$$\Gamma_{\rho_3(1690)\to\rho(770)\,\rho(770)} \lesssim 30 \text{ MeV}\,, \tag{5.53}$$

which implies the following limit for the coupling $g_{w_3v_1v_1}$

$$g^2_{w_3v_1v_1} \lesssim 148\,. \tag{5.54}$$

Table 5.11 contains the predictions for the vector-vector pair decay channels. Next, we add the estimates of the dominant decay channels for the other spin-3 mesons to get the following total decay widths, which are in agreement with the PDG listed in Table 5.3:

$$\Gamma^{\text{sum}}_{K_3^*(1780)} \approx 146 \text{ MeV}\,, \quad \Gamma^{\text{sum}}_{\omega_3(1670)} \approx 175 \text{ MeV}\,, \quad \Gamma^{\text{sum}}_{\phi_3(1850)} \approx 80 \text{ MeV}\,. \tag{5.55}$$

| Parameters | Numerical Values | Sections |
|---|---|---|
| $g^2_{w_3pp}$ | $(1.5 \pm 0.1) \cdot 10^{-10}\,\text{MeV}^{-4}$ | Sec. 5.3.1 |
| $g^2_{w_3vp}$ | $(9.2 \pm 1.9) \cdot 10^{-16}\,\text{MeV}^{-6}$ | Sec. 5.3.2 |
| $g^2_{w_3tp}$ | $(2.8 \pm 1.2) \cdot 10^{-9}\,\text{MeV}^{-4}$ | Sec. 5.3.4 |
| $g^2_{w_3b_1p}$ | $(0.008 \pm 0.003)\,\text{MeV}^{-2}$ | Sec. 5.3.5 |
| $g^2_{w_3a_1p}$ | $\approx 3 \cdot 10^{-4}\,\text{MeV}^{-2}$ | Sec. 5.3.6 |
| $g^2_{w_3v_1v_1}$ | $\lesssim 148$ | Sec. 5.3.7 |

**Table 5.13:** Parameters of Lagrangians in Table 5.1.

## 5.4 Spin-3 Glueball

In this section, we extend the model results to a hypothetical glueball with $J^{PC} = 3^{--}$. This glueball has not yet been observed experimentally, just like all the other gluonic bound states, but its mass is predicted to be around 4.13 GeV [131], 4.33 GeV [132], or 4.20 GeV [130] by LQCD calculations using the quenched approximation. General works on glueballs can be found in *e.g.* Refs. [127, 128, 177, 178], and within effective models in Refs. [22, 83, 85, 101, 179]. All decay ratios in this section should be viewed as first indicative results.

### 5.4.1 Effective Lagrangians for $G_3$

Each glueball is a flavor singlet object, which is invariant under $SU_{\mathrm{V}}(3)$ transformations. We use this fact to determine the Lagrangian analogous to chapter 4 to describe the decays of the spin-3 glueball. In practise we can obtain the needed interaction terms via the simple replacement

$$W^{\mu\nu\rho} \to G_3^{\mu\nu\rho} \cdot \mathbb{1}_{3\times 3}\,, \tag{5.56}$$

where $G_3^{\mu\nu\rho}$ is the glueball field with $J^{PC} = 3^{--}$ and $\mathbb{1}_{3\times 3}$ is the $3 \times 3$ identity matrix. The interaction terms that go along with a commutator consequently disappear, and the only nonzero contributions to an effective action for the $3^{--}$ glueball can be obtained from the previous interaction terms involving the anticommutator in Table 5.14.

| decay process | Interaction Lagrangians |
|---|---|
| $G_3 \to V_1 + P$ | $\mathcal{L}_{g_3 v_1 p} = c_{g_3 v_1 p}\, G_{3,\mu\alpha\beta}\, \varepsilon^{\mu\nu\rho\sigma}\, \mathrm{tr}\big[\{(\partial_\nu V_{1,\rho}),\, (\partial^\alpha \partial^\beta \partial_\sigma P)\}_+\big]$ |
| $G_3 \to \gamma + P$ | $\mathcal{L}_{g_3 \gamma p} = c_{g_3 \gamma p}\, \frac{e}{g_\rho}\, G_{3,\mu\alpha\beta}\, (\partial_\nu a_\rho)\, \varepsilon^{\mu\nu\rho\sigma}\, \mathrm{tr}\big[\{Q,\, (\partial^\alpha \partial^\beta \partial_\sigma P)\}_+\big]$ |
| $G_3 \to V_1 + A_1$ | $\mathcal{L}_{g_3 v_1 a_1} = c_{g_3 v_1 a_1}\, G_{3,\mu\alpha\beta}\, \varepsilon^{\mu\nu\rho\sigma}\, \mathrm{tr}\big[\{(\partial_\nu V_{1,\rho}),\, (\partial^\alpha \partial^\beta A_{1,\sigma})\}_+\big]$ |
| $G_3 \to B_1 + P$ | $\mathcal{L}_{g_3 b_1 p} = c_{g_3 b_1 p}\, G_3^{\mu\nu\rho}\, \mathrm{tr}\big[\{B_{1,\mu},\, (\partial_\nu \partial_\rho P)\}_+\big]$ |
| $G_3 \to B_1 + A_1$ | $\mathcal{L}_{g_3 b_1 a_1} = c_{g_3 b_1 a_1}\, G_3^{\mu\nu\rho}\, \mathrm{tr}\big[\{B_{1,\mu},\, (\partial_\nu A_{1,\rho})\}_+\big]$ |

**Table 5.14:** Interaction Lagrangians for the hypothetical glueball state $G_3(4200)$.

Furthermore, there are decays of the types $G_3(4200) \to V_1 + A_1$ and $G_3(4200) \to B_1 + A_1$ that were kinematically prohibited and therefore not included in the calculations for the lightest conventional mesons with $J^{PC} = 3^{--}$. Theoretical estimates for decay ratios are provided in the section below and can be computed by using the same procedure as before. The only quantities required to calculate the decay ratios are the particle masses.

| decay ratio | theory |
|---|---|
| $\frac{\Gamma_{G_3(4200)\to\bar{K}^*(892)\,K}}{\Gamma_{G_3(4200)\to\rho(770)\,\pi}}$ | 1.11 |
| $\frac{\Gamma_{G_3(4200)\to\omega(782)\,\eta}}{\Gamma_{G_3(4200)\to\rho(770)\,\pi}}$ | 0.17 |
| $\frac{\Gamma_{G_3(4200)\to\omega(782)\,\eta'(958)}}{\Gamma_{G_3(4200)\to\rho(770)\,\pi}}$ | 0.089 |
| $\frac{\Gamma_{G_3(4200)\to\phi(1020)\,\eta}}{\Gamma_{G_3(4200)\to\rho(770)\,\pi}}$ | 0.098 |
| $\frac{\Gamma_{G_3(4200)\to\phi(1020)\,\eta'(958)}}{\Gamma_{G_3(4200)\to\rho(770)\,\pi}}$ | 0.11 |

**Table 5.15:** Decay ratios of $G_3(4200)$ w.r.t. $\rho\,\pi$.

### 5.4.2 Decay channel $G_3 \to V_1 + P$

The decay of a spin-3 tensor glueball into a vector and a pseudoscalar meson is represented by the following interaction Lagrangian:

$$\mathcal{L}_{g_3 v_1 p} = c_{g_3 v_1 p}\, G_{3,\mu\alpha\beta}\, \varepsilon^{\mu\nu\rho\sigma}\, \mathrm{tr}\big[\{(\partial_\nu V_{1,\rho}),\, (\partial^\alpha \partial^\beta \partial_\sigma P)\}_+\big]\,, \tag{5.57}$$

and the decay rate formula:

$$\Gamma_{G_3\to V_1+P}(m_{g_3}, m_v, m_p) = c^2_{g_3 v_1 p}\, \frac{|\vec{k}_{g_3,v,p}|^7}{105}\, \kappa_i\, \Theta(m_{g_3} - m_v - m_p)\,, \tag{5.58}$$

where $\kappa_i$ are given in Table 5.20. Throughout the remaining of the chapter, we use the glueball mass $m_{g_3} = 4200$ MeV in Table 5.2 to calculate the momenta of the outgoing particles. Since we do not know the coupling constant $c_{g_3 v_1 p}$, we compute the nonzero ratios in Table 5.15. This channel will eventually appear in the experiments (ones $G_3$ will be found) as a decay into three (or four) pseudoscalar mesons, because the vector mesons further decay into two (or three) pseudoscalar mesons.

### 5.4.3 Decay channel $G_3 \to \gamma + P$

We can also study radiative decay terms for the spin-3 glueball using vector meson dominance via the shift of Eq. (3.15):

$$\mathcal{L}_{g_3\gamma p} = c_{g_3\gamma p} \frac{e}{g_\rho} G_{3,\mu\alpha\beta} \left(\partial_\nu a_\rho\right) \varepsilon^{\mu\nu\rho\sigma} \operatorname{tr}\left[\left\{Q, \left(\partial^\alpha\partial^\beta\partial_\sigma P\right)\right\}_+\right],$$

leading to the $\kappa_i^\gamma$ are summarized in Table 5.20, which are used in the following decay rate formula to obtain the results in Table 5.16

$$\Gamma_{G_3\to\gamma+P}(m_{g_3}, m_p) = c^2_{g_3 v_1 p} \left(\frac{e}{g_\rho}\right)^2 \frac{\left|\vec{k}_{g_3,\gamma,p}\right|^7}{105\pi} \kappa_i^\gamma .$$

We present the decays $\Gamma_{G_3\to\gamma\, P}$ w.r.t. the strong decay channel $\Gamma_{G_3(4200)\to\rho(770)\,\pi}$ in Table 5.16.

| decay ratio | theory |
|---|---|
| $\frac{\Gamma_{G_3(4200)\to\gamma\,\pi^0}}{\Gamma_{G_3(4200)\to\rho(770)\,\pi}}$ | $129\cdot 10^{-5}$ |
| $\frac{\Gamma_{G_3(4200)\to\gamma\,\eta}}{\Gamma_{G_3(4200)\to\rho(770)\,\pi}}$ | $37\cdot 10^{-5}$ |
| $\frac{\Gamma_{G_3(4200)\to\gamma\,\eta'(958)}}{\Gamma_{G_3(4200)\to\rho(770)\,\pi}}$ | $1\cdot 10^{-5}$ |

**Table 5.16:** Decay ratios of $G_3(4200)$ w.r.t. $G_3(4200) \to \rho\,\pi$.

### 5.4.4 Decay channel $G_3 \to V_1 + A_1$

By taking into account the Lagrangian in which $G_3$ couples to $V_1\,A_1$, we provide the glueball decay ratios into one vector and one axial-vector

$$\mathcal{L}_{g_3 v_1 a_1} = c_{g_3 v_1 a_1} G_{3,\mu\alpha\beta}\, \varepsilon^{\mu\nu\rho\sigma} \operatorname{tr}\left[\left\{\left(\partial_\nu V_{1,\rho}\right), \left(\partial^\alpha\partial^\beta A_{1,\sigma}\right)\right\}_+\right],$$

which leads to the tree-level decay formula given by

$$\Gamma_{G_3 \to V_1 + A_1}(m_{g_3}, m_v, m_{a_1}) = \tag{5.59}$$

$$= c_{g_3 v_1 a_1}^2 \frac{|\vec{k}_{g_3,v,a_1}|^5}{210\, m_{g_3}^2} \left[ |\vec{k}_{g_3,v,a_1}|^2 \left( 3 + 2 \frac{m_{g_3}^2}{m_{a_1}^2} \right) + 7\, m_v^2 \right] \kappa_i \, \Theta(m_{g_3} - m_v - m_{a_1}) \,,$$

where the coefficients $\kappa_i$ are presented in Table 5.20 and the outcomes are shown in Table 5.17.

| decay ratio | theory |
|---|---|
| $\frac{\Gamma_{G_3(4200) \to \bar{K}^*(892)\, K_{1,A}}}{\Gamma_{G_3(4200) \to \rho(770)\, a_1(1260)}}$ | 0.78 |
| $\frac{\Gamma_{G_3(4200) \to \omega(782)\, f_1(1285)}}{\Gamma_{G_3(4200) \to \rho(770)\, a_1(1260)}}$ | 0.29 |
| $\frac{\Gamma_{G_3(4200) \to \omega(782)\, f_1(1420)}}{\Gamma_{G_3(4200) \to \rho(770)\, a_1(1260)}}$ | 0.0009 |
| $\frac{\Gamma_{G_3(4200) \to \phi(1020)\, f_1(1285)}}{\Gamma_{G_3(4200) \to \rho(770)\, a_1(1260)}}$ | 0.0011 |
| $\frac{\Gamma_{G_3(4200) \to \phi(1020)\, f_1(1420)}}{\Gamma_{G_3(4200) \to \rho(770)\, a_1(1260)}}$ | 0.16 |

**Table 5.17:** Decay ratios of $G_3(4200)$ w.r.t. $G_3(4200) \to \rho\, a_1$.

## 5.4.5 Decay channel $G_3 \to B_1 + P$

The decay of the $J^{PC} = 3^{--}$ glueball into a pseudovector and a pseudoscalar meson is given by the interaction Lagrangian

$$\mathcal{L}_{g_3 b_1 p} = c_{g_3 b_1 p}\, G_3^{\mu\nu\rho} \, \mathrm{tr}\big[ \{ B_{1,\mu},\, (\partial_\nu \partial_\rho P) \}_+ \big] \,, \tag{5.60}$$

which leads to the decay rate formula

$$\Gamma_{G_3\to B_1+P}(m_{g_3}, m_{b_1}, m_p) = c^2_{g_3 b_1 p} \frac{|\vec{k}_{g_3,b_1,p}|^5}{420\, m^2_{g_3}} \left(7 + 3\frac{|\vec{k}_{g_3,b_1,p}|^2}{m^2_{b_1}}\right) \kappa_i\, \Theta(m_{g_3} - m_{b_1} - m_p)\,. \tag{5.61}$$

We consider two mixing angle options ($\beta_{b_1} \approx 0^\circ$ and $\beta_{b_1} \approx -40^\circ$) and report the decay rates in Table 5.18 using $m_{K_{1,B}} \approx m_{K_1(1270)}$.

| decay ratio | theory for $\beta_{b_1} = 0^\circ$ | theory for $\beta_{b_1} = -40^\circ$ |
|---|---|---|
| $\frac{\Gamma_{G_3(4200)\to K_{1,B}\,K}}{\Gamma_{G_3(4200)\to b_1(1235)\,\pi}}$ | 1.15 | 1.15 |
| $\frac{\Gamma_{G_3(4200)\to h_1(1170)\,\eta}}{\Gamma_{G_3(4200)\to b_1(1235)\,\pi}}$ | 0.17 | 0.33 |
| $\frac{\Gamma_{G_3(4200)\to h_1(1170)\,\eta'(958)}}{\Gamma_{G_3(4200)\to b_1(1235)\,\pi}}$ | 0.12 | 0.001 |
| $\frac{\Gamma_{G_3(4200)\to h_1(1415)\,\eta}}{\Gamma_{G_3(4200)\to b_1(1235)\,\pi}}$ | 0.10 | 0.001 |
| $\frac{\Gamma_{G_3(4200)\to h_1(1415)\,\eta'(958)}}{\Gamma_{G_3(4200)\to b_1(1235)\,\pi}}$ | 0.08 | 0.16 |

**Table 5.18:** Decay ratios of $G_3(4200)$ w.r.t. $G_3(4200) \to b_1\,\pi$.

### 5.4.6 Decay channel $G_3 \to B_1 + A_1$

Additionally, for $\beta_{a_1} \approx 0^\circ$, we show in Table 5.19 the results for the decay of the spin-3 glueball into a pseudovector and an axial-vector using the following Lagrangian

$$\mathcal{L}_{g_3 b_1 a_1} = c_{g_3 b_1 a_1}\, G_3^{\mu\nu\rho}\, \mathrm{tr}\big[\{B_{1,\mu}, (\partial_\nu A_{1,\rho})\}_+\big]\,, \tag{5.62}$$

with the resulting decay formula

$$\Gamma_{G_3 \to B_1 + A_1}(m_{g_3}, m_{b_1}, m_{a_1}) = \tag{5.63}$$
$$= c^2_{g_3 b_1 a_1} \frac{|\vec{k}_{g_3,b_1,a_1}|^3}{840\, m^2_{b_1}\, m^2_{a_1}\, m^2_{g_3}} \left[ 6\, |\vec{k}_{g_3,b_1,a_1}|^4 + \right.$$
$$\left. + 35\, m^2_{b_1}\, m^2_{a_1} + 14\, |\vec{k}_{g_3,b_1,a_1}|^2 \left( m^2_{b_1} + m^2_{a_1} \right) \right] \kappa_i\, \Theta(m_{g_3} - m_{b_1} - m_{a_1}) \,,$$

where the $\kappa_i$ are listed in Table 5.20.

| decay ratio | theory for $\beta_{b_1} = 0^\circ$ | theory for $\beta_{b_1} = -40^\circ$ |
|---|---|---|
| $\frac{\Gamma_{G_3(4200) \to K_{1,B}\, K_{1,A}}}{\Gamma_{G_3(4200) \to b_1(1235)\, a_1(1260)}}$ | 0.96 | 0.96 |
| $\frac{\Gamma_{G_3(4200) \to h_1(1170)\, f_1(1285)}}{\Gamma_{G_3(4200) \to b_1(1235)\, a_1(1260)}}$ | 0.34 | 0.20 |
| $\frac{\Gamma_{G_3(4200) \to h_1(1170)\, f_1(1420)}}{\Gamma_{G_3(4200) \to b_1(1235)\, a_1(1260)}}$ | 0 | 0.11 |
| $\frac{\Gamma_{G_3(4200) \to h_1(1415)\, f_1(1285)}}{\Gamma_{G_3(4200) \to b_1(1235)\, a_1(1260)}}$ | 0 | 0.092 |
| $\frac{\Gamma_{G_3(4200) \to h_1(1415)\, f_1(1420)}}{\Gamma_{G_3(4200) \to b_1(1235)\, a_1(1260)}}$ | 0.17 | 0.10 |

**Table 5.19:** Decay ratios of $G_3(4200)$ w.r.t. $G_3(4200) \to b_1\, a_1$ .

# 5.5 Conclusions

In this chapter, we used an effective QFT model based on flavor symmetry to study the decays of the lightest mesonic $\bar{q}q$ nonet with $J^{PC} = 3^{--}$. Only the dominant terms in the large-$N_\mathrm{c}$ expansion and the lowest derivatives were kept in our model. We come to the conclusion that the $\bar{q}q$ assignment for decay widths and some known branching ratios, which are listed in PDG [27], works quite well, as the comparison of theoretical outcomes with the current state of experimental data shows. However, some of the decay channels in our study also merit more in-depth theoretical and experimental work in the future. Moreover, we presented a variety of predictions for strong and radiative decays of conventional $J^{PC} = 3^{--}$ mesons, eventually useful for upcoming experimental tests. Figure 5.1 and Figure 5.2 summarize the comparison between our model results and PDG as well as LQCD predictions.

| decay process | $\kappa_i$ |
|---|---|
| $G_3(4200) \to \rho(770)\,\pi$ | 3 |
| $G_3(4200) \to \bar{K}^*(892)\,K$ | 4 |
| $G_3(4200) \to \omega(782)\,\eta$ | $[\cos(\beta_p - \beta_{v_1})]^2$ |
| $G_3(4200) \to \omega(782)\,\eta'(958)$ | $[\sin(\beta_p - \beta_{v_1})]^2$ |
| $G_3(4200) \to \phi(1020)\,\eta$ | $[\sin(\beta_p - \beta_{v_1})]^2$ |
| $G_3(4200) \to \phi(1020)\,\eta'(958)$ | $[\cos(\beta_p - \beta_{v_1})]^2$ |
| $G_3(4200) \to \gamma\,\pi^0$ | 1 |
| $G_3(4200) \to \gamma\,\eta$ | $\left[\frac{1}{3}\left(\cos\beta_p - \sqrt{2}\sin\beta_p\right)\right]^2$ |
| $G_3(4200) \to \gamma\,\eta'(958)$ | $\left[\frac{1}{3}\left(\sin\beta_p + \sqrt{2}\cos\beta_p\right)\right]^2$ |
| $G_3(4200) \to \rho(770)\,a_1(1260)$ | 3 |
| $G_3(4200) \to \bar{K}^*(892)\,K_{1,A}$ | 4 |
| $G_3(4200) \to \omega(782)\,f_1(1285)$ | $[\cos(\beta_{a_1} - \beta_{v_1})]^2$ |
| $G_3(4200) \to \omega(782)\,f_1'(1420)$ | $[\sin(\beta_{a_1} - \beta_{v_1})]^2$ |
| $G_3(4200) \to \phi(1020)\,f_1(1285)$ | $[\sin(\beta_{a_1} - \beta_{v_1})]^2$ |
| $G_3(4200) \to \phi(1020)\,f_1'(1420)$ | $[\cos(\beta_{a_1} - \beta_{v_1})]^2$ |
| $G_3(4200) \to b_1(1235)\,\pi$ | 3 |
| $G_3(4200) \to K_{1,B}\,K$ | 4 |
| $G_3(4200) \to h_1(1170)\,\eta$ | $[\cos(\beta_p - \beta_{b_1})]^2$ |
| $G_3(4200) \to h_1(1170)\,\eta'(958)$ | $[\sin(\beta_p - \beta_{b_1})]^2$ |
| $G_3(4200) \to h_1(1415)\,\eta$ | $[\sin(\beta_p - \beta_{b_1})]^2$ |
| $G_3(4200) \to h_1(1415)\,\eta'(958)$ | $[\cos(\beta_p - \beta_{b_1})]^2$ |
| $G_3(4200) \to b_1(1235)\,a_1(1260)$ | 3 |
| $G_3(4200) \to K_{1,B}\,K_{1,A}$ | 4 |
| $G_3(4200) \to h_1(1170)\,f_1(1285)$ | $[\cos(\beta_{a_1} - \beta_{b_1})]^2$ |
| $G_3(4200) \to h_1(1170)\,f_1'(1420)$ | $[\sin(\beta_{a_1} - \beta_{b_1})]^2$ |
| $G_3(4200) \to h_1(1415)\,f_1(1285)$ | $[\sin(\beta_{a_1} - \beta_{b_1})]^2$ |
| $G_3(4200) \to h_1(1415)\,f_1'(1420)$ | $[\cos(\beta_{a_1} - \beta_{b_1})]^2$ |

**Table 5.20:** Clebsch-Gordan coefficients $\kappa_i$ for the decay channels of $G_3(4200)$ .

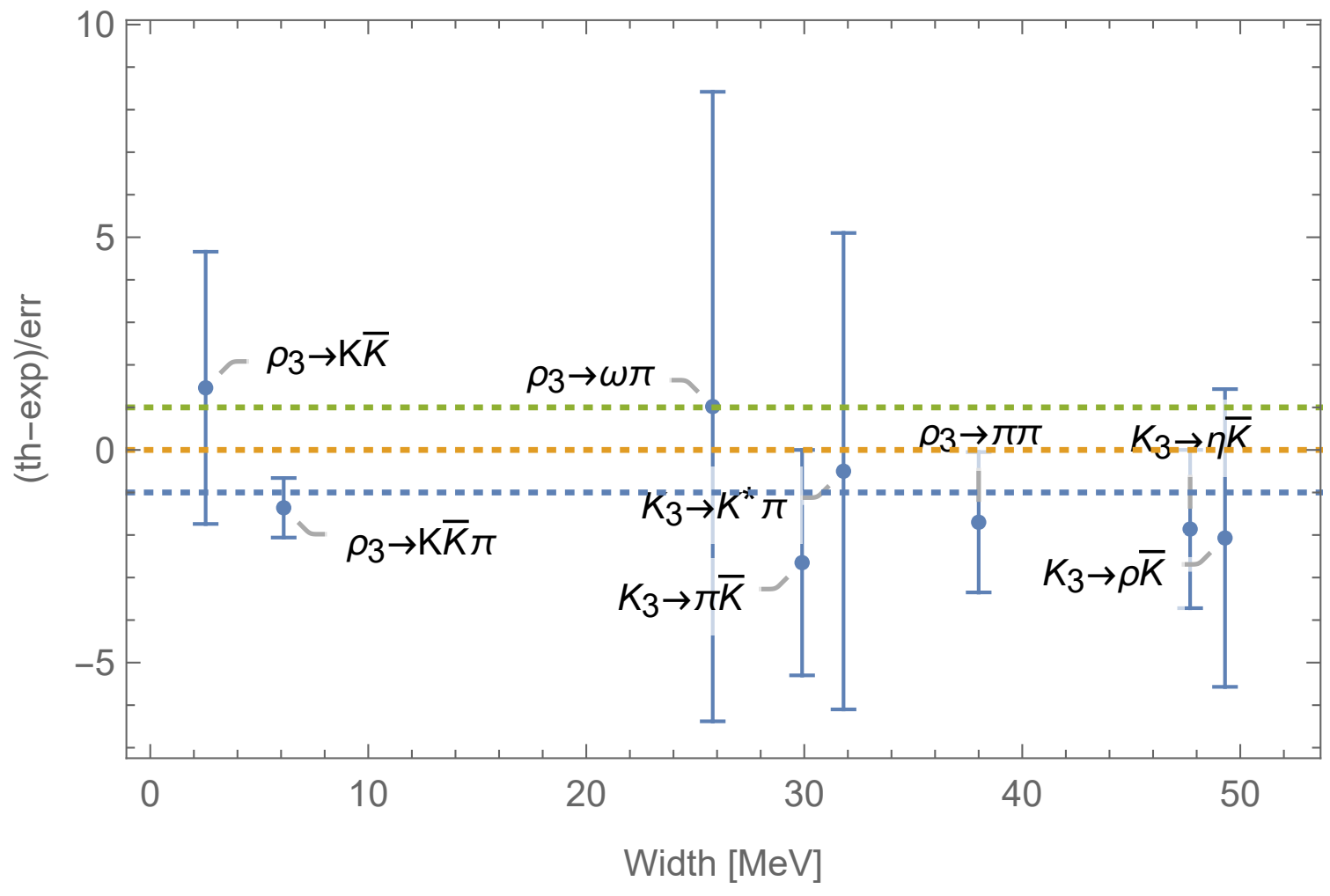

**Figure 5.1:** Model results compared to PDG [27].

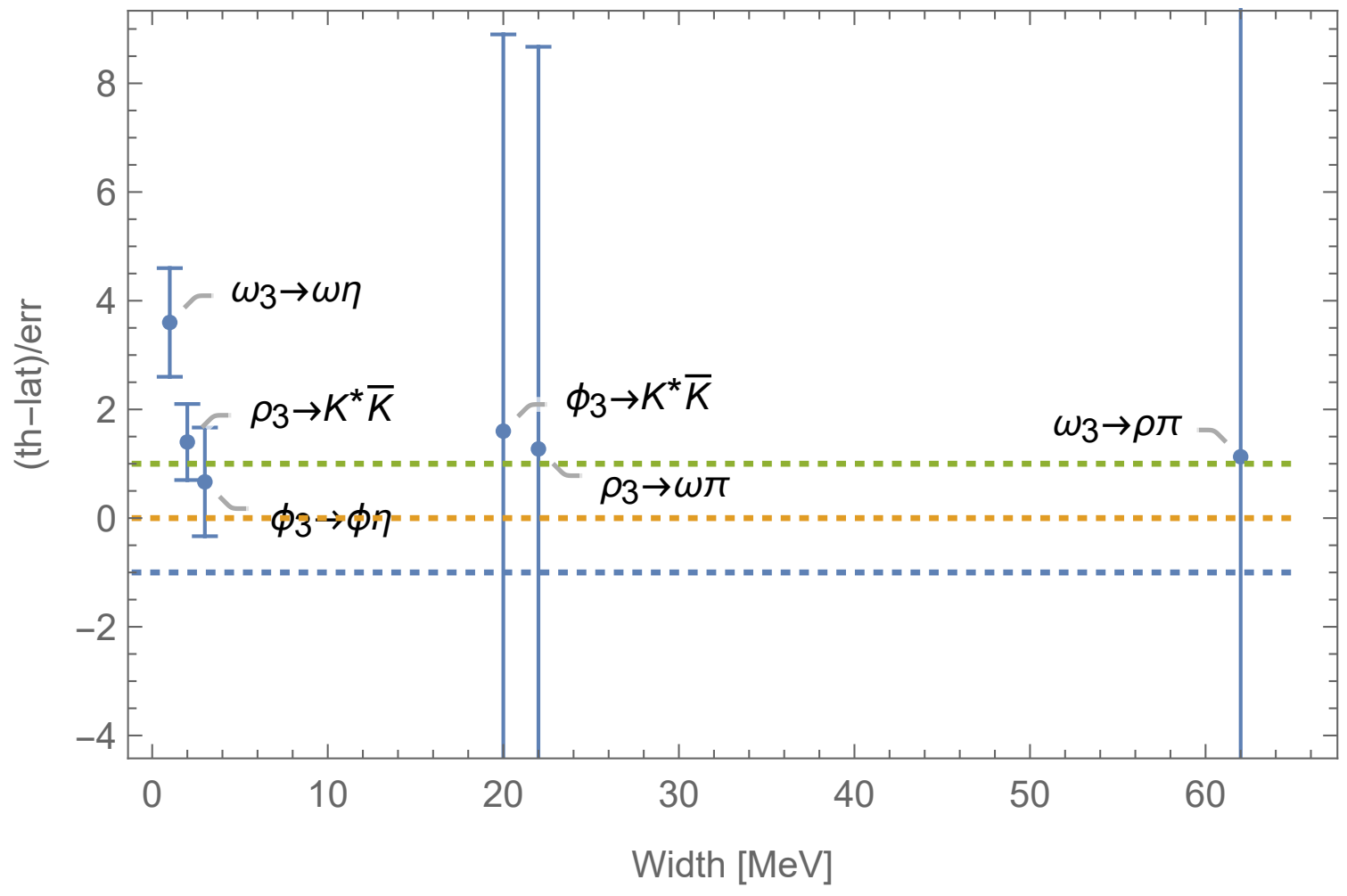

**Figure 5.2:** Model results compared to LQCD [28].

We have then extended the effective model for spin-3 mesons to the case of a glueball. Because of the unknown coupling, we presented only the strong and radiative decay ratios.

# Chapter 6

# Glueball Resonance Gas model

This chapter is about the effect of the glueball spectrum on the pure YM part of the QCD Equation of State (EoS) at nonzero temperature.

## 6.1 Introduction

Some experiments at the RHIC and LHC are devoted to the formation of extremely hot and dense QCD matter through the collision of heavy nuclei. Intriguing non-perturbative characteristics of QCD are realized in the QCD in the medium, which are explored through these experiments. That is why the understanding of the QCD phase diagram is the subject of extensive theoretical studies [180–182]. Theoretical methods for studying the QCD phase diagram are divided into two parts:

1. From first principles:
    - Lattice QCD (see e.g. Refs. [30, 183–188]);
    - Functional methods (functional renormalization group - FRG (see e.g. Refs. [45, 189–191]);
    - Dyson-Schwinger equation (see e.g. Refs. [192, 193]);
    - Perturbative approach ( see e.g. [194] and Refs therein).
2. Model-based approaches:
    - Nambu-Jona-Lasinio (NJL)-type models [112, 113];
    - Hadronic models (see e.g. eLSM [195, 196] );

- Hadron Resonance Gas (HRG) model (see e.g. [197–202]).

At zero quark chemical potential, LQCD shows a smooth cross-over phase transition between high-temperature regime quark-gluon plasma (QGP) and HRG at low temperature.

As we have mentioned in the previous chapters, glueballs are the predictions of the pure YM sector of QCD. They have been investigated using a variety of techniques, such as LQCD, where their spectrum is analyzed; in spite of that, they need a final experimental confirmation (although certain candidates do exist) [29, 132, 203, 204]. They have been investigated in various models [67, 205–208]. Studies on the pomeron and odderon trajectories support the existence of glueballs. [209].

One can use glueballs to construct the so-called Glueball Resonance Gas (GRG) in order to study the properties of the YM Equation of State at low temperatures. Lattice simulations show a first-order phase transition between the gluonic bound states at low temperature (the glueballs) and a gas of gluons at high temperature [30, 210–215]. The inclusion of the Hagedorn contribution (see the original work in [216]) to the GRG with the mass spectrum Ref. [203] is proposed in order to fill the gap between the lattice thermodynamic simulations for the pressure close to the critical temperature $T_c$ [217].

There are other methods to study the thermal properties of QCD, including the T-Matrix approach [218], the Polyakov-loop potential [219], functional methods [220], holographic QCD [221], quasi-particle models [222, 223], bag model [224, 225], hard-thermal-loop perturbation theory [198] and a gas of gluons at high temperature [226–228].

In this chapter, we revisit the role of the glueball spectrum, taken from Refs. [29, 132, 203] within the low-temperature phase diagram of the pure YM sector of QCD by comparing the results to the lattice thermodynamic outcomes of Ref. [30]. The physical masses of the glueball spectrum depend on the lattice parameters, which eventually lead to different $T_c$ for three different glueball gases. The best agreement of the GRG and with thermal LQCD simulation in [30] is obtained by using the masses of Ref. [29]. This agreement implies $T_c \sim 320 \pm 20$ MeV which is somewhat larger than the commonly used value of $T_c \sim 260$ MeV [30]. It is interesting to note that Ref [229] estimates $T_c$ around 300 MeV for pure YM using functional techniques. Furthermore, assuming the Regge trajectories for the glueball states, contributions from radially excited glueballs are also taken into account. Also, the role of the interaction between glueballs will be investigated.

## 6.2 Thermodynamics of the GRG

The energy density $\epsilon_i$ and pressure $p_i$ of the i-th glueball with the total spin $J_i$ are used to explain the thermodynamic properties of the GRG and have the following forms (see Ref. [180] for more details about QCD thermodynamics):

$$\epsilon_i = (2J_i + 1) \int_0^\infty \frac{k^2}{2\pi^2} \frac{\sqrt{k^2 + m_i^2}}{\exp\left[\frac{\sqrt{k^2+m_i^2}}{T}\right] - 1} dk \,, \tag{6.1}$$

and

$$p_i = -(2J_i + 1) T \int_0^\infty \frac{k^2}{2\pi^2} \ln\left(1 - e^{-\frac{\sqrt{k^2+m_i^2}}{T}}\right) dk \,. \tag{6.2}$$

The total energy density and the total pressure of the GRG are calculated by taking into account $N$ glueballs as follows:

$$\epsilon^{GRG} \equiv \epsilon = \sum_{i=1}^{N} \epsilon_i \,, \qquad p^{GRG} \equiv p = \sum_{i=1}^{N} p_i \,. \tag{6.3}$$

Information regarding the trace anomaly $I$ and the entropy density $s$ is provided

$$I = \epsilon - 3p \,, \quad s = \frac{p + \epsilon}{T} \,. \tag{6.4}$$

In the following, it is practical to work with the dimensionless pressure $\hat{p}$, energy density $\hat{\epsilon}$, trace anomaly $\hat{I}$ and entropy $\hat{s}$

$$\hat{p} = p/T^4 \,, \quad \hat{\epsilon} = \epsilon/T^4 \,, \quad \hat{I} = \hat{\epsilon} - 3\hat{p} \,, \quad \hat{s} = s/T^3 \,. \tag{6.5}$$

Using the masses from the lattice simulations provided in Table 6.1 and Table 6.2, we compute the GRG pressure and energy density using the aforementioned equations and compare the results to the relevant lattice calculations below the critical temperature in the continuum limit of Ref. [30]. The thermodynamic quantities are functions of the ratio $T/T_c$ in Ref. [30]. Thus, it is crucial to take into account the value of the critical temperature independently for each given lattice glueball spectrum. To this end, we call that the critical temperature $T_c$ is connected to both the QCD string tension $\sigma$ and the so-called Lambda parameter $\Lambda_{MS}$ as follows [210,230,231]:

$$T_c = 0.629(3) \cdot \sqrt{\sigma}, \tag{6.6}$$

$$T_c = 1.26(7) \cdot \Lambda_{MS} = 1.26(7) \cdot 0.614(2) \cdot r_0^{-1} \,. \tag{6.7}$$

By using the formulas above, we determine the relevant critical temperature (along with an estimation of its error) for each simulation of the glueball mass spectrum, which is summarized in Table 6.1 (only statistical errors are given in [29]).

| LQCD papers | Number of glueballs | Lattice Parameter | $T_c$ (using Eqs. (6.6)-(6.7)) |
|---|---|---|---|
| Chen et.al [203] | 12 | $r_0^{-1} = 410(20)$ MeV | $317 \pm 23$ MeV |
| Meyer [132] | 22 | $\sqrt{\sigma} = 440(20)$ MeV | $277 \pm 13$ MeV |
| Athenodorou and Teper [29] | 20 | $r_0^{-1} = 418(5)$ MeV | $323 \pm 18$ MeV |

**Table 6.1:** The LQCD parameters that have been used to get the glueball mass spectrum.

In Ref. [29], inspired by [232], a new scale $r_0 = 0.472(5)$ fm (corresponding to $r_0^{-1} = 418(5)$ MeV) is used, which results in slightly different glueball masses compared to Ref. [203]. Yet the results are compatible within errors. In contrast, masses and corresponding $T_c$ in Ref. [132] are slightly smaller than Refs. [29,132] .

In Ref. [135] for $N_c = 6$, a new approach is used to determine the masses of the glueballs with quantum numbers $J^{PC} = 0^{++}, 2^{++}, 0^{-+}, 2^{-+}, 1^{+-}$. The results match the values of Ref. [29] with an error of $2 - 5\%$, validating both the large-$N_c$ limit and the masses in that reference. Moreover, it is intriguing that recent developments in functional techniques and Bethe-Salpeter approaches provide a spectrum that is similar to that found from lattice simulations [35,36,233].

We contrast LQCD simulations for the pure YM sector in Ref. [30] to the GRG results for pressure in Fig. 6.1, the trace anomaly in Fig. 6.2 and the entropy density in Fig. 6.3 using the parameters given in Table 6.1. As we see from the plots, the most recent lattice study, [29], generates a GRG that closely resembles the results in Ref. [30]

| $n\,J^{PC}$ | M[MeV] | | | $n\,J^{PC}$ | M[MeV] | | |
|---|---|---|---|---|---|---|---|
| | Chen et.al. [203] | Meyer [132] | A & T [29] | | Chen et.al. [203] | Meyer [132] | A & T [29] |
| $1\,0^{++}$ | 1710(50)(80) | 1475(30)(65) | 1653(26) | $1\,1^{--}$ | 3830(40)(190) | 3240(330)(150) | 4030(70) |
| $2\,0^{++}$ | | 2755(30)(120) | 2842(40) | $1\,2^{--}$ | 4010(45)(200) | 3660(130)(170) | 3920(90) |
| $3\,0^{++}$ | | 3370(100)(150) | | $2\,2^{--}$ | | 3740(200)(170) | |
| $4\,0^{++}$ | | 3990(210)(180) | | $1\,3^{--}$ | 4200(45)(200) | 4330(260)(200) | |
| $1\,2^{++}$ | 2390(30)(120) | 2150(30)(100) | 2376(32) | $1\,0^{+-}$ | 4780(60)(230) | | |
| $2\,2^{++}$ | | 2880(100)(130) | 3300(50) | $1\,1^{+-}$ | 2980(30)(140) | 2670(65)(120) | 2944(42) |
| $1\,3^{++}$ | 3670(50)(180) | 3385(90)(150) | 3740(70) | $2\,1^{+-}$ | | | 3800(60) |
| $1\,4^{++}$ | | 3640(90)(160) | 3690(80) | $1\,2^{+-}$ | 4230(50)(200) | | 4240(80) |
| $1\,6^{++}$ | | 4360(260)(200) | | $1\,3^{+-}$ | 3600(40)(170) | 3270(90)(150) | 3530(80) |
| $1\,0^{-+}$ | 2560(35)(120) | 2250(60)(100) | 2561(40) | $2\,3^{+-}$ | | 3630(140)(160) | |
| $2\,0^{-+}$ | | 3370(150)(150) | 3540(80) | $1\,4^{+-}$ | | | 4380(80) |
| $1\,2^{-+}$ | 3040(40)(150) | 2780(50)(130) | 3070(60) | $1\,5^{+-}$ | | 4110(170)(190) | |
| $2\,2^{-+}$ | | 3480(140)(160) | 3970(70) | | | | |
| $1\,5^{-+}$ | | 3942(160)(180) | | | | | |
| $1\,1^{-+}$ | | | 4120(80) | | | | |
| $2\,1^{-+}$ | | | 4160(80) | | | | |
| $3\,1^{-+}$ | | | 4200(90) | | | | |

**Table 6.2:** The list of the glueball masses in Refs. [29, 132, 203].

for temperatures less than $0.9T_c$. The critical temperature $T_c \approx 323 \pm 18$ MeV turns out to be a larger than the reference value 260-270 MeV estimated in Ref. [30], but is consistent with the FRG result of Ref. [229].

In particular, we see from Figs. (6.1-6.3) that the GRG based on the mass spectrum in Ref. [29] (green dashed line) explains the data of Ref. [30] (blue thick line) better than the mass spectrum in Refs [132, 203] (black dotted and orange dashed lines). The fact that the pressure determined in Ref. [30] agrees with the free GRG with glueballs from [29] also suggests that the additional contribution from Hagedorn spectrum may not be necessary, at least up to $0.9T_c$.

The agreement is not spoiled by the contribution of further states as well as the interaction between the glueballs, see the next sections. Having a negligible contribution from the interaction is consistent with predictions by the large-$N_c$ limit of the YM theory, which implies the glueballs can be seen, in first approximation, as non-interacting bosons since the interaction among glueballs scales as $N_c^{-2}$ [119, 120, 234].

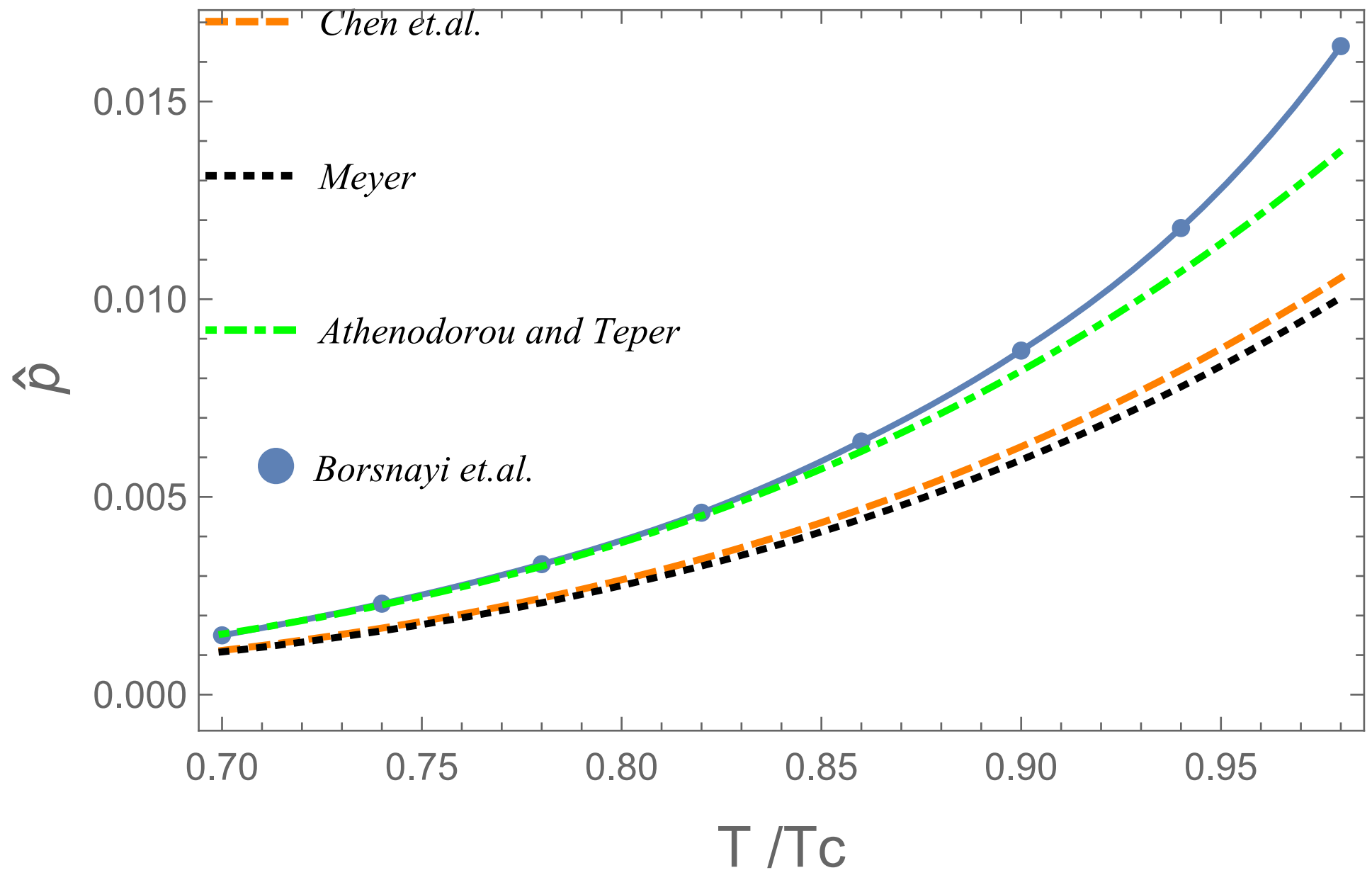

**Figure 6.1:** Comparison of the pressure (continuous limit) calculated in Ref. [30] to the GRG pressures for three different sets of lattice masses [29,132,203].

## 6.3 Contribution of heavier glueballs

In this section, we study the contribution to the thermodynamical quantities of further heavier excited glueball states, which have not been observed in lattice simulations yet. Namely, for each $J^{PC}$, an infinite tower of glueball states is believed to exist [119].

Light conventional quark-antiquark states can be analyzed quantitatively in a three-dimensional space including the mass $M$, the radial quantum number $n$, and the total spin $J$, as indicated in various publications (e.g., Refs. [235,236]). Masses of the conventional mesons fulfill the relation and lie in the so-called Regge trajectories:

$$M^2(n, J^{PC}) = a(n+J) + b_{J^{PC}} \,. \tag{6.8}$$

We extend the Regge trajectories for the conventional mesons to the unconventional ones, such as glueballs. This approximation is limited because of the lack of experimental data. There are other elements that must be taken into account, such as the number of gluons included in a particular $J^{PC}$ glueball, in order to group the states into various planes [132]. Also, despite the remark of Chen et. al. [237] about the ground-state glueballs with the quantum numbers $J^{PC} = 1^{++}$ and $1^{-+}$ that do not exist in the relativistic framework because their corresponding currents disappear, the authors of Ref. [204] obtain the masses for these glueballs. This discrepancy implies that, eventually, not all $J^{PC}$ may be formed in LQCD.

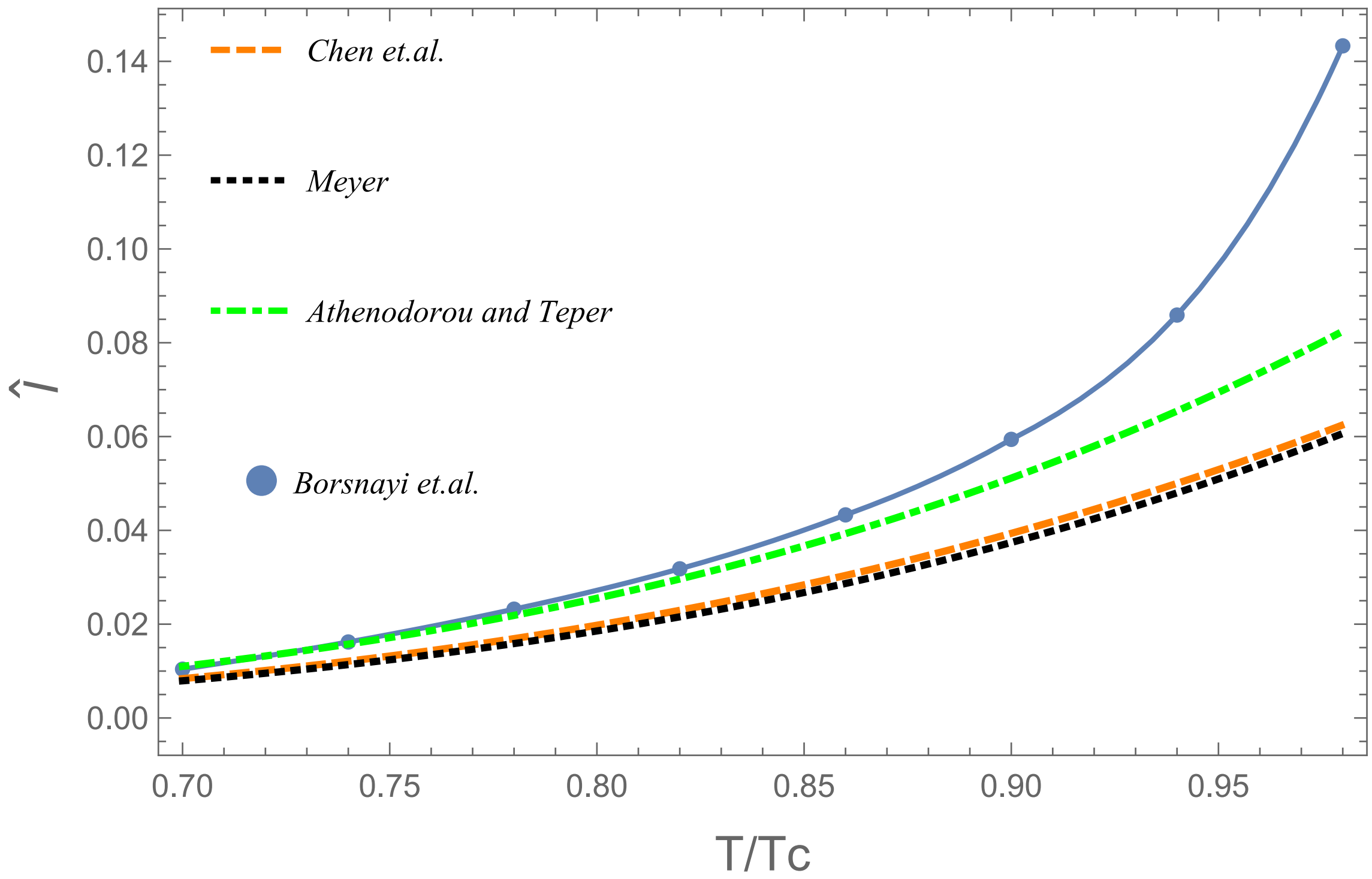

**Figure 6.2:** Comparison of the trace anomaly (continuous limit) calculated in Ref. [30] to the GRG trace anomalies for three different sets of lattice masses [29, 132, 203].

One can approximately construct a Regge trajectory for the glueball states in Ref. [29] by considering the ground states with physical masses lighter than 3 GeV. There are five such states with quantum numbers: $J^{PC} = 0^{++}, 2^{++}, 0^{-+}, 2^{-+}, 1^{+-}$ (Note that these ground states are observed in Refs. [132, 203] as well.) We define the $\chi^2$ as follows:

$$\chi^2(a, b_{0^{++}}, b_{2^{++}}, b_{0^{-+}}, b_{2^{-+}}, b_{1^{+-}}) = \sum_{n=1}^{2} \sum_{J^{PC}} \left( \frac{M(n, J^{PC}) - M^{\text{lat}}(n, J^{PC})}{\delta M^{\text{lat}}(n, J^{PC})} \right)^2, \qquad (6.9)$$

where the parameters of the Regge trajectories are $a$ and $b_{J^{PC}}$; the masses of the glueballs and their errors on them are $M^{\text{lat}}$ and $\delta M^{\text{lat}}$ as shown in the "AT" columns of Table 6.2. Table 6.3 shows the numerical values of the parameters from the fit, along with the masses that were employed. Interestingly, as a result of the fit for the masses in Ref. [29], we obtain a unique slope with $a = 5.49 \pm 0.17\,\text{GeV}^2$ and $\chi^2_{\text{d.o.f.}} = 0.84$.

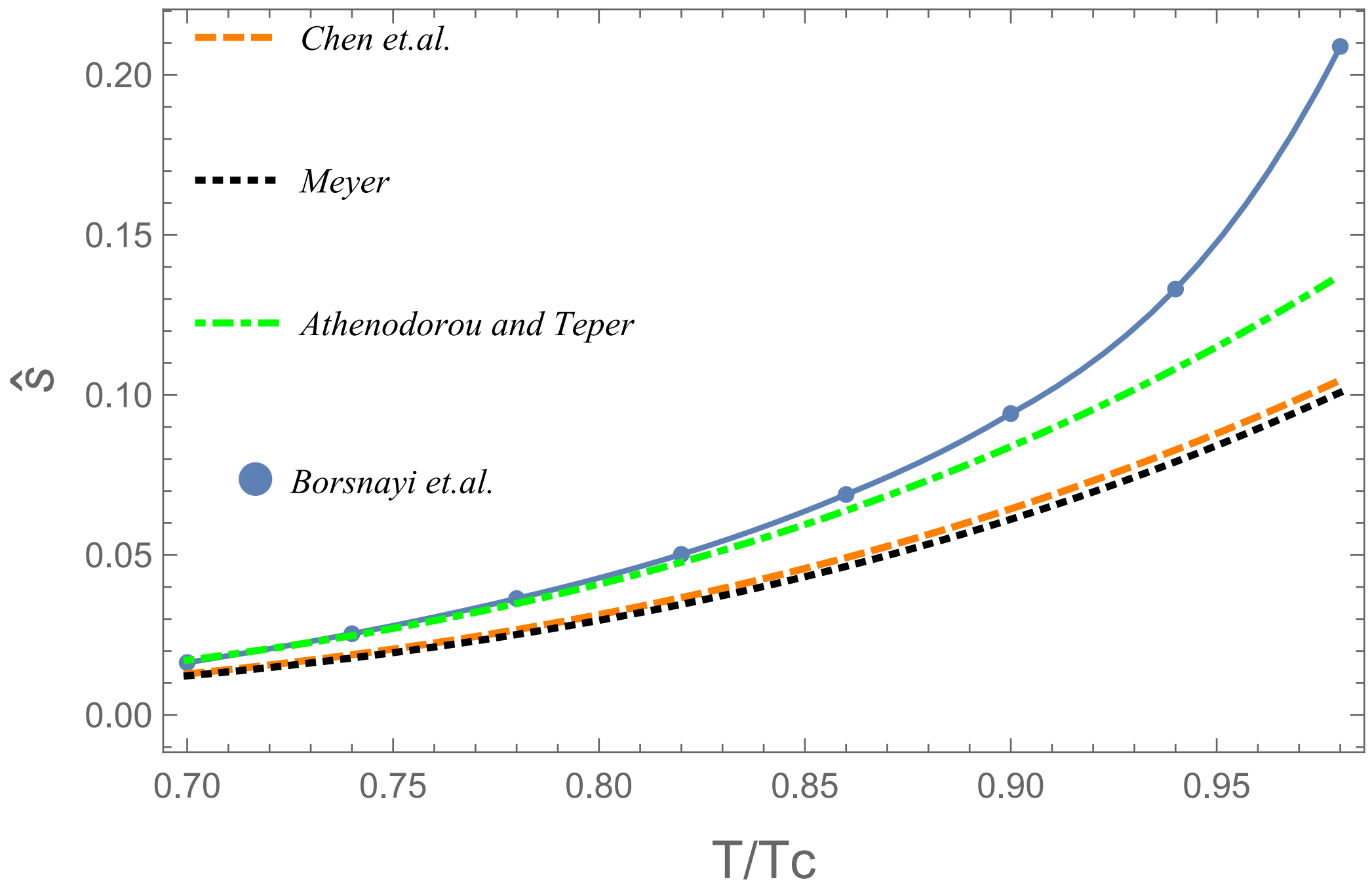

**Figure 6.3:** Comparison of the entropy (continuous limit) calculated in Ref. [30] to the GRG entropies for three different sets of lattice masses [29, 132, 203].

The pressure and trace anomaly of the GRG are then calculated using these Regge trajectories by including glueball states up to $n = 10$ for each quantum number specified in Table 6.2 (i.e., assuming the same slope $a$ for all observed ground states in Ref. [29], not only those used in the fit of Eq. (6.9)). From Fig. 6.4, we see that the effect of further radially excited glueball states on the pressure and trace anomaly (red full lines) becomes visible very near $T_c$, but it is still very small [1].

It should be stressed tat, the technique used in this section may be considered as a way to estimate the magnitude of the impact of heavy glueballs on the thermodynamic variables, but it does not provide aprecise way to predict the masses of further radially exited glueballs.

Before finishing this section, we present the pressure of GRG in Fig. 6.5, in which the substantial impact of pomeron (C=+1) states becomes evident when compared to odderon (C=-1) states. The pomeron states dominate, but the odderon ones are non-negligible. For studies of pomeron/odderon, see Refs. [209, 238, 239].

[1] If we take into account an infinite number of glueballs with Hagedorn-like properties, the picture could eventually alter and improve the agreement around $T_c$.

| Glueball spectrum compared to the fit in Eq. (6.9) | | | | Parameters [GeV$^2$] |
|---|---|---|---|---|
| $n\ J^{PC}$ | $m$ [GeV] (from [29]) | Fit [GeV] | $\chi_i^2$ | |
| $\mathbf{1}\ \mathbf{0}^{++}$ | 1.653(26) | 1.647(25) | 0.04 | $b_{0^{++}} = -2.78 \pm 0.21$ |
| $\mathbf{2}\ \mathbf{0}^{++}$ | 2.842(40) | 2.865(30) | 0.3 | |
| $\mathbf{1}\ \mathbf{2}^{++}$ | 2.376(32) | 2.367(30) | 0.08 | $b_{2^{++}} = -10.87 \pm 0.57$ |
| $\mathbf{2}\ \mathbf{2}^{++}$ | 3.30(5) | 3.33(3) | 0.38 | |
| $\mathbf{1}\ \mathbf{0}^{-+}$ | 2.561(40) | 2.572(38) | 0.08 | $b_{0^{-+}} = 1.12 \pm 0.27$ |
| $\mathbf{2}\ \mathbf{0}^{-+}$ | 3.54(8) | 3.48(4) | 0.57 | |
| $\mathbf{1}\ \mathbf{2}^{-+}$ | 3.07(6) | 3.11(5) | 0.52 | $b_{2^{-+}} = -6.79 \pm 0.66$ |
| $\mathbf{2}\ \mathbf{2}^{-+}$ | 3.97(7) | 3.90(4) | 1.10 | |
| $\mathbf{1}\ \mathbf{1}^{+-}$ | 2.944(42) | 2.955(37) | 0.07 | $b_{1^{+-}} = -2.25 \pm 0.45$ |
| $\mathbf{2}\ \mathbf{1}^{+-}$ | 3.80(6) | 3.77(3) | 0.23 | |
| | | | $\chi^2_{tot}$=3.38 | $a = 5.49 \pm 0.17$ |

**Table 6.3:** Fit summary obtained from Eq. (6.9) by assuming Eq. (6.8).

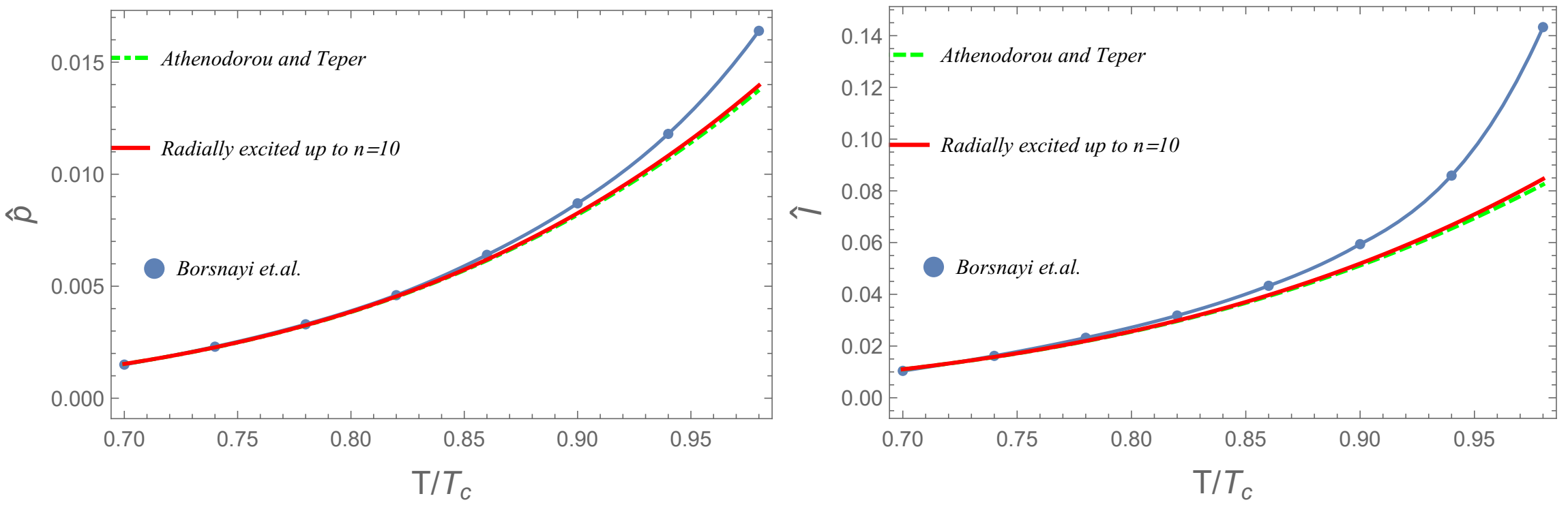

**Figure 6.4:** The effect of the radially excited glueball states up to $n = 10$ within GRG on the pressure (left) and trace anomaly (right).

## 6.4 Glueball-glueball interactions

The GRG was approximated as a free gas in the previous section, which is a valid approximation at lowest order. This section is devoted to studying the role of the interaction between glueballs in GRG. Importantly, the interactions of two lightest states in the glueball spectrum, scalar and tensor glueball, are investigated.

One of the common methods is the so-called S-matrix or phase-space formalism [240] applied to the investigation of the nonzero properties of hadrons (see e.g. Refs. [241–253]). The derivative of the scattering phase shift, which is proportional to the density of states, enables us to determine how the interaction affects the pressure. Hence, it is possible to properly consider both attraction or repulsion. The corresponding pressure within this approach is equivalent to the free gas in the small interaction limit. The GRG free gas in the YM sector of QCD is therefore justified by this method, which also gives the possibility of going beyond the free gas approximation.

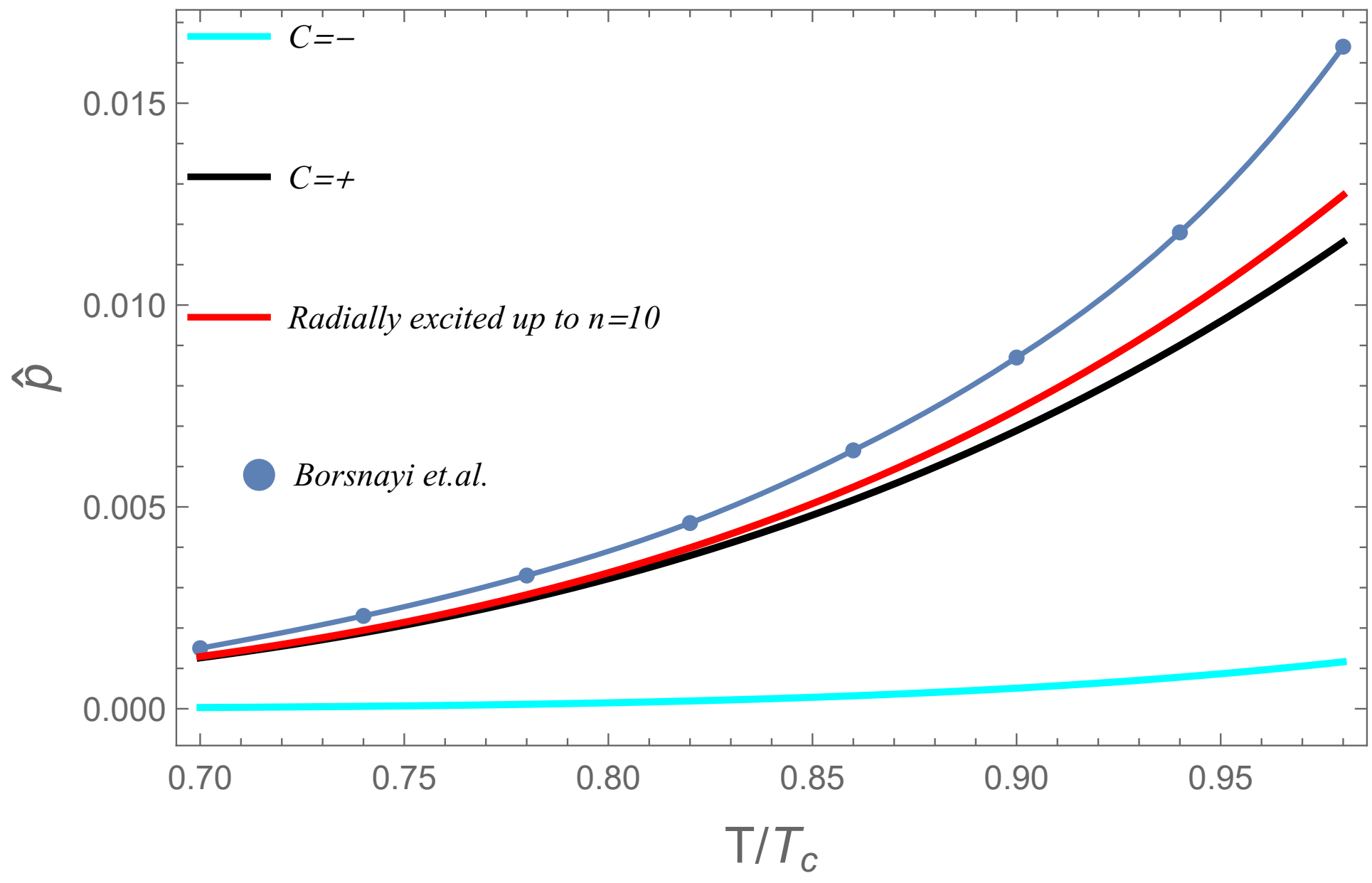

**Figure 6.5:** The GRG pressure depending on the negative and positive charge conjugation of the glueball spectrum compared to Ref. [30].

Considering the two lightest glueballs, the scalar glueball $G$ and the tensor glueball $G_2 \equiv G_2^{\mu\nu}$ within the following potential

$$V_{\text{eff}}(G, G_2) = V_{\text{dil}}(G) - \frac{\alpha}{2} G^2 G_{2,\mu\nu} G_2^{\mu\nu} \,, \tag{6.10}$$

where the dilaton potential $V_{\text{dil}}(G)$ describes the scalar glueball (see Eq. (6.13)). Recall the scale $G_0 \sim 0.4$ GeV, which resembles the trace anomaly of the YM theory, is the only dimensional parameter present in the effective potential. There are two dimensionless parameters $\lambda$ and $\alpha$ which are fixed by using the LQCD predictions for $m_G \sim 1.7$ GeV and $m_{G_2} \sim 2.4$ GeV. As a result of the following shift

$$G \longrightarrow G + G_0 \,, \tag{6.11}$$

we obtain the relation between the parameters of the potential and masses of the scalar and the tensor glueball fields in the following forms:

$$m_{0^{++}}^2 = m_G^2 = \lambda G_0^2 , \qquad m_{2^{++}}^2 = m_{G_2}^2 = \alpha G_0^2 . \tag{6.12}$$

The expansion of the dilaton potential around its minimum $G_0$:

$$V_{dil}(G) = -\frac{1}{16}G_0^4 + \frac{1}{2}m_G^2 G^2 + \frac{1}{3!}\left(5\frac{m_G^2}{G_0}\right)G^3 + \frac{1}{4!}\left(11\frac{m_G^2}{G_0^2}\right)G^4 + \cdots . \tag{6.13}$$

leads to the tree-level scattering amplitude based on the three-leg and four-leg interactions. The corresponding amplitude has the following form:

$$A(s,t,u) = -11\frac{m_G^2}{G_0^2} - \left(5\frac{m_G^2}{G_0}\right)^2 \frac{1}{s - m_G^2} - \left(5\frac{m_G^2}{G_0}\right)^2 \frac{1}{t - m_G^2} - \left(5\frac{m_G^2}{G_0}\right)^2 \frac{1}{u - m_G^2} . \tag{6.14}$$

The $l$-th amplitude, therefore, reads as follows after being rewritten as a function of the channel $s$ and the scattering angle $\theta$:

$$A_l(s) = \frac{1}{2}\int_{-1}^{1} d\cos\theta A(s,\cos\theta)P_l(\cos\theta) \ , \tag{6.15}$$

which is related to the phase shift $\delta_l(s)$ using the 3-momentum $k$ of any particle in the center of mass frame as:

$$\delta_l(s) = \frac{1}{2}\arg\left[1 + 2i\frac{k}{16\pi\sqrt{s}}A_l(s)\right] . \tag{6.16}$$

Two unitarization procedures (on-shell and $N/D$) are used to determine the phase shift in Refs. [253–256]. We consider these results within the following expression for pressure:

$$\hat{p}^{\text{int}}_{0^{++}0^{++}} = -\frac{1}{T^3}\sum_{l=0}^{\infty}\int_{2m_{0^{++}}}^{\infty} dx \frac{2l+1}{\pi}\frac{d\delta_l^{0^{++}0^{++}}(x)}{dx}\int\frac{d^3k}{(2\pi)^3}\ln\left(1 - e^{-\frac{\sqrt{k^2+x^2}}{T}}\right) + \hat{p}_B , \tag{6.17}$$

where the summation is over the angular momentum $l$, the integration variable $x$ is the square root of the Mandelstam variable $s$, and $k$ is the length of the three-momentum of the scalar glueball. The phase shift of l-th wave $\delta_l^{0^{++}0^{++}}(x)$ describes the scattering of two scalar glueballs with the quantum number $J^{PC} = 0^{++}$.

The pressure of a bound state of two scalar glueballs $\hat{p}_B$ that can eventually exist is also considered in Eq. (6.17). Using the dilaton potential to represent the interaction of scalar glueballs, as suggested by Ref. [254, 255, 257], a bound state of two scalar glueballs known as glueballonium does in fact exist, even though the pressure is not significantly affected by this state. This is because the pressure is a continuous function of the interaction strength; as a result, when a bound state emerges, there is no abrupt change in the pressure. The jump in the phase-shift contribution above the threshold is correlated with the jump in pressure that results from the formation of a bound state. This partial cancellation occurs whenever a bound state forms [252, 253, 258].

The phase shifts computed in Ref. [254] are implemented in Figure 6.6, which depicts the $l = 0, 2, 4$ pressure terms resulting from the interaction of two scalar glueballs. We use $m_{0^{++}} = 1.7$ GeV, which is consistent with Refs. [29,203] and $G_0 = 0.4$ GeV, which delivers a glueballonium with a mass of 3.34 GeV (for further information see Ref. [254]). Namely, the critical value to form a bound state with a mass of $2m_G$ is $G_{0,\text{crit}} = 0.504$ GeV. The mass of the bound state is decreasing for decreasing $G_0$ and exceeds $m_B = 3.34$ GeV for $G_0 = 0.4$ GeV. On the other hand, no glueballonium exists for $G_0 > G_{0,\text{crit}}$. This fact is included in the following expression of the pressure for the bound state:

$$\hat{p}_B = -\theta(\Lambda_{G,crit} - G_0)\frac{1}{T^3}\int \frac{d^3k}{(2\pi)^3} \ln\left(1 - e^{-\beta\frac{\sqrt{k^2+m_B^2}}{T}}\right). \tag{6.18}$$

The results for the s-wave ($l = 0$) are evidently the dominating ones. When increasing the value of $l$, the contribution rapidly becomes smaller and smaller, allowing us to approximate the total interaction with only the s-wave. While significantly greater than the other waves, the $l = 0$ wave contribution is still negligible when compared to the pressure of the $0^{++}$ non-interacting glueballs.

In order to describe the interaction between two tensor glueballs, we use an analogous method to two-pion scattering in Ref. [259]. A schematic description of the $s$-channel scattering is given below:

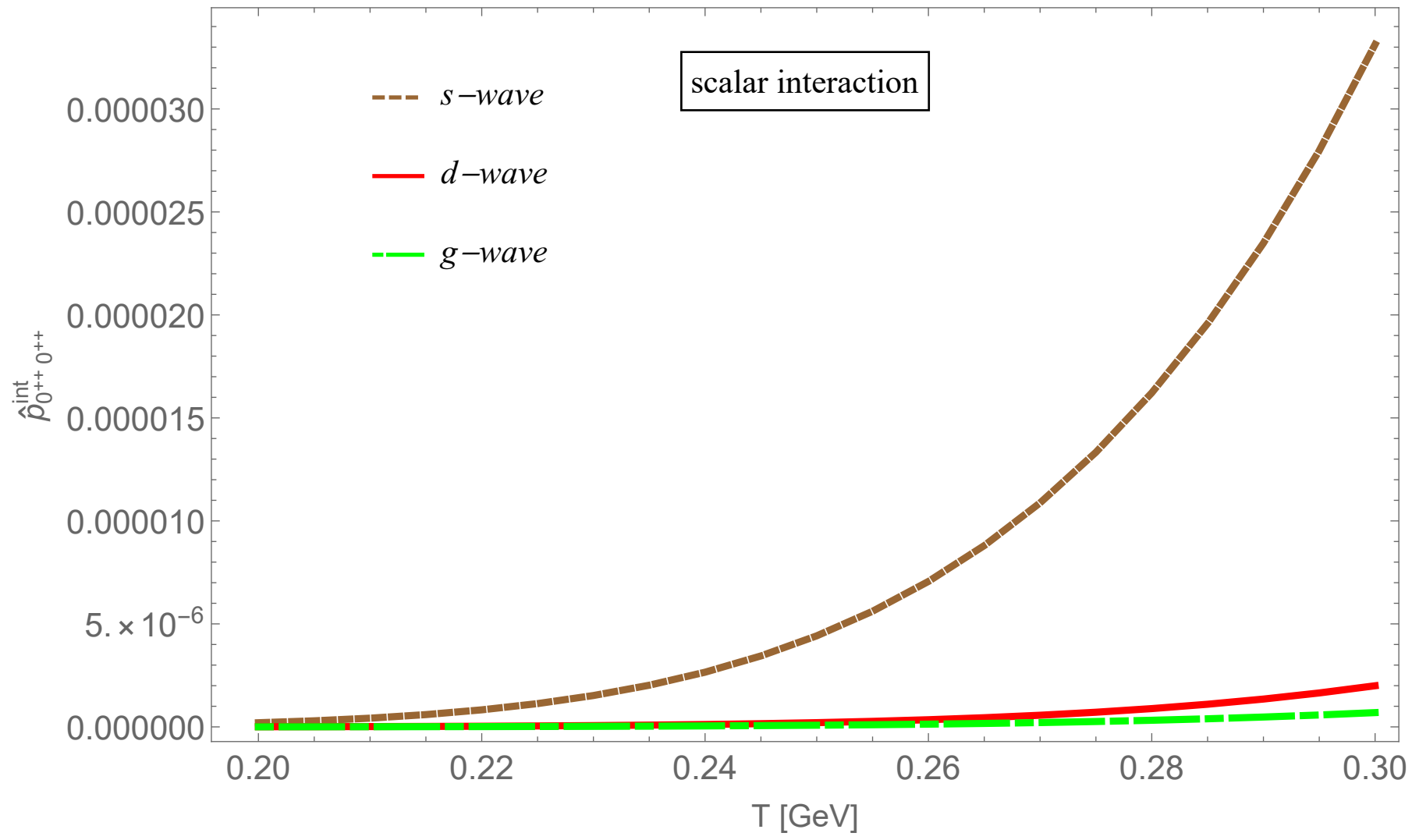

**Figure 6.6:** Contributions of three different waves to the pressure from scalar-scalar glueball interactions.

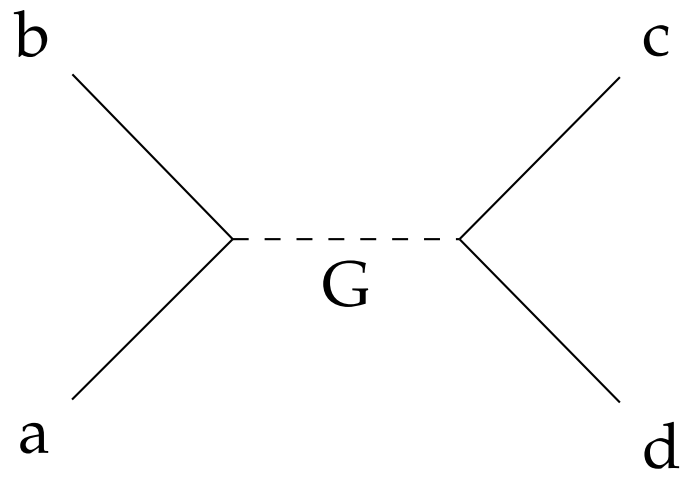

$$a + b \longrightarrow c + d$$

where the third component of the spin $J$ of the incoming and outgoing tensor glueballs is denoted by $a, b, c, d$. To be more specific, four particles can each have a unique value for the third spin component, $m_J$, provided that

$$m_J^{ab} \equiv m_J^a + m_J^b = m_J^{cd} \,. \tag{6.19}$$

There are analogous diagrams for the $t$ and $u$ channels.

A general expression for the total amplitude with the transition matrix $T$ reads:

$$\langle cd|T|ab\rangle = A\delta^{ab}\delta^{cd} + B\delta^{ac}\delta^{bd} + C\delta^{ad}\delta^{bc}, \tag{6.20}$$

and

$$A := \frac{-4(\alpha G_0)^2}{s - m_G^2}, \quad B := \frac{-4(\alpha G_0)^2}{t - m_G^2}, \quad C := \frac{-4(\alpha G_0)^2}{u - m_G^2} \,. \tag{6.21}$$

We obtain the expression for the initial state using the Clebsh-Gordan coefficients

$$|J^{ab}, m_J^{ab}\rangle = \sum_{m_J^a, m_J^b} \langle J^a\ m_J^a\ J^b\ m_J^b | J^{ab}\ m_J^{ab}\rangle |J^a, m_J^a\rangle \otimes |J^b, m_J^b\rangle\ . \tag{6.22}$$

Considering $J^a = J^b = 2$ and $m_J^a \in [-2, +2]$, $m_J^b \in [-2, +2]$ leads to

$$J^{ab} \in [0, 4]\,, \quad m_J^{ab} \in [-J^{ab}, +J^{ab}]\,. \tag{6.23}$$

Similarly, the outgoing state is expressed in the same compact form.

There are 25 nonzero amplitudes $\langle J^{cd}, m_J^{cd}|T|J^{ab}, m_J^{ab}\rangle$, for which $J^{cd} = J^{ab}$ and $m_J^{cd} = m_J^{ab}$ are satisfied. We introduce $T^q$, where the total incoming (and outgoing) spin $q \equiv J^{cd} = J^{ab} = 0, 1, 2, 3, 4$ in Eq. (6.20) as:

$$T^q = \sum_{i_a, i_b, i_c, i_d} \rho_{abcd} \langle i_c|\langle i_d|T|i_a\rangle|i_b\rangle\,, \quad \text{where} \quad i_a, i_b, i_c, i_d \equiv (I, II, III, IV, V). \tag{6.24}$$

Here $\rho_{abcd}$ is defined in terms of the Clebsh-Gordan coefficients

$$\rho_{abcd} := \langle 2\ m_J^a\ 2\ m_J^b | J^{ab}\ m_J^{ab}\rangle \cdot \langle 2\ m_J^c\ 2\ m_J^d | J^{cd}\ m_J^{cd}\rangle\ . \tag{6.25}$$

Note that we have used the short-hand notation

$$|i_a\rangle \otimes |i_b\rangle = |i_a\rangle|i_b\rangle, \tag{6.26}$$

where the value of any $\langle i_c|\langle i_d|T|i_a\rangle|i_b\rangle$ are represented in terms of the parameters $A$, $B$ and $C$, given in Eq. (6.21), and the relations between two basis read:

$$\begin{aligned}
|2, +1\rangle &= -\sqrt{\frac{1}{2}}\Big(|I\rangle - i|II\rangle\Big)\,, \\
|2, -1\rangle &= \sqrt{\frac{1}{2}}\Big(|I\rangle + i|II\rangle\Big)\,, \\
|2, +2\rangle &= -\sqrt{\frac{1}{2}}\Big(|III\rangle - i|IV\rangle\Big)\,, \\
|2, -2\rangle &= -\sqrt{\frac{1}{2}}\Big(|III\rangle + i|IV\rangle\Big)\,, \\
|2, 0\rangle &= |V\rangle\ .
\end{aligned} \tag{6.27}$$

We obtain the following results, which will be used to describe the interaction pressure between the tensor glueballs:

$$\boxed{T^4 = T^2 = B + C\,, \qquad T^3 = T^1 = B - C\,, \qquad T^0 = 5A + B + C}\,.$$

For tensor glueballs, we use the the analogous expression to Eq. (6.17):

$$\hat{p}^{\text{int}}_{2^{++}2^{++}} = -\frac{1}{T^3}\sum_{J=0}^{4}\sum_{l=0}^{\infty}\int_{2m_2^{++}}^{\infty} dx(2J+1)\frac{2l+1}{\pi}\frac{d\delta_l^{2^{++}2^{++},J}(x)}{dx}\int\frac{d^3k}{(2\pi)^3}\ln\left(1-e^{-\frac{\sqrt{k^2+x^2}}{T}}\right), \tag{6.28}$$

where the total spin $J$ of the system varies between 0 and 4 and $m_{2^{++}} = 2.4$ GeV istaken from Ref. [29, 203].

The phase shifts $\delta_l^{2^{++}2^{++},J}(x)$ are calculated by utilizing the effective theory for the tensor glueball we discussed above. Similar to the scalar glueball case, the result for the $s$-channel is larger than the others (see Fig. 6.7). We expect a small contribution in the case of interaction terms between tensor glueballs (e.g., a four-leg interaction).

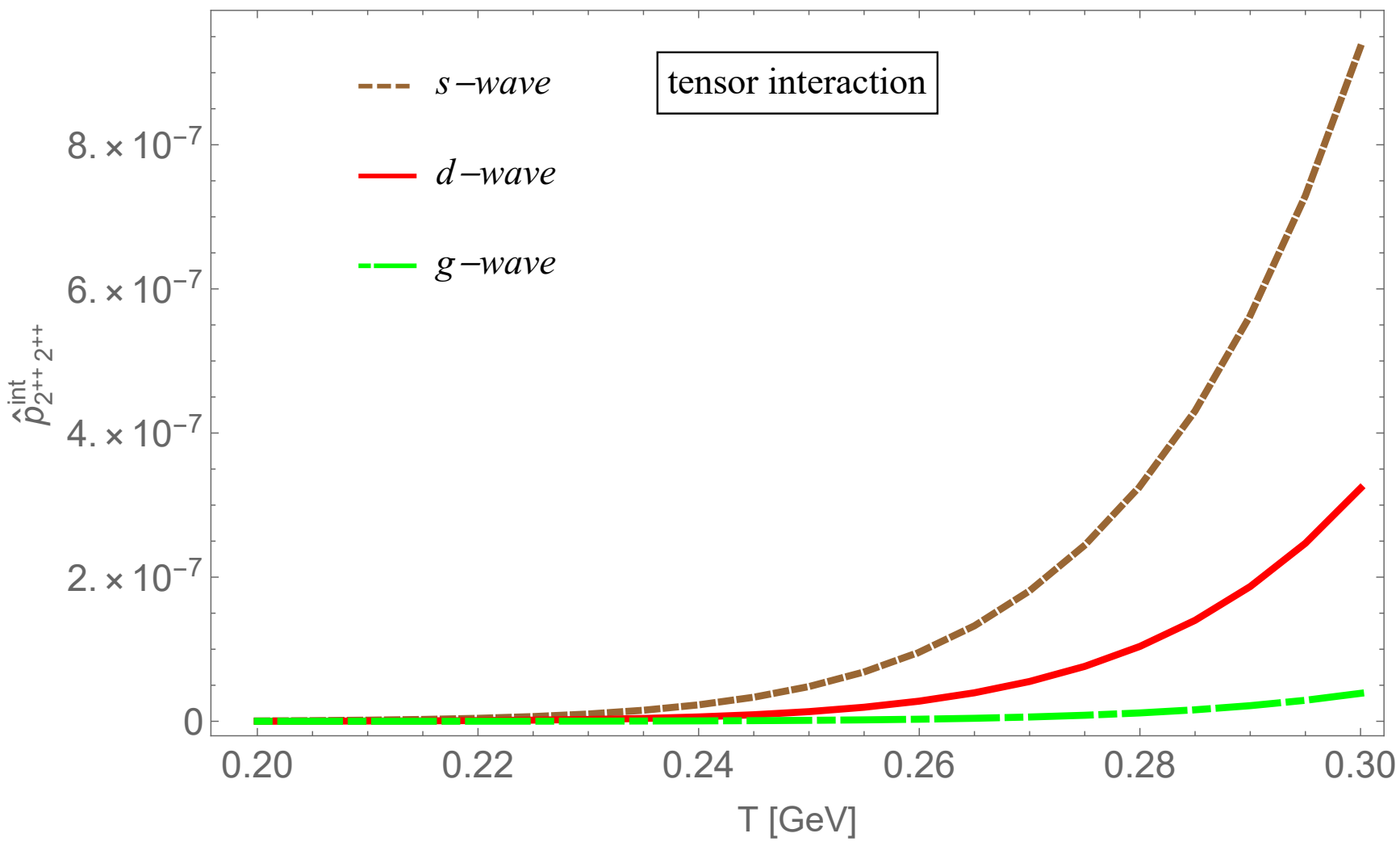

**Figure 6.7:** Contributions of three different waves to the pressure from tensor-tensor glueball interactions.

By adding together all possible terms, it is possible to determine the entire contribution to the pressure of interacting glueballs:

$$\hat{p}^{\text{int}} \approx \hat{p}^{\text{int}}_{0^{++}0^{++}} + \hat{p}^{\text{int}}_{2^{++}2^{++}} + \cdots\,, \tag{6.29}$$

where the first two terms are related to the interactions between the two lightest glueballs, while the additional terms are denoted by the dots referring to smaller contributions from interaction terms of heavier glueballs. We have seen that the tensor-tensor contribution is considerably smaller than that of the scalar-scalar one. By making use of the proper treatment of inelasticities and unstability of the states (see Refs. [245, 251, 253, 260]), one could extend the results to the scattering of any glueball states with $J_1$ and $J_2$.

In summary, the plots for the pressure and energy density would essentially show no impact from the combined effects of the scalar and tensor interaction components. In light of this, it appears that the free gas approximation (together with the eventual small contribution of heavy glueball states) is a reliable approach to the YM pressure for temperatures below $T_c$.

## 6.5 Conclusion

In this chapter, we revisited the GRG model by considering the recent glueball mass spectrum in Ref. [29] (as well as the previous results in Refs. [132, 203]). Including the further glueball states from Regge trajectories as well as the interaction between two light glueball states led to a small change in the thermodynamic quantities. We observe that the lattice data of Ref. [30] below the critical temperature (predicted to be around $T_c = 323 \pm 18$ MeV) is well explained by the GRG model.

# Chapter 7

# Conclusions and discussions

In this thesis, I have mainly explored two aspects of QCD: high-spin mesons within the low-energy mesonic spectrum and the thermodynamics of the Yang-Mills (YM) theory.

For the first part, I used an effective model based on chiral symmetry, called the extended Linear Sigma Model (eLSM). This model, previously used for lower spin mesons, was extended to spin-2 conventional $\bar{q}q$ mesons (tensor and axial-tensor mesons, linked by chiral transformations) and a tensor glueball. Finally, the eLSM applies also to $J = 3$ mesons, but the lack of information about $J^{PC} = 3^{++}$ implies that only flavor symmetry is implemented in this case.

Results for the ground state tensor mesons (with $J^{PC} = 2^{++}, 3^{--}$) confirm their quark-anti-quark structure. Qualitative agreement with PDG as well as LQCD results is obtained. Further interesting hadronic and radiative decay channels could be predicted and need to be tested in ongoing experiments (e.g., GlueX and CLAS12) and compared with LQCD simulation results.

Thanks to the eLSM, we were able to estimate the masses and decay widths of the axial tensor mesons (with $J^{PC} = 2^{--}$). Our model outcomes, as well as the LQCD results, predict that they are broad. The main decay channels are:

- decay widths of $\rho_2$ with the mass $m_{\rho_2} = 1663\,\text{MeV}$

$$\Gamma(\rho_2 \to \rho(770)\,\eta) = 99 \pm 50\,\text{MeV}\,,$$
$$\Gamma(\rho_2 \to \bar{K}^*(892)\,K + \text{c.c.}) = 85 \pm 43\,\text{MeV}\,,$$
$$\Gamma(\rho_2 \to \omega(782)\,\pi) = 419 \pm 210\,\text{MeV}\,;$$

- decay widths of $\omega_2$ with the mass $m_{\omega_2} = 1663$ MeV

$$\Gamma(\omega_2 \to \rho(770)\,\pi) = 1314 \pm 657\,\text{MeV}\,,$$
$$\Gamma(\omega_2 \to \bar{K}^*(892)\,K + \text{c.c.}) = 85 \pm 43\,\text{MeV}\,,$$
$$\Gamma(\omega_2 \to \omega(782)\,\eta) = 93 \pm 47\,\text{MeV}\,;$$

- decay widths of $\phi_2$ with the mass $m_{\phi_2} = 1971$ MeV

$$\Gamma(\phi_2 \to \bar{K}^*(892)\,K + \text{c.c.}) = 510 \pm 255\,\text{MeV}\,,$$
$$\Gamma(\phi_2 \to \phi(782)\,\eta) = 101 \pm 51\,\text{MeV}\,.$$

Next, we estimated the decay ratios for the spin-2 tensor glueball, which shows that the dominant decay channels are the ones into two vector mesons. Various spin-2 resonances listed in PDG are tested as possible tensor glueball candidates. Despite the large decay width, our model estimations agree with PDG for the resonance $f_2(1950)$. These outcomes could be interesting for experiments such as BESIII, and LHCb searching for the tensor glueball.

For what concerns the thermodynamics of the YM theory, we revisit the Glueball Resonance Gas (GRG) model by using the glueball spectrum obtained in 3 different LQCD simulations. We have compared the model results for the pressure (and other thermodynamic quantities) with the corresponding LQCD results of the pressure for pure YM theory below the critical temperature of deconfinement (glueballs $\to$ gluons). The GRG model was improved by including the contribution from further excited states and from the interaction between the two lightest glueballs (scalar and tensor ones). These additional contributions turned out to be negligible, which supports the validity of the free gas description of the glueball states as a valid approximation.

As an outlook of this thesis, one can go beyond the hadronic degrees of freedom. Namely, one can study the effect of quarks (quark zero modes) to investigate some interesting problems for mesons. For instance, we know that the large mixing between $\eta$ and $\eta'(958)$ is related to the instantons. Analogously, one can study the role of the instanton in the mixing between the spin-2 mesons ($J^{PC} = 2^{-+}$) $\eta_2(1645)$ and $\eta_2(1870)$. In this case, the instanton-induced interactions will lead to hadronic effective Lagrangians of the type considered in this thesis. Yet, the coupling constant could be computed by using a microscopic instanton-based calculation. This approach can be useful to estimate the size and signature of the mixing angle between the isoscalar mesons with $J^{PC} = 1^{+-}, 2^{-+}$, as well as to evaluate novel decay modes.

# List of figures

# List of tables